\documentclass[twocolumn,citeautoscript,pra,superscriptaddress,amsmath,amssymb,longbibliography]{revtex4-1}
 
\usepackage{graphicx}
\usepackage{color}
\usepackage{upgreek}
\usepackage{float}
\usepackage{lipsum}
\usepackage{mathrsfs}
\usepackage{booktabs}
\usepackage{enumitem}
\usepackage{url}
\usepackage[colorlinks,citecolor=blue,linkcolor=blue]{hyperref}
 
\newcommand{\ket}[1]{\ensuremath{\left| #1 \right\rangle}}

\DeclareMathOperator*{\argmin}{\arg\!\min}
\newcommand{\Tr}[0]{\ensuremath{\text{Tr}}}
\newcommand{\kket}[1]{\ensuremath{\left| #1 \right\rangle\!\rangle}}
\newcommand{\bbra}[1]{\ensuremath{\left\langle\!\langle #1 \right|}}

\newcommand{\pt}{$\Upsilon_{k:0}$}

\newtheorem{definition}{Definition}

\usepackage[most]{tcolorbox}
\usepackage{xparse}
\usepackage{lipsum}

\begin{document}

\title{Unifying non-Markovian characterisation\\with an efficient and self-consistent framework}
\author{G. A. L. White}
\email{greg.al.white@gmail.com}
\affiliation{Dahlem Center for Complex Quantum Systems, Freie Universit\"at Berlin, 14195 Berlin, Germany}
\affiliation{School of Physics and Astronomy, Monash University, Clayton, VIC 3800, Australia}
\affiliation{School of Physics, University of Melbourne, Parkville, VIC 3010, Australia}

\author{P. Jurcevic}
\affiliation{IBM Quantum, IBM T. J. Watson Research Center, Yorktown Heights, NY 10598, USA}

\author{C. D. Hill}
\affiliation{Silicon Quantum Computing, The University of New South Wales, Sydney, New South Wales 2052, Australia}
\affiliation{School of Physics, University of Melbourne, Parkville, VIC 3010, Australia}
\affiliation{School of Mathematics and Statistics, University of Melbourne, Parkville, VIC, 3010, Australia}

\author{K. Modi}
\email{kavan\_modi@sutd.edu.sg}
\address{Science, Mathematics and Technology Cluster, Singapore University of Technology and Design, \\8 Somapah Road, 487372 Singapore}\vspace{0.25em}
\address{School of Physics and Astronomy, Monash University, Clayton, Victoria 3800, Australia}

\begin{abstract}
Noise on quantum devices is much more complex than it is commonly given credit. Far from usual models of decoherence, nearly all quantum devices are plagued both by a continuum of environments and temporal instabilities. These induce noisy quantum and classical correlations at the level of the circuit. The relevant spatiotemporal effects are difficult enough to understand, let alone combat. There is presently a lack of either scalable or complete methods to address the phenomena responsible for scrambling and loss of quantum information. Here, we make deep strides to remedy this problem. We establish a theoretical framework that uniformly incorporates and classifies all non-Markovian phenomena. Our framework is universal, assumes no parameters values, and is written entirely in terms of experimentally accessible circuit-level quantities. We formulate an efficient reconstruction using tensor network learning, allowing also for easy modularisation and simplification based on the expected physics of the system. This is then demonstrated through both extensive numerical studies and implementations on IBM Quantum devices, estimating a comprehensive set of spacetime correlations. Finally, we conclude our analysis with applications thereof to the efficacy of control techniques to counteract these effects -- including noise-aware circuit compilation and optimised dynamical decoupling. We find significant improvements are possible in the diamond norm and average gate fidelity of arbitrary $SU(4)$ operations, as well as related decoupling improvements in contrast to off-the-shelf schemes.
\end{abstract}

\maketitle


\section{Introduction}
As quantum devices inch closer to fault tolerance~\cite{postler2022demonstration,kim2023evidence,abobeih2022fault,noiri2022fast,google2023suppressing,bluvstein2023logical}, it becomes increasingly necessary to tame the dynamics that will be most relevant to error correction and the performance of quantum algorithms: correlated and coherent noise~\cite{proctor2022measuring,aharonov1997fault,huang2020classical,Clader2021,correlated-qec}. Not all errors are equal, and rather than treating them as such, it is shrewd to consider hardware-aware approaches for the realisation of clean quantum computers~\cite{bonilla2021xzzx,tuckett2018ultrahigh,PRXQuantum.4.040311,nautrup2019optimizing,farrelly2021tensor}. 
One necessary aspect here is to have a robust pipeline to translate pathological dynamics from a noisy quantum device into interpretable and useful information for the stated goals. \par
Owing to the diversity of approaches in the development of high-fidelity quantum devices, it is essential to have robust strategies for \emph{operational} characterisation that do not depend on the specifics of the hardware. That is, an operational model should capture the essential logical information in a manner independent of the underlying physics. Such characterisation techniques will play a central role in the development of practical quantum error correction schemes, which will be subjected to noise spanning several orders of magnitude in time scales.
Moreover, operational characterisation can also be used for reporting device quality~\cite{eisert2020quantum}, help to diagnose fabrication issues~\cite{white2023filtering}, inform optimal control techniques~\cite{PhysRevLett.126.200401,butler2023optimizing,Ball_2021,PRXQuantum.2.030315}, and be fed forward to noise-aware quantum error mitigation strategies and error correction protocols~\cite{PRXQuantum.3.010345, tuckett2018ultrahigh,wang2023dgr}. 
It is equally important for these models to be expressive. They should capture the real dynamics seen on real quantum devices.
Currently, there is a disparity between behaviour that be accurately modelled, and the malignant behaviours that are exhibited in actuality. Chief among these complex effects is noise that may be correlated temporally or spatiotemporally -- otherwise known as non-Markovian.

Superconducting qubits, for example, tend to have high fidelity gates with stable power spectra, by virtue of microwave frequency control. However, qubits interact substantially with environmental features, including higher energy levels, two-level system defects, $ZZ$ coupling crosstalk, and electric field fluctuations~\cite{wei2022hamiltonian,muller2019towards,wilen2021correlated, Dial_2016}.
In contrast, ion traps are nearly completely isolated from their surrounding environment, but much of the noise experienced is control-based owing to the instability of optical addressing~\cite{RevModPhys.87.1419,ParradoRodriguez2021crosstalk}. 
Spin qubits, meanwhile, have both $1/f$ charge noise as well as possible hyperfine coupling to nearby nuclear spins~\cite{kuhlmann2013charge,yoneda2022noise,rojas2023spatial}. 
Similar can be said for interacting with a complex environment in cold atoms~\cite{wu2021concise}, bosonic systems~\cite{Terhal_2020}, and even Majorana qubits~\cite{PhysRevB.97.054508}.

When considering non-standard error channels, the phrase ``non-Markovianity'' has run somewhat rampant in the quantum characterisation, verification, and validation (QCVV) literature, signifying any dynamics that might \emph{not} fit a Markov model, including time-dependent Markovian processes, within the practicalities of characterisation. Although semantically correct in most instances, this coarse-graining of information can be problematic, since both tools and formalisms may be aimed at drastically different physical mechanisms. 
Employing the same terminology gives the illusion that various techniques may be inapplicably mixed and matched. The problem is compounded when considering non-Markovian notions from a master equation vantage point~\cite{PhysRevLett.101.150402,Rivas_2014}.

This raises the question of whether different non-Markovian errors are discernible, (easily) detectable, and practically controllable. The downstream effects from results are varied. For instance, memory effects due to an interaction with an environment can (in principle) be removed with carefully considered live control~\cite{berk2021extracting}, supplied to decoders~\cite{chubb2021statistical}, or may prompt the design of bespoke error correction codes around the hardware~\cite{mauron2023optimization,su2023discovery}.
Long-time device instability, in contrast, is less pertinent to single-shot error reduction methods and more relevant -- for instance -- to error mitigation techniques where data about observables are aggregated over an extended period of time~\cite{cai2023quantum}.
All, of course, are interesting to know from a device fabrication point of view.

Process tensor tomography (PTT) has emerged as a method to characterise a large class of non-Markovian dynamics, and has been employed in a variety of scenarios to understand noisy quantum devices~\cite{White-NM-2020,White-MLPT,white2021many,white2023filtering,giarmatzi2023multitime,PhysRevA.102.062414,PhysRevA.106.022411,Li_NMGST}. A process tensor is an operational description of quantum stochastic processes, and its reconstruction provides all of the relevant information about multi-time correlations arising from uncontrollable dynamics~\cite{Pollock2018a,Milz2021PRXQ}. 
However, in its present form, PTT assumes perfect gates and characterises only the correlated interactions between a system and its bath. 
The effects of gate-dependent noise -- such as coherent error -- are hence neglected. This can be addressed by including gates within the model in addition to the process tensor~\cite{Li_NMGST}. But more significantly, neither non-Markovian correlations within the control electronics, nor active leakage are captured by this technique. Moreover, experimental requirements are currently too expensive for the diagnostic results to be scalably fed forward into the removal of noise.

Here, we aim to corral these separate notions of non-Markovianity into a single \emph{self-consistent} framework, with a solution to characterise their dynamics in both an efficient and a modular manner. 
The results herein are not only applicable to the study of quantum noise, but any open quantum system. For example, a quantum sensor participating in a non-Markovian interaction will benefit from this method of learning. 
Conceptually, our work also contributes to an expanded definition of quantum stochastic processes which incorporates (unknown) stochastic effects within probes of quantum dynamics. This can include correlated parameters at the control level, or process-control interplay, whereby the application of a probe affects the bath dynamics itself. \par Thus, we systematically broaden the notion of non-Markovian quantum stochastic processes to cover the realistic case where at least some part of the set of instruments available to an experimenter form an unknown environment in their own right.
These claims are demonstrated in our results whereby we perform high-fidelity and large-scale characterisation of a variety of simulated non-Markovian noise modesl. We then showcase the ability to map out spatiotemporal correlations on IBM Quantum devices emergent entirely from the noise. The modularity of our methods are also demonstrated, where different components of the noise model may be straightforwardly rearranged according to hardware idiosyncrasies, such as expected physics or native gate sets. Lastly, we collect substantial evidence on the extent to which this detailed characterisation can improve control of superconducting quantum devices via two-qubit gate decompositions and optimised dynamical decoupling sequences.
It is our hope that this detailed and comprehensive framework of non-Markovian noise will be conducive not only in refining error-suppression strategies across various quantum computing platforms, but also in guiding future theoretical and experimental investigations into the subtle interplay between quantum systems and their complex, correlated environments.

\section{Background and \\
Summary of Results}
\label{sec:background}
In this work, we present a series of results that may appeal to a broad church of the quantum computing community. At a conceptual level, we introduce a new and comprehensive theory of non-Markovian processes as it pertains to controlled quantum systems -- e.g., quantum computers. Our theory is said to be self-consistent as it accounts for both the background and the control noise with testable hypotheses.
Using this framework, we derive concrete methods for determining all relevant spatiotemporal noisy effects, which can be partitioned into experimentally relevant components. We make this framework efficient with the use of tensor networks; this advances the state-of-the-art for scalability not only in non-Markovian noise characterisation, but in all self-consistent Markovian noise estimation. These claims are supplemented with extensive numeric and experimental evidence as to their efficacy, and we demonstrate how their implementation can be used to achieve significantly enhanced control of noisy quantum devices. Our work therefore encompasses a gamut of results in theoretical, numeric, and experimentally-relevant cases pertaining to open quantum systems.

However, it is unlikely that any single reader would desire to read this work in a start-to-finish thread. We therefore start off with the present section, which is intended to serve as a bird's eye view of our work as well as its appropriate context. In particular, we proceed by detailing the structure of our work; provide relevant background regarding process tensors and multi-time processes; offer an overview of our conceptual and methodological advances; a sketch of implementation details; and lastly, a reading guide to indicate where different members of the quantum information communities might find this manuscript most helpful to parse.

The paper organised as follows, and designed to be partitioned appropriately for readers of varying backgrounds and interests. In particular: Part 1 (Section~\ref{sec:applied-results}) of our main body presents the applied results, to showcase the expressiveness and efficacy of our framework, as well as its hardware requirements and experiment design. This is then followed by Part 2 (Section~\ref{sec:technical-methods}), wherein we elucidate technical methods and the conceptual details of our non-Markovian framework. 
\par

Within Section~\ref{sec:applied-results}, we demonstrate that it is possible to accurately and efficiently characterise malignant non-Markovian noise in a host of synthetic and experimental settings. These go beyond what has previously been possible both in scale and in expressivity of non-Markovianity.
In Section~\ref{sec:synth-demonstrations}, we demonstrate this self consistent characterisation across a variety of synthetic noise models, including exchange interactions with nuclear spins; coherent and $1/f$ gate noise; and spillage from control operations into environment dynamics.
This showcases an ability to characterise a wide variety of noise models for which there is no operational equivalent in the literature. Indeed, we claim that these results paint a more or less complete picture of non-Markovian noise.
This is followed up in Section~\ref{ssec:nisq-tn} by a series of demonstrations on IBM Quantum devices, where we capture and quantify spatiotemporally correlated dynamical effects. 
We lastly turn these tools to applied considerations in Section~\ref{sec:applications}. Specifically, we consider two relevant problems in circuit design (i) optimal noise-aware decomposition from high level unitaries into primitive gate sets, and (ii) optimised dynamical decoupling sequences in the presence of unknown noise. 
It is our intention to demonstrate that non-Markovian process characterisation can be both lightweight and comprehensive, while maintaining all of the critical information relevant to benchmarking and optimally controlling a quantum device. 

\begin{figure}[!t]
	\centering
	\includegraphics[width=\linewidth]{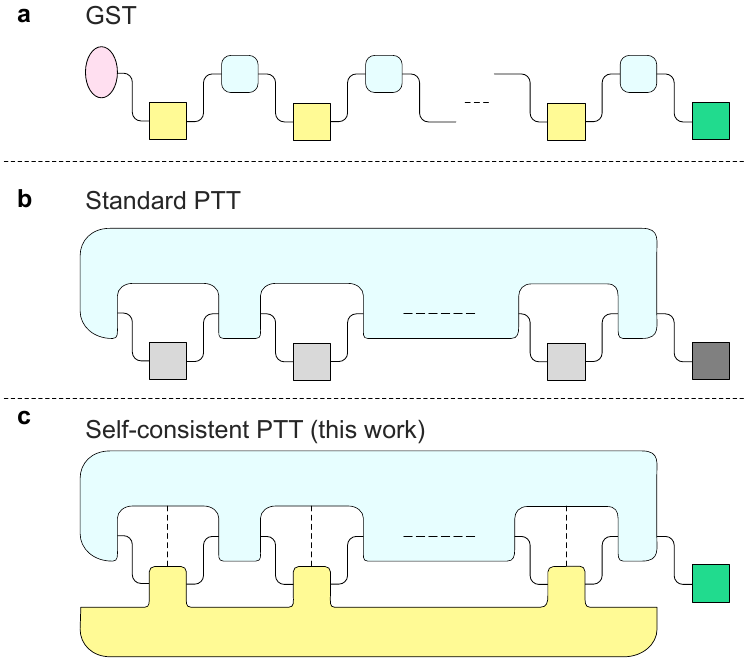}
	\caption[A comparison of various models for estimating dynamics Markovian and non-Markovian dynamics ]{A comparison of various models for estimating dynamics. Colour indicates an object which is estimated, grey indicates that object is taken to be known. \textbf{a} GST estimates a set of time-local gates with some time-local background process. \textbf{b} PTT models a generically correlated background process, but assumes gates to be known. \textbf{c} The self-consistent protocol introduced in this work estimates all objects simultaneously, and allows for temporal correlations in both the background process and in the control instruments. }
	\label{fig:NMGST-buildup}
\end{figure}

These applied results are possible to the technical advances detailed in Section~\ref{sec:technical-methods}, where we go into the analytical methods and design of our framework.
These details are presented for the interested reader who wishes to understand the implementation, or replicate or build on our results. 
The core conceptual component comprises Section~\ref{sec:categories}. This is a set of definitions motivating the philosophy of our work. We aim to establish a categorisation of correlated dynamics. We intend this to be useful both to distill and sub-divide the meaning of `non-Markovian' in practice. 
Note that this is not intended to be a definitive hierarchy as such, but rather a categorisation that shall serve as a useful structure in anticipation of our estimation techniques. Our categorisation stems from intuition and detailed studies of the literature on quantum noise. Its power is highlighted by the subsequent technical results that we reach.
In Sections~\ref{sec:PT-LPDOs} and~\ref{sec:self-consistency}, we construct an efficient, modular, and self-consistent characterisation procedure for estimating arbitrary correlated quantum dynamics by first deriving it in full generality, and then compressing it using tensor networks. This renders the total model size in palatable forms without sacrificing the above generality or interpretability.

Launching into Part 1 -- without discussing the background and the technical achievements of the Part 2 of this paper -- would be simply confusing, to say the least. Thus, we will first provide a high-level overview of the background, which include the \textit{gate set tomography} (GST) framework for self-consistent characterisations of quantum noise and the \textit{process tenor tomography} (PTT) framework to characterise non-Markovian quantum noise. These are depicted in panels a and b of Figure~\ref{fig:NMGST-buildup}, respectively. Notably, GST is self-consistent but assumes Markovian and time-independent noise. PTT, on the other hand, can handle non-Markovian noise but is not self-consistent. This paper closes that gap by developing a self-consistent PTT procedure, as depicted in Figure~\ref{fig:NMGST-buildup}c. We also, at a high level, summarise the technical tools and the advances that enable self-consistent non-Markovian tomography to be practically implemented in this section. With that, we will be able to dive into Part 1 of this paper to display the high efficacy of our tools. Part 2 will then supply a deeper dive into the conceptual and technical details of our work for the interested reader. The structure of our work is sketched out more comprehensively in Figure~\ref{fig:noise-categorisation}.

\subsection{Background on Multi-time Processes}

The core contributions in this work build on a formalisation of quantum stochastic processes within the framework of process tensors~\cite{Pollock2018a}, also sometimes known in the literature as quantum combs~\cite{Chiribella2009TheoreticalNetworks} or process matrices~\cite{1367-2630-18-6-063032}, among other names in different contexts. To start, we give an overview of these objects and their estimation within the PTT framework~\cite{White-MLPT}.
\subsubsection{Process Tensors}
Quantum channels account for two-time correlations, between the input and output states. In general, a quantum experiment may have a sequence of control operations applied to the system at times $\mathbf{T}_k:=\{t_0,\cdots, t_k\}$. Such a process is represented as a multilinear map from a sequence of controllable operations on the system to a final state density matrix. Here, the map, known as a process tensor~\cite{Pollock2018a, 1367-2630-18-6-063032}, represents all of the uncontrollable dynamics of the process, depicted as the blue comb in Figure~\ref{fig:NMGST-buildup}b/c. 

Process tensors formally generalise quantum channels to many-time processes.
Namely, we are interested in multi-time quantum correlations to generalise the two-time correlations embodied by quantum channels. To be precise, we consider the situation where a $k$-step process is driven by a sequence $\mathbf{A}_{k-1:0}$ of control operations, each represented mathematically by completely positive (CP) maps: $\mathbf{A}_{k-1:0} := \{\mathcal{A}_0, \mathcal{A}_1, \cdots, \mathcal{A}_{k-1}\}$, after which we obtain a final state $\rho_k(\mathbf{A}_{k-1:0})$ conditioned on this choice of interventions. 
These controlled dynamics have the form:
\begin{equation}\label{eq:multiproc}
        \rho_k\left(\textbf{A}_{k-1:0}\right) = \text{tr}_E [U_{k:k-1} \, \mathcal{A}_{k-1} \cdots \, U_{1:0} \, \mathcal{A}_{0} (\rho^{SE}_0)],
\end{equation}
where $U_{k:k-1}(\cdot) = u_{k:k-1} (\cdot) u_{k:k-1}^\dag$ and $\mathcal{A}_{j}$ is the CP map applied at time $j$.

The process tensor $\mathcal{T}_{k:0}$ is defined as the mapping from past controls $\mathbf{A}_{k-1:0}$ to future states $\rho_k\left(\textbf{A}_{k-1:0}\right)$:
\begin{equation}
\label{eq:PT}
\mathcal{T}_{k:0}\left[\mathbf{A}_{k-1:0}\right] = \rho_k(\mathbf{A}_{k-1:0}).
\end{equation}
Note that although this is suggestive of only obtaining the state of a system at time $t_k$, since the $\{\mathcal{A}_j\}$ are generalised quantum instruments, the process tensor maps to measurements and states at any time defined on the process.

It is usually convenient to work with the Choi state of the process tensor. A generalisation of the Choi-Jamio\l kowski isomorphism (CJI) allows for process tensors to be represented as many-body quantum states~\cite{Pollock2018a}. This is in connection with Choi states of the control operations, which we denote with a caret $\hat{\mathcal{A}}_j$. 
In this picture, at each time, one half of a fresh Bell pair is swapped to interact with the environment. The mixed state at the end, $\Upsilon_{k:0}$ is the process tensor Choi state. This can be used to produce the action of the process tensor on any controlled sequence of operations (consistent with the time steps) by projection:
\begin{equation}\label{eq:PToutput}
    \rho_k(\mathbf{A}_{k-1:0}) \!=\! \text{Tr}_{\overline{\mathfrak{o}}_k} \! \left[ (\mathbb{I}_{\mathfrak{o}_k}\otimes \hat{\mathcal{A}}^\text{T}_{k-1}\otimes \cdots \otimes \hat{\mathcal{A}}_0^\text{T} ) \Upsilon_{k:0} \right],
\end{equation}
where $\overline{\mathfrak{o}}_k$ indicates every index except $\mathfrak{o}_k$. This equation is inclusive of all intermediate $SE$ dynamics as well as any initial correlations, and illustrates how sequences of operations constitute observables of the process tensor.

The Choi state is an operator on multipartite Hilbert spaces as
\begin{equation}
    \Upsilon_{k:0} \in \mathcal{B}(\mathcal{H}_{\mathfrak{o}_k} \otimes \mathcal{H}_{\mathfrak{i}_k} \otimes \mathcal{H}_{\mathfrak{o}_{k-1}} \otimes \ldots \otimes \mathcal{H}_{\mathfrak{i}_{1}} \otimes \mathcal{H}_{\mathfrak{o}_0}).
\end{equation}
Each time has an associated output ($\mathfrak{o}$) and input ($\mathfrak{i}$) leg from the process. If the dynamical process is non-Markovian, then the system-environment interactions will distribute temporal correlations as spatial correlations between legs of the process tensor at different times. Moreover, each index at a single time may be separated into subsystems to denote the purely spatial separation (for example, a register of qubits). This allows spatiotemporal correlations to then be probed using any number of established quantum or classical many-body tools. 

Although a control sequence $\{\mathcal{A}_0,\mathcal{A}_1,\cdots,\mathcal{A}_{k-1}\}$ forms a dual object to process tensors, it is not the most general form of this dual object. Indeed, control sequences themselves may embed (in principle, known) multi-time correlations.
For a given $\mathbf{A}_{k-1:0}$, we use $\hat{\mathbf{A}}_{k-1:0}$ as shorthand for the collective Choi state of the full sequence. This can then be expressed as
\begin{equation}
    \hat{\mathbf{A}}_{k-1:0} = \sum_{\sigma}\alpha_\sigma \hat{\mathcal{A}}_{k-1}\otimes \hat{\mathcal{A}}_{k-2}\otimes\cdots\otimes \hat{\mathcal{A}}_0,\quad\alpha_\sigma\in\mathbb{R},
\end{equation}
where the index $\sigma$ and the amplitudes $\alpha_\sigma$ account for correlations in time.
Such an object is known as a \emph{tester} and can be viewed physically as controlled repeated interactions with an auxiliary system as well as possible recorded measurements.

\subsubsection{Process Tensor Tomography}
\label{methods:PTT}

The process tensor framework is built to access many-time quantum correlations through the experimental access of operational quantities.
Recently, some of the authors introduced PTT to facilitate the estimation of many-time correlations on real devices~\cite{White-NM-2020, White-MLPT, white2021many}. In essence, sequence of control operations constitute linear observables on multi-time processes. A quorum of observables hence uniquely determines the process tensor. PTT is a methodology prescribing both the sequence of experiments and the procedure to estimate the process tensor representing the dynamics observed in the laboratory.

For exposition, we express $\mathbf{A}_{k-1:0}$ as a simple sequence of operations: $\hat{\mathbf{A}}_{k-1:0} = \bigotimes_{i=0}^{k-1}\hat{\mathcal{A}}_i$. Here, the tensor product structure is a consequence of selecting these operations independently. From this, one obtains a conditional final state $\rho_k(\mathbf{A}_{k-1:0})$ as per Equation~\eqref{eq:PToutput}. Suppose now that the final state of the system is measured with an informationally complete (IC) apparatus, described by positive operator-valued measure (POVM) $\{\Pi_x\}$. The output distribution, conditioned on these operations and final measurement, is then given by the spatiotemporal generalisation of Born's rule~\cite{Shrapnel_2018}:
\begin{equation}
    \label{eq:PT_action}
    p_{x|\mathbf{A}_{k-1:0}} = \text{Tr}\left[\Upsilon_{k:0}(\Pi_x \otimes  \hat{\mathbf{A}}_{k-1:0}^{\text{T}})\right].
\end{equation}
This generalises joint probability distributions of a classical stochastic process to the quantum regime~\cite{Milz2020}.
Indeed, from Equation~\eqref{eq:PT_action}, one can see that a sequence of ccontrol operations constitutes a `measurement' on the multi-time process. 
Operations at each step may be deterministically applied -- such as a unitary operation -- or stochastic, such as a measurement and feed-forward or more general quantum instruments. 
Note that many-time processes suffer from the same dimensionality curse as states and multipartite classical distributions, and indeed may be as complex as quantum states~\cite{aloisio-complexity}. Here, with the number of timesteps, the number of histories to account for grows exponentially. 

To use this information to reconstruct a process tensor $\Upsilon_{k:0}$, the first step is pick an informationally complete basis for the terminating measurement and $\{\Pi_x\}$ and the quantum operations $ \{ \mathcal{B}_j^{\mu_j}\}$. Here, the $j$ index denotes the relative time of operation, and $\mu_j$ indexes the basis elements at that particular time. The $\mathcal{B}_j^{\mu_j}$ can be any CP map applied to the system, such as a unitary gate or a measurement and re-preparation.
If we can estimate the corresponding outcome probabilities, $p_{x,\vec{\mu}}$, then, via a set of linear equations, uniquely fixes $\Upsilon_{k:0}$. That is, applying Eq.~\eqref{eq:PT_action} to the basis elements yields
\begin{equation}
    p_{x,\vec{\mu}} = \text{Tr}\left[ \Upsilon_{k:0} (\Pi_x\otimes \hat{\mathcal{B}}_{k-1}^{\mu_{k-1}\text{T}}\otimes \cdots \otimes \hat{\mathcal{B}}_0^{\mu_0\text{T}}) \right].
\end{equation}
Alternatively, if the probabilities are known only for some subspace of operations (such as the vector space of the span of unitary operations), then the process tensor will be uniquely fixed on that subspace. 

In reality, the measured outcomes $n_{x,\vec{\mu}}$ are noisy estimates of the `true' probabilities, and there is often no physical process tensor that completely matches the data. For this reason -- as is common -- we treat estimation of a process as an optimisation problem where a model for the process is fit to the data. Specifically, a unique $\Upsilon_{k:0}$ is found by minimising the log-likelihood
\begin{equation}
    f(\Upsilon_{k:0}) = \sum_{x,\vec{\mu}}-n_{x,\vec{\mu}}\ln p_{x,\vec{\mu}},
\end{equation}
where $\Upsilon_{k:0}$ is a positive matrix obeying a set of causality conditions. This optimisation is carried out using a projected gradient descent algorithm, where at each step the model is iterated in a direction that decreases the log-likelihood, but always constrained to lie on the manifold of positive, causal states. More details can be found in Ref.~\cite{White-MLPT}.

Once an estimate of the process tensor is found, a prediction can be made for the resulting state conditioned on some arbitrary sequence as per Equation~\eqref{eq:PToutput}. This defines a natural measure for goodness-of-fit for an estimate, which we call reconstruction fidelity. Effectively, this is testing data for the model. One can generate a set of gate sequences $\{\mathbf{G}_{k-1:0}^j\}$ and experimentally reconstruct $\{\rho^{\text{(exp)}}_k(\mathbf{G}_{k-1:0}^j)\}$ from the quantum device. 
An estimated $\Upsilon_{k:0}$ is then used to predict the set of states $\{\rho^{\text{(pred)}}_k(\mathbf{G}_{k-1:0}^j)\}$. Reconstruction fidelities are then the element-by-element Uhlmann fidelities between $\rho^{(\text{exp})}$ and $\rho^{(\text{pred})}$. These are used to certify that the model does indeed explain the multi-time dynamics.

\subsubsection{Self-Consistent Tomography}
\label{subsec:sctomo}

Tomographic protocols in quantum devices may be broadly categorised on the premise of whether or not they are self-consistent. A method is termed self-consistent if it does not require perfect \emph{a priori} knowledge about control elements. Note that this does not imply that no assumptions are required at all. Randomised benchmarking (RB) is a simple example where this is true: properties from a series of unknown gate elements are extracted purely from an experimentally-determined decay curve. 
In contrast, quantum process tomography (QPT) aims to estimate an unknown channel $\Lambda$ via the measurement of a \emph{known} IC set of states $\{\rho_{\text{in}}^{(i)}\}$ and IC measurement $\{\Pi_x\}$. In the presence of substantial state preparation and measurement (SPAM) errors, the estimation of $\Lambda$ can differ drastically from experimental reality~\cite{greenbaum2015introduction}.

Most famously in this context, gate set tomography (GST), introduced across Refs.~\cite{PhysRevA.87.062119,RBK2017,Nielsen2021gatesettomography}, is a self-consistent extension to QPT. The model features a \emph{gate set} $\mathcal{G} := \{\kket{\rho},\bbra{\Pi_0},\cdots, G_0, G_1,\cdots\}$ which incorporates state preparation, measurement operators, and an IC set of gates as each unknown starting points. The experiment design supplies enough data to estimate the entire gate set. It is worth noting, however, that GST employs a completely Markovian model, where elements of the gate set are composed together via matrix multiplication. In this context, GST is depicted in Figure~\ref{fig:NMGST-buildup}a. Each element of a circuit is modelled (colourised), but no temporal information is fed forward beyond propagation of the state. 

\begin{figure*}[t]
	\centering
	\includegraphics[width=\linewidth]{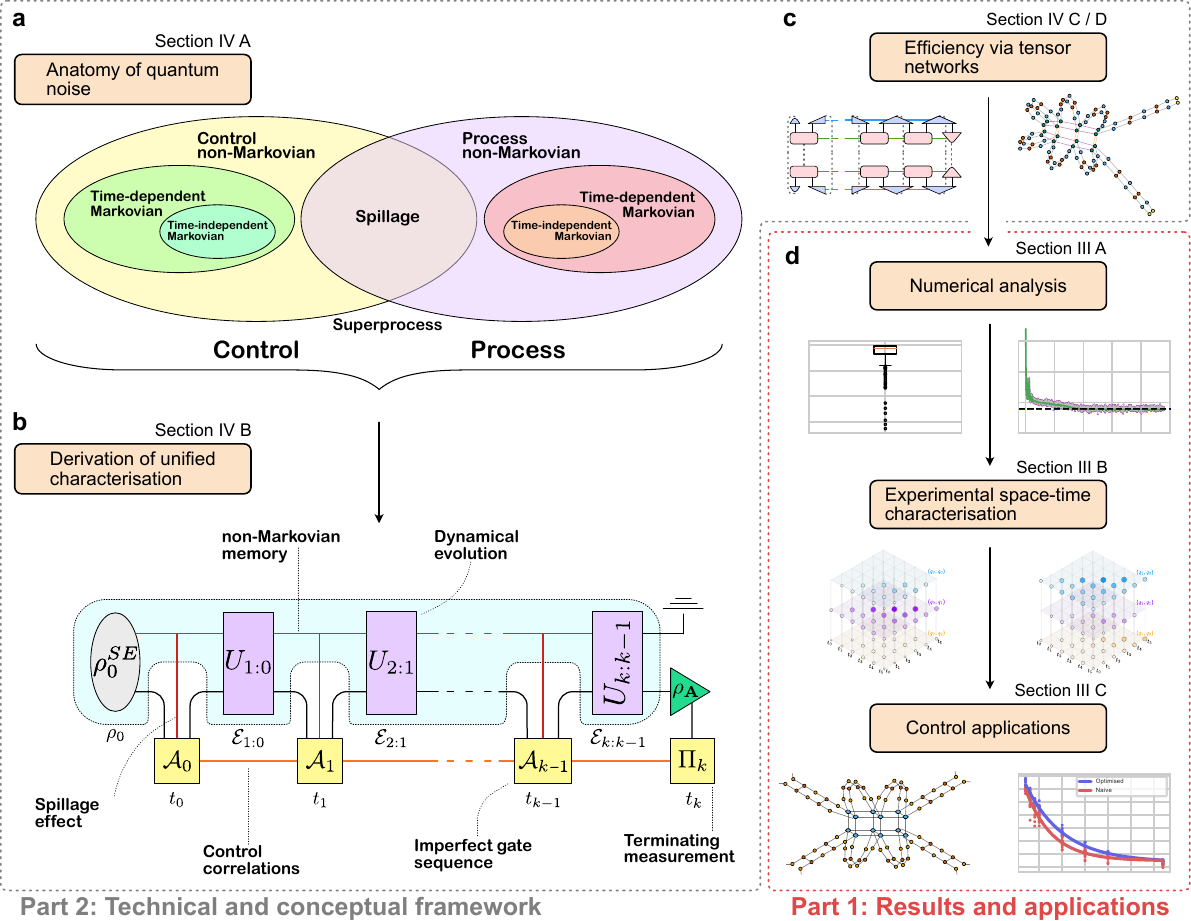}
	\caption{Overview of our work and conceptual outline of the ideas herein. Note that the conceptual order presented here is reversed from the sectioning of this manuscript to emphasise concrete and accessible results first. \textbf{a} Our foundation is on a classification of noisy effects which can be separated into pure process, pure control, and an interplay between the two. All quantum noise can be unambiguously categorised into one of these camps. \textbf{b} We relate our noise classification to a quantum stochastic description of multi-time dynamics, where correlations are described based on their causal origin. Here, a sequence of control operations probe a quantum system: the control electronics themselves may be correlated, the system may couple to a correlated bath, and the control may manipulate that bath (spillage). We derive a self-consistent estimation technique capable of estimating these models. \textbf{c} These models are cast in an efficient format using tensor network methods. \textbf{d} Layout of the results sections of our work, numerically demonstrate our methods, experimentally determine spatiotemporal correlations, and provide examples of control applications.}
	\label{fig:noise-categorisation}
\end{figure*} 

PTT, as discussed above and depicted Figure~\ref{fig:NMGST-buildup}b, falls under the category of self-inconsistent protocols. Both gate sequence and final measurement in this context are assumed to be known. Their perfect forms are used to probe temporal information about the interaction of the system with its environment. Moreover, it is assumed that the `known' gates possess no temporal correlations. Below will construct a self-consistent PTT, i.e., a model that assumes nothing in advance about the gate sequence or the measurement. Moreover, we open up to possible temporal correlations into the control sequence, resolving a large gap in the non-Markovian literature.

\subsection{Conceptual Contribution}

We depict the core results of our work---and their respective sections---in Figure~\ref{fig:noise-categorisation}. In particular, Figure~\ref{fig:noise-categorisation}a captures the major conceptual milestone of this paper. Namely, the anatomy of quantum noise both conceptually and expressively. In real experiments, noise arises from both a background process and the imperfections in the control system. Panel a conceptually separates these two contributions, which is detailed in Sec.~\ref{sec:categories}. This includes the notion of `spillage error', where a control operation affects the background environment. Thus, spillage errors belong to both categories at once. While the conceptual separation of noise is appealing, it is not at the face of it useful.

For practical progress, we need to expressively separate the two quantum noise sources, which is not at all straightforward. To overcome this, we derive a fully self-consistent procedure in Sec.~\ref{app:nmgst-derivation} to characterise all spatiotemporal correlations that arise (i) as a result of uncontrollable system-environment interactions, (ii) as a result of memory effects within the control itself, and (iii) from direct undesirable interactions between the control and the environment. Each result also includes temporal instability of those facets. This constitutes a conceptual advance in the characterisation of non-Markovian quantum systems. Specifically, Figure~\ref{fig:noise-categorisation}b depicts how different facets of temporally correlated noise are represented by process-tensor-like objects, and how they interface with one another. 

Central to this construction is the notion of a `generalised gateset' consisting of: a fixed process tensor to represent the unchanging background dynamics, and a tester to represent each unique gate sequence. The process tensor captures all ``fixed" memory, in the sense that it is present in each experiment. For example, if a qubit were to couple to some defect in its environment, it would be an always-on interaction. We call this `process non-Markovianity'. The testers---one for each gate sequence, in principle---are then shown to be expressive enough to capture both `control non-Markovianity' and `spillage'. In the former, errors in each gate are time correlated. For example, a laser driving a system with a quasistatic fluctuating power spectrum will induce the same coherent error across a circuit, where the coherent error is drawn from a distribution. In the latter, the application of a gate modifies the always-on environment itself---such as if a drive tone were also on resonance with the defect in the process non-Markovian example. These objects are mathematically expressive enough to represent all non-Markovian noise. And moreover, we show how to estimate them in a series of physical experiments.

However, the price of full generality is the demand of prohibitively many resources. We resolve this by employing a sparse correlation assumption and representing each noisy object as a tensor network. Thus, the characterisation is made efficient---assuming only that the non-Markovian memory does not grow exponentially complex in time, or that the noise does not arbitrarily entangle spatially separated qubits. 
We will revisit the estimation of these qualitatively different noise sources in Section~\ref{sec:synth-demonstrations}.

Our theory encompasses the previous understanding of quantum noise, namely it reduces to the two state-of-the-art methods in the appropriate limit. When background noise is assumed to be Markovian, we recover GST. When self-consistency is dropped we regain PTT. We will use these features to test our theory against GST and PTT in the next section on synthetic data. To do this seamlessly, we make use of an additional advantage of tensor networks methods, which is that we can turn on and off various features of our model as desired.
Moreover, the characterisation requires only randomised benchmarking (RB)-type experiments, and hence the experiment design is consistent with current standard laboratory settings, similar to Ref.~\cite{zhang2023randomised}.

\subsection{Technical Approach}

Although our work incorporates a fully general procedure for estimating multi-time processes in Section~\ref{sec:self-consistency}, it is too expensive to be performed in practice and must hence be reduced. The archetypal method for compressing quantum states into more efficient representations has been to employ the use of tensor networks, as depicted in Figure~\ref{fig:noise-categorisation}c. The philosophy behind tensor networks is that quantum states, which live on a series of tensor products of Hilbert and dual spaces, may be broken down across these spaces using well-known decompositions. Then, if these decompositions contain sparse features, they can be zeroed out to build more efficient representations. The most famous example is a matrix product state (MPS) which can exactly represent any pure quantum state~\cite{vidal-MPS}. If that state has exponentially decaying correlations in its geometry, then the MPS can represent it in only a polynomial number of parameters~\cite{Orus2014}.

Tensor networks are significant for a variety of reasons~\cite{Cirac2020MatrixPS}, stemming from the famous density matrix renormalisation group technique locating ground-states of local gapped Hamiltonians~\cite{PhysRevLett.69.2863}.
They highlight the necessary role played by entanglement in pure-state quantum computation~\cite{Eisert2021EntanglingPA,vidal-MPS,entanglement-jozsa}. They also play a crucial role in the advancement of classical computers for the simulation and emulation of quantum algorithms, furthering the barrier through which quantum supremacy may be realised~\cite{huang2020classical,china-sim,pan2021simulating}. 
More recently, tensor networks have also been realised as a useful tool to extend the ensembles on which classical shadows can operate~\cite{akhtar2022scalable,bertoni2022shallow}. Here, we will focus on the application of tensor networks to sparse characterisation, which has recently been achieved with quantum process tomography~\cite{torlai2020quantum} and Hamiltonian learning~\cite{wilde2022scalably}.

Our methods for dealing with tensor networks are predominantly based on the results of Torlai et al. in Ref.~\cite{torlai2020quantum}. The works therein develop a method to perform quantum process tomography in an efficient manner, reconstructing the matrix product operator representation of a quantum channel in its Choi form. The Choi matrix $\hat{\Lambda}$ of a quantum channel $\Lambda$ is uniquely determined by the action of $\Lambda$ on one half of an unnormalised entangled state -- for $N$ qubits, where $d=2^N$ -- $|\Phi^+\rangle = \sum_{i=1}^d|ii\rangle$, with identity map $\mathcal{I}$ on the other half:
\begin{equation}
	\hat{\Lambda} := (\Lambda \otimes \mathcal{I})[|\Phi^+\rangle\!\langle \Phi^+|] = \sum_{i,j=1}^d \Lambda[|i\rangle\!\langle j|]\otimes |i\rangle\!\langle j|.
\end{equation}
Ref.~\cite{torlai2020quantum} employs a locally-purified density operator (LPDO) parametrisation of quantum channels. This is a subset of MPOs representing positive operators. In this tensor network representation, instead of a single tensor per site, each qubit $q_j$ acted on by the channel is associated with a tensor $A_j$ and its element-wise conjugate $A_j^\ast$. The size of these tensors dictates both the expressiveness and complexity of its state representation.
To learn some parametrised model of $\hat{\Lambda}$, the authors randomly sample product stabiliser states as inputs to the channel and perform single-shot measurements in random Pauli bases. The model is then optimised to the data via gradient descent, where gradients are obtained via automatic differentiation with respect to the chosen loss function. \par 
We employ much of this approach in our estimation of multi-time quantum processes. 
This requires first a generalisation of channel tensor networks to multi-time and multi-qubit processes. The structure of these tensor networks is exposed in Section~\ref{sec:PT-LPDOs}.
To accommodate the additional physical constraints, we carefully account for the causality and positivity constraints of the process tensor in our tensor network ansatz. We may then say (opaquely, for the present) that a vector of parameters $\vec{\theta}$ represents our tensor network guess $\Upsilon_{k:0}^{\vec{\theta}}$ for the process tensor $\Upsilon_{k:0}$ in the lab, and that the set of vectors $\{\vec{\phi_j}\}$ parametrise the tester tensor networks modelling each gate sequence $\{\hat{\mathbf{A}}_{k:0}^{\vec{\phi}_{j}}\}$.\par
Although we do not wish to pursue the minor details too far here, our estimation of the noisy dynamics is broadly speaking found by minimising the following objective function with respect to the gateset parameters $\vec{\theta}$ and $\{\vec{\phi_j}\}$:
\begin{equation}
\begin{split}
\label{eq:tn-objective-intro}
	f(\vec{\theta},\{\vec{\phi_j}\}) = &\overset{\text{Model log-likelihood}}{\overbrace{\sum_{d\in \mathcal{D}}-n_d\ln p_d^{\vec{\theta}\vec{\phi_{j}}} }}
	+ \overset{\text{Process causality}}{\overbrace{\sum_{P\in \mathcal{C}_0} \Tr\left[P\cdot [\Upsilon_{k:0}^{\vec{\theta}}]\right]}}\\
	+ &\overset{\text{Control trace preservation}}{\overbrace{\sum_{d\in \mathcal{D}}\sum_{P\in\mathcal{C}_d} \Tr\left[P\cdot \sum_x \hat{\mathbf{A}}_{k:0}^{\vec{\phi}_{(j,x)}}\right]}}.
\end{split}
\end{equation}
Here, the sums are respectively performed over $\mathcal{D}$, representing the dataset, as well as $\mathcal{C}_0$ and $\mathcal{C}_d$, which are sets of specifically chosen Paulis to regularise the causal conditions of the process tensor and the trace preservation of the instruments.
The details of this method are revisited in Sections~\ref{sec:PT-LPDOs} and~\ref{sec:self-consistency}.

\subsection{Reading Guide}
This Section has so far focused heavily on the background, structure, and high-level details of our (admittedly, quite long and sometimes esoteric) work. We will therefore conclude it by providing a quick guide to the reading of this paper for various audiences in the quantum information community.

\begin{itemize}
    \item For those who are interested only in the capabilities of our models, see Sections~\ref{sec:synth-demonstrations},~\ref{ssec:nisq-tn}, and~\ref{sec:applications}. Assuming the high-level understanding of our setting and aims, this validates the performance and expressiveness of our methods across both numerical and experimental settings. It also provides insight into potential applications thereof.
    \item For those who are interested in implementing our techniques in their own setting, see additionally Section~\ref{sec:self-consistency} and the code repository~\cite{PTT_github_repo}.
    \item For those interested in replicating or building on our techniques, see Sections~\ref{sec:PT-LPDOs} and~\ref{app:nmgst-derivation}. For performance benchmarks, see Appendix~\ref{app:benchmarking}.
    \item Lastly, those who are interested in the conceptual developments in the theory of non-Markovianity as well as our philosophy on quantum stochastic processes, see Section~\ref{sec:categories} and Appendix~\ref{app:definitions}.
\end{itemize}

\section{Numerical and Experimental Results}
\label{sec:applied-results}
In this section, we present our results as they pertain to the estimation of non-Markovian dynamics in both synthetic and experimental settings, as well as further exploring control-based applications of the estimations. We first validate our models across the discussed categories of non-Markovianity in a self consistent way. This spans the settings of gate-dependent Markovian error with a non-Markovian process background; non-Markovian control with a non-Markovian background, and lastly non-Markovian spillage on a non-Markovian background. \par
We then perform a range of multi-time and multi-qubit characterisations of IBM Quantum devices. This shows not only that noisy experimental data can be explained, but we additionally quantify the spatiotemporal correlations present. Lastly, we show what sorts of control improvements can be expected in the framework of this characterisation, by showing how it may be used for the task of optimally implementing arbitrary two-qubit unitaries as well as optimising dynamical decoupling sequences.

\subsection{Synthetic Demonstrations}
\label{sec:synth-demonstrations}
To start, we shall consider exotic noise models which are either operationally uncharacterisable by any techniques in the literature or violate the model previously assumed with GST and PTT. In particular, this means involving different forms of control noise. In each instance, we simulate a two-qubit, five-step process where each qubit is coupled to a common bath defect via a random Heisenberg interaction: $H_{\text{int}} = J_x XX + J_y YY + J_z ZZ$, where $J_i\cdot t_{\text{evolve}} = \mathcal{O}(1)$. Heisenberg interactions constitute a good testing ground for non-Markovian models, since they implement a genuine exchange of quantum information. This allows, among other effects, the distribution of temporal entanglement across many times~\cite{milz2021genuine}.

We conduct two phases to the numerics: a training phase, and a validation phase. Each characterisation uses 6000 circuits of random Clifford operations followed by a projective measurement in a random Pauli basis, with each circuit repeated for 1024 shots. Note that this is more circuits than strictly necessary; we are targeting precision over efficiency. For benchmarking with respect to size of data, see Appendix~\ref{app:benchmarking}. 
The training phase consists of fitting the tensor network model to the set of frequencies obtained from the random Clifford circuits via stochastic gradient descent of the objective function in Equation~\eqref{eq:tn-objective-intro}. At each iteration of the optimiser, the 6000 circuits are batched randomly into groups of 1000, with gradients obtained via autodifferentation. 

On top of this, we have a set of 100 validation circuits, which are run with completely random unitary gates. The model is not fit to this validation data, but instead is evaluated to make a prediction of these circuit outcomes. The relative difference between the prediction on this validation data and the actual frequencies (in contrast to the training data) is recorded as a measure of the generalisability of the model. That is to say, it is a heuristic for reconstruction fidelity that the process tensor is able to capture dynamics outside of the characterisation circuits. Typically, because we have a linear model with an informationally complete set of measurements, we do not observe any overfitting, and the validation follows the exact same trend as the training likelihoods, up to a difference in the number of shots collected per circuit for each respective dataset.
As we explore, this generalisability is not only a certification, but can be used to optimise control of the noisy dynamics.

We also implement a hardware-specific simplification of the model. This is more for the purpose of convenience in having a compact gateset for analysis purposes than a requirement. On most present hardware platforms, it is possible to implement `virtual' $Z$ rotations through a frame shift of the hardware~\cite{McKay_2017}. These gates have significantly lower error than their physical counterparts.
On a single qubit, this means that only a single physical gate is necessary. Here, one can select a unitary decomposition of a sequence of physical $X$ pulses and virtual $Z$ rotations. An arbitrary single-qubit unitary can be taken as a function of three parameters, $u(\theta,\phi,\lambda)$: 
\begin{equation}
	\label{eq:u3-decomp-intro}
	R_Z(\phi + 3\pi)\cdot R_X(\pi/2)\cdot R_Z(\theta + \pi)\cdot R_X(\pi/2)\cdot R_Z(\lambda).
\end{equation}
This way, only a single physical gate needs tuning up. Consequently, only a single physical gate needs to be estimated. We adopt this model in both of our simulations and our demonstrations on IBM Quantum devices.
The circuits are hence decomposed into a standard basis $\{R_Z(\theta), \sqrt{X}\}$ as per Equation~\eqref{eq:u3-decomp-intro}.
We take the effects to apply to the physical $\sqrt{X}$ gate. We now permit the control itself (i.e. the Clifford operations) to be noisy.
Previous literature has already established the ability of PTT to capture these effects of process non-Markovianity~\cite{White-NM-2020,White-MLPT,white2021many}. Here, we will see (i) how standard PTT breaks down when exposed to these classes of control noise, and (ii) how our collective approach is able to efficiently and effectively characterise all of these effects in tandem.

\begin{figure*}[!t]
	\centering
	\includegraphics[width=0.98\linewidth]{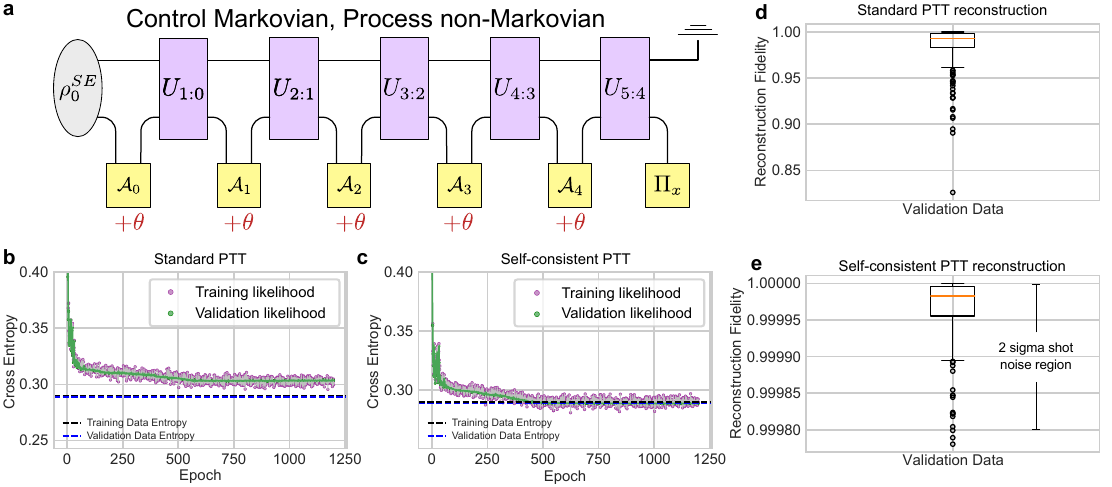}
	\caption[Simultaneous characterisation of coherent gate noise in a non-Markovian environment ]{Simultaneous characterisation of coherent gate noise in a non-Markovian environment. We compare standard PTT (modelling only the process) with estimation of a self-consistent model. 
	\textbf{a} Circuit diagram of the simulated error model. Each $\sqrt{X}$ operation in the control decomposition is coherently rotated by $\theta = \pi/16$ in comparison to the ideal operation. This generates gate-dependent noise. Every detail here is unknown to the characterisation model.
	\textbf{b} and \textbf{c} indicate the convergence graphs for both the stochastic data fit and the quality with respect to validation data. \textbf{d} and \textbf{e} summarise reconstruction fidelities for each of the fit models (their ability to predict other dynamical sequences): we see a significant improvement in the quality of the model by incorporating gate errors as well as background non-Markovianity. 
	}
	\label{fig:coherent_error}
\end{figure*}

\subsubsection{Coherent Gate Error with a non-Markovian Bath}

We start with the most straightforward addition to our model with coherent over-rotations. Specifically, we decompose every gate in accordance with the decomposition given in Equation~\eqref{eq:u3-decomp-intro}. The virtual $Z$ gates are simulated to be perfect, but the supposed $R_X(\pi/2)$ gate is really $R_X(\pi/2 + \pi/16)$. This is situation is sketched in Figure~\ref{fig:coherent_error}a. The effective gate then is a coherently transformed unitary. Importantly, because the $R_X(\pi/16)$ error does not commute through the $Z$ rotations, the effect of this error model on the generic unitaries is entirely gate-dependent. The reason this is important is that any gate-independent noise channels can factor outside and be incorporated into the process tensor estimate. A gate-dependent coherent model, however, changes the linear relationship between the various basis elements. Although coherent error alone could be characterised by GST, introducing the non-Markovian bath both violates the model there and introduced in standard PTT.

We simulate and estimate the five-step process with controls as described above. We employ both the standard PTT model (assumed perfect inputs) and the self-consistent model introduced in this work (which additionally models the process inputs). The process tensor requires only a bond dimension of $\chi=2$ to model the effects of the stray two-level system.
The results are summarised in Figure~\ref{fig:coherent_error}. 
Panels \textbf{b} and \textbf{c} display the convergence graphs of both estimators. Specifically, the purple line is the stochastic loss function given in Equation~\eqref{eq:tn-objective-intro} evaluated on randomised minibatches of the data (1/6 of the total circuit count) as the model is trained. Plotted in green is also the cross entropy of the process tensor model as it is evaluated on the set of 300 random validation circuits. The cross entropy is the first term in Equation~\eqref{eq:tn-objective-intro}, and specifically measures the goodness-of-fit without including regularisation terms.

The convergence of the green curve to the blue dashed line benchmarks the model's ability to capture the entire dynamics. There is a clear disparity between the two models shown. The former is unable to fit the data and suffers from substantial error in predicting the dynamics for arbitrary sequences of unitaries thereafter. 
We illustrate this specifically for the converged models in panels \textbf{d} and \textbf{e}. Here, the set of probability distributions predicted by model $m$, labeled as $p_{x|m}$, is compared to the exact simulated set of probability distributions $p_{x|0}$ via the Hellinger fidelity
\begin{equation}
    H(m,0):=\sqrt{\frac{1}{2}\sum_x \left(\sqrt{p_{x|m}}-\sqrt{p_{x|0}}\right)^2}.
\end{equation}
The self-consistent estimate fits the dynamics correctly to a cross entropy of less than $10^{-5}$, with a median reconstruction fidelity of $1 - 10^{-5}$. This demonstration shows that we can characterise both the non-Markovian background process and noise in the control operations to a very high accuracy.

This numerical setup is a first principles demonstration of a self-consistent non-Markovian characterisation in the simplest possible scenario where both PTT and GST fail. This is a significant reduction in the number of required experiments (from $\mathcal{O}(10^5)$ circuits instead of $\mathcal{O}(10^3)$) between tensor network and dense models, whose separation scales exponentially in $k$. In what follows, we shall see how this model handles more complex noise models, larger qubit registers, and much larger multi-time processes. 

\begin{figure*}[!t]
	\centering
	\includegraphics[width=0.98\linewidth]{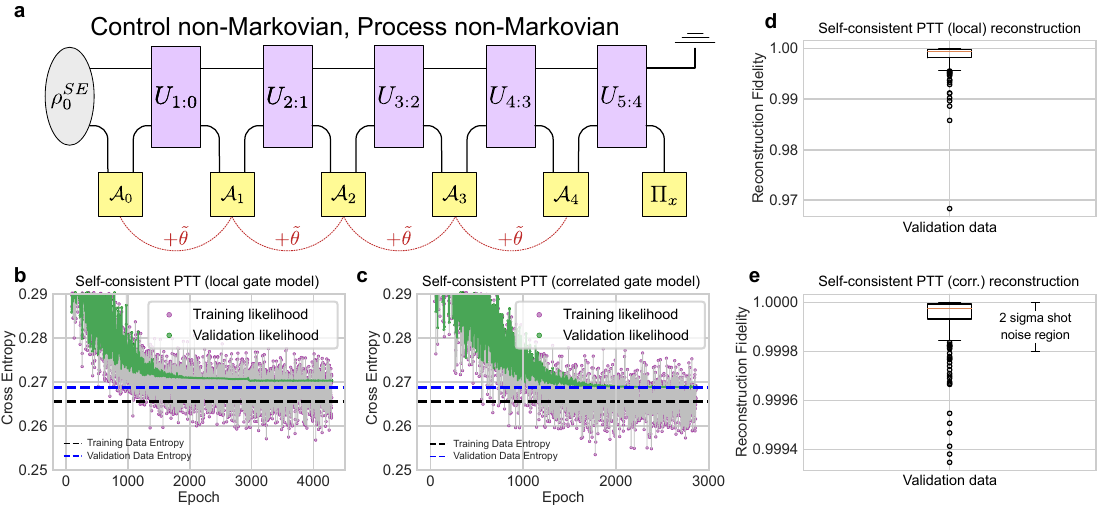}
	\caption[Characterisation of $1/f$ correlated gate noise in a non-Markovian environment ]{Characterisation of $1/f$ correlated gate noise in a non-Markovian environment. We compare the time-local self-consistent PTT model with a fully fledged non-Markovian GST, where both process tensor and testers are estimated. 
	\textbf{a} Circuit diagram of the simulated error model. Each gate in the circuit is coherently rotated by the same $\tilde{\theta}$, which is a stochastic variable resampled in each circuit. Every detail here is unknown to the characterisation model.
	\textbf{b} and \textbf{c} indicate the convergence graphs for both the stochastic data fit and the quality with respect to validation data. \textbf{d} and \textbf{e} summarise reconstruction fidelities for each of the fit models: we see a significant improvement in the quality of the model by estimating correlated gate errors as well as background non-Markovianity. }
	\label{fig:corr_coherent_error}
\end{figure*}

\subsubsection{Control and Process non-Markovianity}
The previous example showed that self-consistent PTT is possible, and that our approach robustly estimates noise in the controls. Let us now turn to a more sophisticated error model, which cannot currently be captured by methods in the QCVV literature. 
Let us consider a circuit where every implementation of a gate $R_X(\pi/2)$ is shifted by some $\varepsilon$ so that the rotation is really $R_X(\pi/2 + \varepsilon)$. If $\varepsilon$ were fixed, then this would be the same coherent case as before. Suppose, however, that at the start of each circuit, $\varepsilon$ were drawn probabilistically from a distribution. Each gate rotates by the same amount, but the amount changes from circuit to circuit so that the coherent error is fully correlated. This scenario is depicted in Figure~\ref{fig:corr_coherent_error}a.

This models the effects of, for example, quasistatic laser noise. It can also be reflected in fluctuating electric and magnetic fields. 
We implement this scenario where each $\varepsilon$ is sampled according to $1/f$ noise. $1/f$ noise is the phenomenon where a stochastic process has power spectrum $S(f) \propto 1/f^\alpha$; here we select $\alpha = 1$. In the context of signal processing, the power spectral density of a signal is the Fourier transform of its autocorrelation function. For $1/f$ noise, this implies that the autocorrelation function has a long-range dependence, meaning that the noise exhibits long-term (polynomially decaying) memory, with correlations persisting over a wide range of time scales.
This archetypal spectrum models myriad physical processes in nature with long-time correlations~\cite{RevModPhys.53.497,west1989ubiquity,PhysRevE.54.2154}, including in almost all quantum hardware~\cite{paladino20141,aquino2023model,wilen2021correlated,PhysRevApplied.17.034074}. The overall process we sample from therefore has process non-Markovianity due to a stray two-level system, as well as instruments plagued with $1/f$ noise. This is the underlying phenomenological model. We note similar model (minus the environment) was studied in Ref.~\cite{gullans2023compressed} where the authors factored the microscopic model into the QPT estimation, to high accuracy and resource reduction.

Without tailoring anything to the underlying physics, we represent the processes with a generic process tensor network, and the controls with a generic tester tensor network, each with bond dimension $\chi = 2$. The results of this are depicted in Figure~\ref{fig:corr_coherent_error}. Once more, panels \textbf{b} and \textbf{c} indicate the model-fitting for each, while panels \textbf{d} and \textbf{e} illustrate the reconstruction fidelities of the converged models on validation data. Here, in particular, we compare the fully self-consistent estimate to the one from the previous results: self-consistent non-Markovian estimate with the controls modelled as time-local. This is intended to eliminate the possibility that the improvement is replicated simply by accounting for some coherent error. The delineation in this instance entirely stems from finding the best model estimate of the temporal correlations in the instrument, for which we see two orders of magnitude improvement in the reconstruction fidelities.
This thus demonstrates the ability to estimate compressed models representing physically realistic quantum correlated noise both in the background dynamics of the system, and in the control instruments of the experimenter.

\begin{figure*}[!t]
	\centering
	\includegraphics[width=0.98\linewidth]{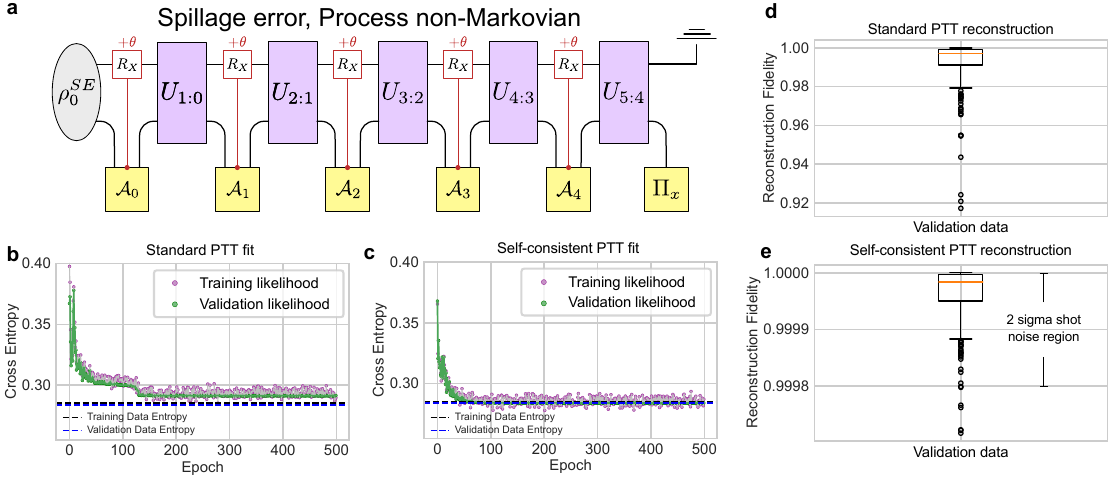}
	\caption[Simultaneous characterisation of spillage error in a non-Markovian environment ]{Simultaneous characterisation of spillage error in a non-Markovian environment. We compare standard PTT (modelling only the process) with estimation of a self-consistent model. 
	\textbf{a} Circuit diagram of the simulated error model. At each application of control operations, a controlled $X$ rotation by $\theta = \pi/16$ is applied to the environment two-level system. 
	Every detail here is unknown to the characterisation model. 
	\textbf{b} and \textbf{c} indicate the convergence graphs for both the stochastic data fit and the quality with respect to validation data. \textbf{d} and \textbf{e} summarise reconstruction fidelities for each of the fit models: we see a significant improvement in the quality of the model by incorporating gate errors as well as background non-Markovianity. }
	\label{fig:spillage_error}
\end{figure*}

\subsubsection{Spillage}
The third exotic noise model we consider is that of spillage, where the radius of the control operation leaks over into environment systems. This can typically happen in two ways: 
the pulse shape produces (in the frequency domain) spurious resonsances with some parts of the environment, driving a stochastic transition; or, in the case of optically addressed systems, there may be overlap between the finite extent of the laser and the remainder of the environment. 
In either case, the effect is the same: the act of performing a control operation modifies the environment state. If we naively considered the control operation to act on the system alone, this could lead to the puzzling observation of non-linear outcomes in the control. 
One could dilate the system to account for this fact, but it would leave a large gauge freedom and be generally unsatisfactory. In effect, this expands the model to fit the data, but is not indicative of the physics. We have already argued with the use of superprocesses, however, that this particular type of driven dynamical effect is fully captured under the framework where process tensors and testers are both employed, it simply remains to estimate it. Further, we have already argued that the superprocesses used, while conceptually useful, are not strictly necessary for the estimation.

The model we consider for proof-of-principle once more has a non-Markovian environment coupled by exchange interaction. Following each $\sqrt{X}$ gate on the system, a controlled-$R_X$ gate is applied between system and environment, controlling a rotation by $\theta = \pi/16$.
This gate dependent error does not factor outside each of the gates, and so cannot be absorbed into the process tensor. It is hence incompatible with the standard model of PTT. We illustrate this in Figure~\ref{fig:spillage_error}a. This model is less physics-inspired than others and more intended to be a stress-test for the described spillage effects. In essence, however, it is intended to behave similarly to active leakage of control operations to the persistent environment that generate entanglement.

To respectively account both for non-Markovian correlations and the superprocess transform, we require a process tensor with bond dimension $\chi = 2$ and tester with bond dimension $\chi = 4$. The former encapsulates the single-qubit environment, and the latter encapsulates the two-qubit gate operation. Though, we do not choose these values in advance, we start from a Markov model and expand until the cross validation is sufficiently well-explained without overfitting. 
The results comparison between standard PTT and self-consistent PTT are shown in Figure~\ref{fig:spillage_error}. Although the noise model is pathological, we are able to completely capture the effects in practice, and attain a complete convergence. This closes the gap on every category of noise considered in Section~\ref{sec:categories}. We argue, therefore, that we have demonstrated an effectively-universal protocol for performing quantum tomography. This can be applied in real experimental scenarios and generalised also to multi-qubit settings.

\subsection{Demonstrations on IBM Quantum devices}
\label{ssec:nisq-tn}

\begin{figure*}[t!]
	\centering
	\includegraphics[width=\linewidth]{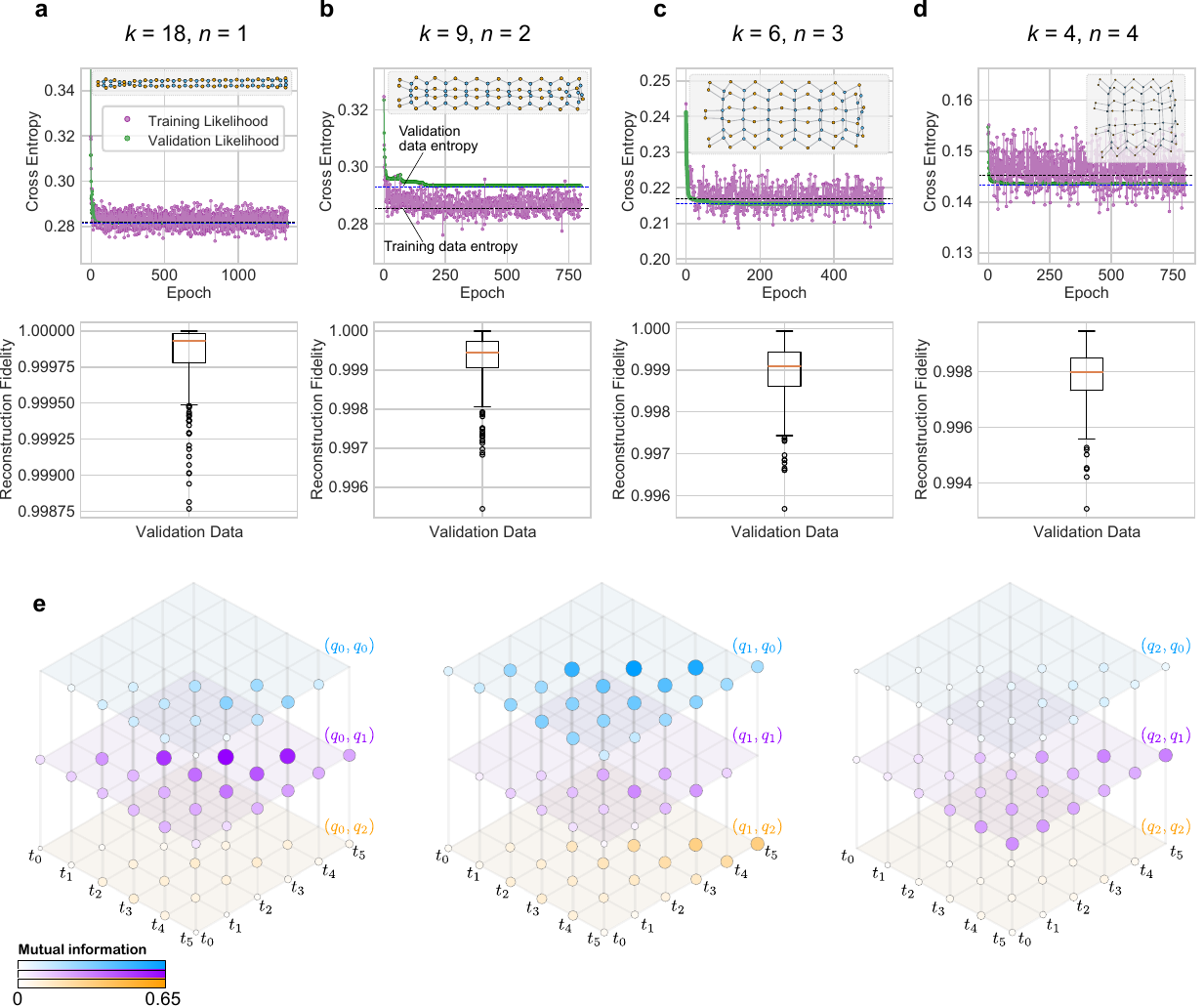}
	\caption[Benchmark estimating large spatiotemporal tensor networks on IBM Quantum devices]{Benchmark estimating large spatiotemporal tensor networks on IBM Quantum devices. We present the convergence of estimation as well as reconstruction statistics of a series of large tensor network reconstructions at a bond dimension of $\chi=2$. In particular, the validity of the model is given by the convergence of the validation cross entropy to the validation data entropy (given as a blue dashed line). Inset in each plot is a depiction of the tensor network model (spatially in the vertical direction and temporally in the horizontal).
	\textbf{a} 18 steps on a single qubit on \emph{ibm\_cairo}; \textbf{b} 9 steps across two qubits on \emph{ibm\_perth}; \textbf{c} six steps across three qubits on \emph{ibm\_perth}; and \textbf{d} four steps across four qubits on \emph{ibm\_kolkata}. This encodes all Markovian, crosstalk, and naturally occurring non-Markovian dynamics from these instances to a high fidelity. \textbf{e} We analyse these spatiotemporal correltions for the three qubit, six step scenario. We illustrate the mutual information between the dynamical maps evolving qubits $q_i$ and $q_j$ from times $t_k$ and $t_l$. Numbers for two copies of the same qubit indicate pure non-Markovianity, diagonals of a qubit-pair indicate pure spatial interactions, while off-diagonals are space-time correlations as mediated by a shared bath. $q_1$ is identified as the most problematic, while $q_2$ has the least complex noise. }
	\label{fig:tn-ibm-benchmarks}
\end{figure*}

We now have a robust method for efficiently estimating process tensors with sparse memory structures in practice. We have tested this on synthetic data, but it remains to be seen how this approach performs in a real device setting. On three different IBM Quantum devices, we perform a total of four demonstrations to capture the underlying process. Each one of these is far larger than previously achievable: a single qubit captured across 18 time steps, two qubits across nine time steps, three qubits across six time steps, and four qubits across four time steps. 
We elected to use 512 shots per circuit in each of the characterisation settings. 
Using a total of $10^7$ shots, we reconstructed each of the above processes with a wait time of $800$ ns -- or about the duration of two CNOT gates -- between each step. This is well within the coherence times of each of the qubits used. 

The resulting estimates are given in Figure~\ref{fig:tn-ibm-benchmarks}a--d. The convergence graphs show the cross entropy, or log likelihood, between the predictions made by the estimate and the observed data. We plot this for the estimation data (purple), as well as overlaid with the validation data (green). In dashed lines, we also indicate the entropy of each data set, which is the theoretical minimum of the cross entropy. Note that at each iteration, a random batch of size 1000 of the characterisation data is taken. These data points naturally vary from the entropy of the whole data set. Consequently, it is the convergence of the green curve to the dashed blue line which is the best indicator of goodness-of-fit. In each instance, we see a difference in log-likelihood of less than $5\times 10^{-3}$.
Interestingly, choosing $\chi=2$ is sufficient to capture all crosstalk, non-Markovian, and Markovian dynamics on these devices up to our desired precision. In the respective cases, we achieve a median reconstruction infidelity of $10^{-4}$; $5\times 10^{-4}$; $10^{-3}$ and $2\times 10^{-3}$. Thus, the non-Markovian noise is present, but it is not too large. Only a small amount of quantum memory is required to explain these effects.

To use the resulting estimates for in-depth diagnostics, we can use the estimated models to compute spatiotemporal correlations (combined with the methods of Ref.~\cite{white2021many} to overcome informational incompleteness afforded by employing only unitary operations). We illustrate this using the three-qubit/six-step process estimated on \textit{ibm\_perth}. From $\Upsilon_{6:0}^{(q_1,q_2,q_3)}$, we compute the correlations between all conditional marginals $\hat{\mathcal{E}}_{j:j-1}^{q_i}$ and $\hat{\mathcal{E}}_{m:m-1}^{q_l}$. That is to say, 
\begin{equation}
	\begin{split}
	&\hat{\mathcal{E}}_{j:j-1,m:m-1}^{(q_i,q_l)}\\ 
	&= \text{Tr}_{\bar{\mathfrak{o}}_j\bar{\mathfrak{i}}_j\bar{\mathfrak{o}}_m\bar{\mathfrak{i}}_m}\left[\Upsilon_{6:0}^{(q_1,q_2,q_3)}\rho^{0,(q_i)}_{\mathfrak{i}_{j+1}}\mathbb{I}^{(q_i)}_{\mathfrak{o}_{j-1}}\rho^{0,(q_l)}_{\mathfrak{i}_{m+1}}\mathbb{I}^{(q_l)}_{\mathfrak{o}_{m-1}}\right],
	\end{split}
\end{equation}
where we have implicitly have projections onto the $X$-gate superoperator $\hat{\mathcal{X}}$ at every other step. The purpose of applying an $X$ operation on every remaining qubit-time location is to refocus the dynamics as much as possible and see what remains. 
The mutual information is the relative entropy between this conditional process tensor and the product of its marginals. That is, 
\begin{equation}
	S\left[\hat{\mathcal{E}}_{j:j-1,m:m-1}^{(q_i,q_l)}\mid\mid \hat{\mathcal{E}}_{j:j-1}^{(q_i)}\otimes \hat{\mathcal{E}}_{m:m-1}^{(q_l)}\right].
\end{equation}
This quantifies the amount of information gained about one step of the dynamics from knowledge of the other. Or, alternatively, can be related to the probability of confusing a correlated with uncorrelated model~\cite{Pollock2018,bengtsson_zyczkowski_2006}. The results of this analysis are shown in Figure~\ref{fig:tn-ibm-benchmarks}e, and are illustrative of a predominant problem qubit $(q_1)$ experiencing significant spacetime-non-local errors. Meanwhile the dynamics of $(q_2)$, for instance, are much closer to a Markov model. \par
Methods to characterise spatiotemporal correlations both rigorously and in full generality do not exist elsewhere in the literature to our knowledge. Here, we have demonstrated a highly efficient and highly versatile technique for capturing all models of naturally occurring dynamics. Moreover, we have shown that it functions in the practical test-bed of IBM Quantum devices. The resulting diagnostics provide direct insight into the location and nature of correlated errors. This information is highly relevant to device fabrication and benchmarking, and could be fed forward into tensor network decoders responsible for maximising logical error rate given a specific noise model~\cite{piveteau2023tensor}.
Moreover, as we shall now investigate, their characterisation can be fed-forward directly into control applications which mitigate the effects.

\subsection{Control Applications}
\label{sec:applications}

Much like we built upon our characterisation protocols to be more efficient and more robust, so too can we focus ourselves on control tools that readily integrate into circuit compilation software. We first introduce our tensor network characterisation methods to the task. This permits a generalisation to arbitrary step numbers, and is cognisant of any errors in the control hardware. We propose two protocols to this effect: noise-aware decomposition of two-qubit unitaries, and tailored dynamical decoupling sequences. By performing characterisation across a wide range of IBM Quantum devices, we can map out where the most benefit would be found in terms of aiming to cancel the effects of non-Markovian noise. The key point here is that we have a model mapping our control operations to outcomes. Hence, we can classically analyse the characterisation results to achieve optimal control outcomes.

Note that at present we have not integrated these methods into live feed-forward protocols, such as with online learning that feeds into control. Our results hence present a \emph{forecast} of what should be possible in terms of marrying characterisation with control. Each characterisation comes equipped with certified high-fidelity predictive capabilities to lend trust to the forecast (to a reconstruction fidelity error of $10^{-4}$). Feeding the results forward directly into optimal control is the subject of ongoing work, but we believe our present results to be a reliable indicator of how optimised gates will perform, since the reconstruction fidelities are much tighter than the difference between the native gate and the optimised gate. This type of analysis gives an informed view of where structured improvements are (or are not) available in the control of a quantum device.

\subsubsection{Logically Improving Arbitrary Two-Qubit Gates}

In the search for non-Markovian optimal control techniques, a key required property is generalisability. We cannot characterise each step of each quantum algorithm to run it optimally. Instead, we must focus on profiling the building blocks to quantum circuits so that the amount of required characterisation is fixed. For example, brickwork-type circuits composed of arbitrary $SU(4)$ gates have each $SU(4)$ itself decomposed into the native gates of the hardware -- typically CNOT gates and single-qubit rotations. These decompositions are usually chosen based on the ideal parameters of the gates. Here, we show how with some knowledge of the noise, the compilation strategy can be such that the gate parameters either compensate for or cancel out the correlated error. These strategies use only the tensor networks estimated via our methods so far.

\begin{figure*}[!htb]
	\centering
	\includegraphics[width=\linewidth]{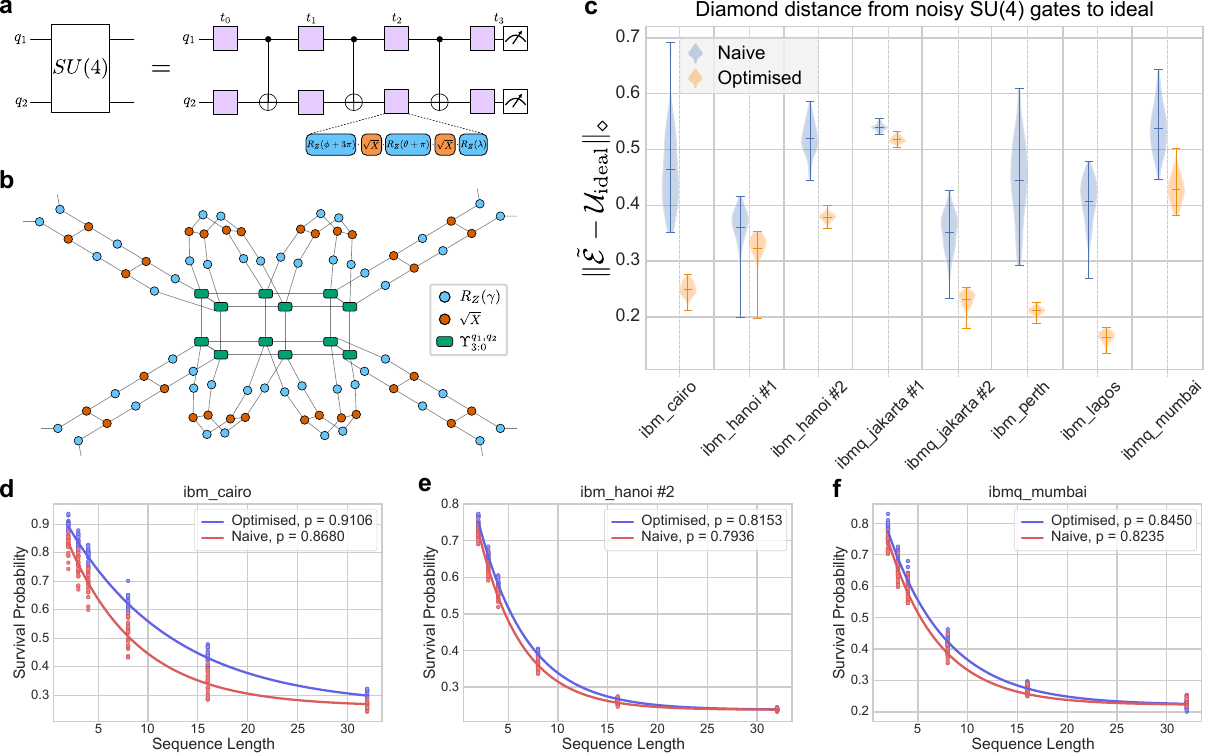}
	\caption[Scheme to characterise and improve arbitrary $SU(4)$ gate decompositions. ]{Scheme to characterise and improve arbitrary $SU(4)$ gate decompositions. We apply our characterisation tools to understand and optimise generic $SU(4)$ gates for the underlying device noise, accounting for crosstalk and non-Markovianity. \textbf{a} Any unitary operation in $SU(4)$ may be decomposed into three CNOT gates and eight local unitaries on each of the two qubits. For IBM Quantum devices, the local unitaries may be further decomposed into physical $\sqrt{X}$ gates and virtual $R_Z(\lambda)$ rotations. This naturally defines a three-step, two-qubit process tensor. The measurements are then used in the characterisation process. \textbf{b} We show a tensor network representation of the two qubit gate. Each CNOT can be viewed as a step of the process, and the $R_Z$ gates as inputs to that process. Once the complex noise is learned, we may then optimise the $R_Z$ parameters so that the effective channel is as close as possible to any two-qubit gate of our choosing. 
	\textbf{c} We characterise the described scenario on a host of IBM Quantum devices. The single characterisation is then used to optimise 100 randomly generated two-qubit unitaries. The plot shows predicted optimal diamond distances between the noisy gates and the ideal across the distribution. We see significant improvements across many of the devices.
	\textbf{d}--\textbf{f} A sample of resulting predicted randomised benchmarking curves from the devices in \textbf{c}. Although average gate fidelity improvements are not as stark as diamond distances, we still see meaningful corrections in the fidelity are possible from our noise-aware choice of parameters.
	}
	\label{fig:su4-opt}
\end{figure*}

Consider, then, the problem of constructing an arbitrary two-qubit gate in terms of CNOTs and local rotations. This decomposition is given in Figure~\ref{fig:su4-opt}a. We can recognise this gate decomposition as a three step, two qubit process tensor. Here, the CNOTs are a fixed part of the process, and the process accepts the single qubit gates as its input. We have so far seen that this is well within our capabilities to characterise. Let $\Upsilon_{3:0}^{q_1,q_2}$ be the two-qubit process tensor across three steps as represented by the given circuit. Each gate is decomposed into virtual $Z$ rotations and $R_X(\pi/2)$ pulses and the resulting process tensor estimated using the tensor network methods therein. This gives us an estimate for the \emph{effective} noisy two-qubit gate for any values of local unitary operations:
\begin{equation}
	\begin{split}
		\mathcal{E}^{q_1,q_2}(\vec{\theta},\vec{\phi},\vec{\lambda}) &= (u_1 \otimes u_2) \circ \mathcal{E}_{2:1;1:0}^{q_1,q_2 \ (u_3,u_4,u_5,u_6)} \circ (u_7 \otimes u_8),\quad \\&\text{where }
		\hat{\mathcal{E}}_{2:1;1:0}^{q_1,q_2 \ (u_3,u_4,u_5,u_6)} =\\
		 &\Tr_{\mathfrak{o}_0,\mathfrak{o}_1,\mathfrak{i}_2,\mathfrak{o}_2,\mathfrak{i}_3}\left[\Upsilon_{3:0}^{q_1,q_2} \cdot \hat{u}_3\otimes \hat{u}_4\otimes \hat{u}_5\otimes\hat{u}_6\right].
	\end{split}
\end{equation}
Here, half of the local unitaries ($u_1$, $u_3$, $u_5$, $u_7$) are applied to the first qubit, and the other half $u_2$, $u_4$, $u_6$, $u_8$ applied to the second qubit. These are interleaved between the CNOTs, as shown in Figure~\ref{fig:su4-opt}a.
The construction here allows for a noise-aware compilation of quantum circuits without needing to be repeated for the specific circuit parameters. It also permits, for instance, immediate quantum process tomography for whatever set of parameters we choose. We can hence investigate the properties of arbitrary $SU(4)$ operations, and optimise the choice of local gates such that the generic two-qubit unitary is as precise as possible. 

Once the dynamics are characterised to estimate $\Upsilon_{3:0}^{q_1,q_2}$ for the stated circuit in Figure~\ref{fig:su4-opt}a, the aforementioned optimisation is performed:
\begin{equation}
	\label{eq:su4-opt}
	\argmin_{\vec{\theta},\vec{\phi},\vec{\lambda}} 1 - \Tr(\hat{\mathcal{E}}^{q_1,q_2}(\vec{\theta},\vec{\phi},\vec{\lambda})\cdot \hat{U}_{\text{target}}^\dagger),
\end{equation}
to find the optimal decomposition of any two-qubit gate on our quantum device. The tensor network for this expression is shown in Figure~\ref{fig:su4-opt}b. Illustrated here is (i) the process, which encodes the action of the noisy CNOTs and any other background dynamics; (ii) the noisy physical $\sqrt{X}$ gates; and (iii) the virtual $R_Z$ gates whose rotation values are the parameters of the $SU(4)$. For any choice of parameters, the tensor network contracts to provide the effective mapping of that noisy $SU(4)$ implementation. Hence, we can extract complex properties, such as the diamond distance from the noisy gate to the ideal, for any choice of parameter value without re-doing quantum process tomography for each gate. Only a single iteration of PTT is required. It is these parameter values over which we perform the optimisation in practice. 

We characterised the stated process across a set of eight implementations on IBM Quantum devices. Each demonstration used $3\times 10^6$ shots divided across 3000 different circuits, and achieved a converged characterisation well within shot noise as tested on validation data with an average reconstruction error of $\lessapprox 10^{-4}$. Where the same device is used twice, a different set of qubits is chosen. To analyse the controllable noise profile, we generated 100 random two-qubit unitary operations and performed the optimisation given in Equation~\eqref{eq:su4-opt}. We then compared these noisy gates to the effective map one would get if they were to chose ideal gate parameters (a naive scenario). 

Our first comparison is to compute the diamond distance between the naive and the optimised implementation of each noisy gate $\tilde{\mathcal{E}}$ and its ideal counterpart $\mathcal{U}_{\text{ideal}}$. Recall that the diamond distance is a measure of maximal distinguishability of two channels when applied to some possibly entangled reference state~\cite{wilde_2013}
\begin{equation}
	\|\tilde{\mathcal{E}} - \mathcal{U}_{\text{ideal}}\|_\diamond = \frac{1}{2}\sup_{\rho}\|(\tilde{\mathcal{E}}\otimes \mathcal{I})[\rho] - (\mathcal{U}_{\text{ideal}}\otimes \mathcal{I})[\rho]\|_1.
\end{equation}
Importantly, this is an honest measure of gate performance inclusive of coherent errors and relevant to the actual error rates enforced in quantum error correction thresholds~\cite{aharonov1997fault,PhysRevA.99.022313}. In general, the diamond error of a gate can be an order of magnitude larger than its average gate infidelity. We compute this measure for each of the 100 randomly generated unitaries on both the naive and optimised parameter choices for each IBM demonstration. The results are displayed in Figure~\ref{fig:su4-opt}c. In two cases, ibm\_hanoi \#1 and ibmq\_jakarta \#1, the error reduction is only marginal, indicating that the noise profile on these qubit pairs is dominated by stochastic, Markovian effects. In most cases, however, the ability to deterministically suppress both correlated and coherent noise results in substantial improvements of both the mean and the variance across the distribution of gates. This highlights the pathology of noise in present-day quantum devices: it can be correlated, coherent, and complex -- contrary to the typical expectations of error correction. Our results highlight the need to incorporate bespoke noise profiling in order to gain the best performance out of noisy quantum devices. 

Since our models can compute expected distributions for sequences of gates, we can also evaluate average gate fidelity improvements by way of randomised benchmarking (RB) models. That is, we can optimise each unitary and compose these together into sequences to simulate the effects of RB experiments. 
Some representative curves are shown in Figure~\ref{fig:su4-opt}d-f. For expected improvement $r$ in the diamond error, the corresponding RB fidelity improvement is $\mathcal{O}(r^2)$. Nevertheless, these average-case gains are still tangible and can improve on a few percent of these gates' error budgets.

Although we do not have demonstrations illustrating such, we also point out that our characterisation can offer a much more accurate estimate of when a gate can be improved with the removal of one or more CNOTs, trading off a reduction in stochastic error for a less expressive circuit. 
We can characterise $\Upsilon_{3:0}^{q_i,q_j}$ whilst measuring at each step. This simultaneously estimates $\Upsilon_{2:0}^{q_i,q_j}$ and $\Upsilon_{1:0}^{q_i,q_j}$. The advantage to this is that we can explore a depth-approximation trade-off. Let $U$ be a Haar random unitary gate from $SU(4)$. Now, let $U_i$ be the best possible decomposition of $U$ into $i$ CNOTs and local unitary rotations. Moreover, let $\alpha_i = \|U_i - U\|$ be the distance between the $i$-CNOT decomposition and the target unitary. Clearly, $\alpha_3 = 0$, since any two-qubit gate can be perfectly decomposed with three CNOTs. Suppose, now, given a noisy device, we have $\tilde{U}$, the noisy implementation of $U$. Let $\epsilon_i = \|\tilde{U}_i - U_i\|$. In general, this means that $\|\tilde{U}_i - U\|$ will roughly depend on both $\alpha_i$, the approximation cost; and $\epsilon_i$, the noise cost. There is a tension in the following:
\begin{equation}
	\begin{split}
		&\alpha_3 \leq \alpha_2 \leq \alpha_1 \leq \alpha_0\\
		&\epsilon_3 > \epsilon_2 > \epsilon_1 > \epsilon_0.
	\end{split}
\end{equation}
An extra CNOT allows one to get closer to the mathematical ideal of $U$, but also introduces extra noise in the noisy decomposition. Whether $\alpha_i > \epsilon_i$ or $\epsilon_i > \alpha_i$ is highly dependent on the specific $U$. For example, a two-qubit unitary which is close to not entangling will not benefit from having a three CNOT decomposition. The extra CNOTs will only serve to introduce extra noise. Our two-qubit process tensor allows us to accurately assess this trade-off. We have already seen that we can optimally find the best $\tilde{U}_3$ for a given unitary, now we see that we can compute the smallest $\delta_i$, where $\delta_i = \|\tilde{U}_i - U\|$. This allows us to select between a 0, 1, 2, or 3 CNOT decomposition of the unitary. Hence, we are optimising both for the least noise, and the most favourable approximation error. 

This notion of approximating two-qubit gates in this fashion was explored in Ref.~\cite{Cross-QV}, but adopted the coarse approximation that each gate incurred the same error of a depolarising form. Here, where we know the noise to be more structured, the process characterisation will concretely determine what the noise looks like and what the effects are of approximating. That is to say, that both $\alpha_i$ and $\epsilon_i$ will depend both on the noise of the device, and the unitary being compiled. 

\begin{figure*}[!t]
	\centering
	\includegraphics[width=\linewidth]{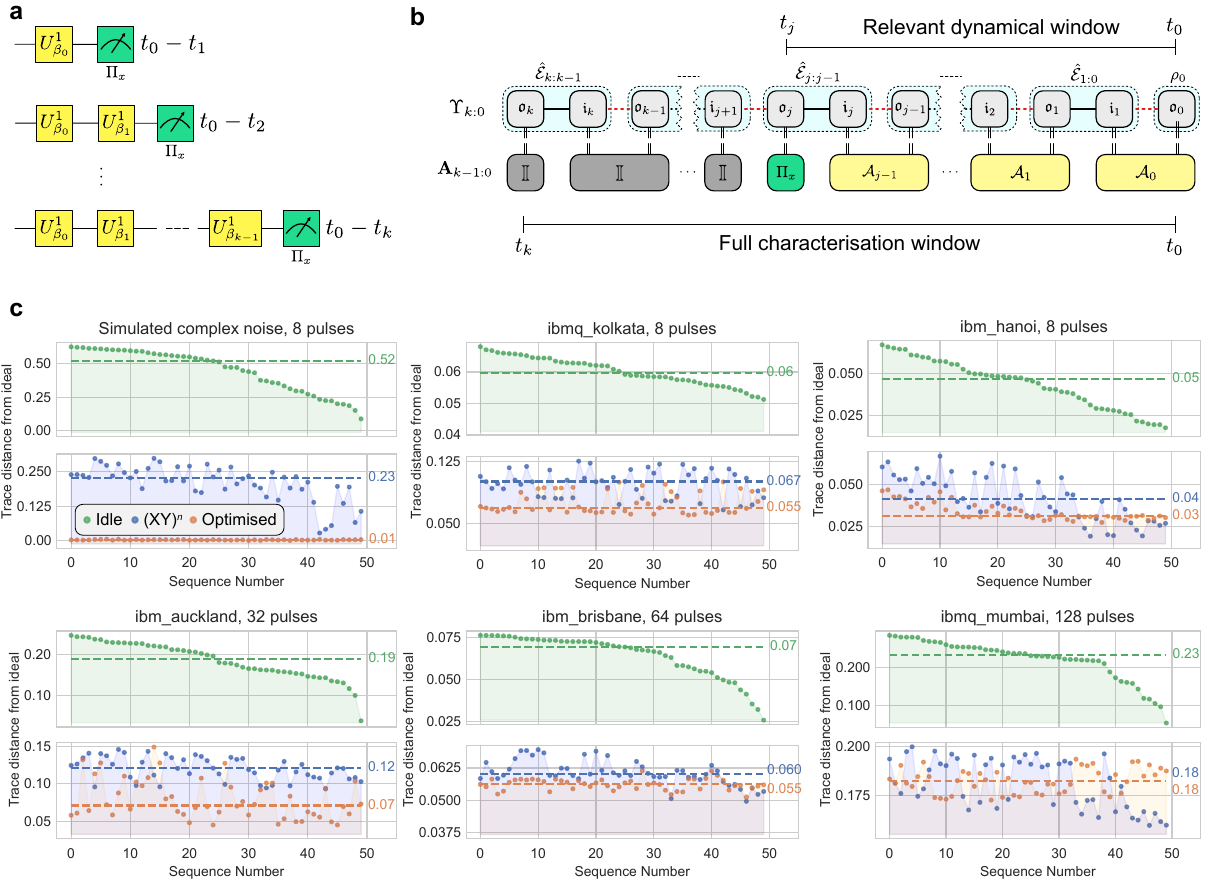}
	\caption[Protocol and predictions for optimisation of dynamical decoupling sequences. ] {Protocol and results for optimisation of DD sequences. To characterise a variety of differently sized windows, a maximum duration can be selected, from which a particular window is used where necessary. \textbf{a} Characterisation takes place using tensor network characterisations, where random Clifford operations are applied. Now, however, the sequence is terminated at different points so differently sized windows can also be characterised. 
	\textbf{b} The process tensor for $t_0$ to $t_k$ is reconstructed, but when a relevant target window $t_0-t_j$ is selected, the remainder of the process can be traced out. The remaining window is then used to optimise a DD sequence for that specific timing.
	\textbf{c} Simulated and experimental results using process tensors of varying lengths to predict optimised DD sequences.
	For the given window, a random state is created, then DD sequence applied, and then compared with the perfect state at the end. We compare the case of completely idle dynamics, a standard $XY4$ sequence, and our noise-aware compilation. On the $y$-axis, we display the trace distance between ideal states and the noisy state. For each of idle dynamics, $XY4$-protected sequences, and optimal DD, we plot the median values as a horizontal line. The results displayed indicate a variety of different behaviours: simulation shows in principle how bespoke methods can perform better; and IBM devices show a range of behaviours where standard DD and bespoke sequences both do and do not perform better, giving insight into control opportunities. 
 }
	\label{fig:dd-results}
\end{figure*}

\subsubsection{Optimising Dynamical Decoupling Sequences}

Another straightforward control application of our methods is in the maintenance of idle qubits. For any device without complete connectivity, a quantum algorithm performed will invariably leave some of the qubits sitting idle for some amount of time, where they may interact with the environment around them. A well-known method to combat this is the application of a dynamical decoupling (DD) sequence~\cite{PhysRevLett.82.2417}, whereby control pulses are tailored to universally cancel out the noise from some Hamiltonian to a certain order. Although optimal sequences have been well-characterised with respect to universally cancelling out errors up to a desired order~\cite{PhysRevA.75.062310, PhysRevA.84.042329}, the trade-off in generality is performance. It has been shown that optimised DD sequences can outperform off-the-shelf sequences, accounting for the specific decoherence modes and imperfect pulses~\cite{PhysRevA.88.052306,biercuk2009optimized}. 

Here, we present a method by which efficient characterisation can inform noise-aware DD sequences on a device. The approach is naturally hardware agnostic, and allows the optimisation process to be performed entirely offline rather than with live feedback.
Moreover, since the characterisation is self-consistent, it accounts not only for any system-bath couplings, but any imperfections in the control sequence or incorporations of dynamically corrected gates~\cite{PhysRevLett.102.080501}. 

When we test this on simulated multi-axis noise, we find that our method can nearly perfectly cancel out the effects of the complex Hamiltonian, and significantly out-performs the well-known $XY8$ sequence~\cite{gullion1990new}.
We also test it across a range of IBM Quantum devices for 8, 32, 64, and 128 gate pulse sequences. Although we always find optimal decoupling from the non-Markovian environment, there is little variation between the improvement over idle (doing nothing) sequences, and applying repetitions of $XY4$. An improvement over doing nothing is indicative of non-Markovian correlations that may be cancelled out. An improvement over $XY4$ indicates that the device has some level of complex (multi-axis) non-Markovian noise, since $XY4$ sequences perfectly cancel correlated $Z$ errors. 

Suppose an experimenter has a qubit that they wish to keep coherent for a fixed time window, from $t_0$ to $t_k$. We can use process tensor information to achieve this. Given a process estimate $\Upsilon_{k:0}$, one can trace over the initial state to obtain a sequence of possibly correlated CPTP maps 
\begin{equation}
	\Tr_{\mathfrak{o}_0}[\Upsilon_{k:0}] = \hat{\mathcal{E}}_{k:k-1;k-1:k-2;\cdots;1:0}.
\end{equation}
Applying a sequence of $k-2$ gates $\{D_1, D_2,\cdots,D_{k-1}\}$ generates the conditional dynamical map 
\begin{equation}
	\hat{\mathcal{E}}_{k:0}^{(D_1,D_1,\cdots,D_{k-1})} = \Tr_{\bar{\mathfrak{o}}_k \bar{\mathfrak{i}}_k}\left[\hat{\mathcal{E}}_{k:k-1;k-1:k-2;\cdots;1:0}\cdot \bigotimes_{i=1}^{k-1} \hat{\mathcal{D}}_i^{\text{T}}\right].
\end{equation}
That is, it is the dynamical map describing evolution from time $t_0$ to $t_k$ while accounting for the control operations applied by an experimenter.
If 
\begin{equation}
	\|\hat{\mathcal{E}}_{k:0}^{(D_1,D_1,\cdots,D_{k-1})} - \hat{\mathcal{I}}\| < \|\hat{\mathcal{E}}_{k:0}^{(I_1, I_2, \cdots , I_{k-1})} - \hat{\mathcal{I}}\|,
\end{equation}
for some norm $\|\cdot\|$, then the set of operations $\{D_i\}$ has (possibly imperfectly) decoupled the system from the environment: it has left the effective dynamics cleaner than doing nothing. If we find $\{D_i\}$ such that 
\begin{equation}
	\label{eq:dd-opt}
	\|\Tr_{\mathfrak{o}_0}[\Upsilon_{k:0}^{(D_0,D_1,\cdots,D_{k-1})}]- \hat{\mathcal{I}}\|
\end{equation}
is minimised, then we say a given set of gates is optimally decoupling. That is to say, the channel from time $t_0$ to time $t_k$ can be optimised to be as close to the identity as possible. In our work we optimise for the trace distance between the conditional process tensor and the identity channel.

It may seem that this approach is overly specific to the window $[0,t_k]$ that has been characterised. Often the idle time a qubit experiences in a circuit is highly variable. 
But, if measurements are implemented at earlier times (i.e., data collecting in the form of Figure~\ref{fig:dd-results}a) then the final state is known at any time up to and including $t_k$.
Hence, when a circuit has idle time shorter than this, we can still optimise for the relevant window and trace over the remainder, as depicted in Figure~\ref{fig:dd-results}b. From this, we have a protocol which is both modular, and extendable or retractable to arbitrary numbers of times. This reduces the burden of characterisation, and means that only a single round is required to be able to discern the DD sequence for any time window. 

We analyse this optimisation both in simulation and on IBM Quantum devices. In the IBM demonstrations, wait times of 800~ns were factored in between each gate with random two-qubit unitaries applied to each neighbouring qubit with that qubit's neighbour. This is with the exception of the 64 and 128 pulse sequences, which had zero wait time between gates.
In our simulation, we take a qubit coupled via exchange interaction to a another two-level system. This models potential defects in superconducting devices~\cite{muller2019towards} as well as nuclear spins and other qubits in solid-state hardware~\cite{He2019,yoneda2022noise}. 
For a circuit of $k$ steps, we first reconstruct the process tensor $\Upsilon_{k:0}$ using our self-consistent tensor network representation. The circuit is structured so that the first gate will be used to create some state, and then the following $k-1$ gates apply a decoupling sequence.
Each process tensor is estimated using 3000 circuits of random Clifford operations run at 1024 shots. Each one is additionally validated against 300 circuits from outside the characterisation set to a reconstruction error of $\lessapprox 10^{-4}$.
The optimisation in~\eqref{eq:dd-opt} is then performed on a classical computer. Next, 50 random unitary operations are selected to initialise a quantum state, the decoupling sequence is applied to that random state, and then the error at the end of the circuit is determined. By error, we use the trace distance between the noisy $\frac{1}{2}\|\tilde{\rho}_i^{\text{(DD)}} - \rho_i^{(\text{ideal})}\|_1$. Note that the DD sequence is applied equally to all states. 

The results are shown in Figure~\ref{fig:dd-results}c. In top panels, we plot the state errors for the idle dynamics ordered from highest to lowest. In the bottom panels, we plot the corresponding information for repetitions of $XY4$ as well as the optimal decoupling sequence found. The median values of each sequence is plotted as a horizontal line. In all cases, the optimised sequence is able to better protect a state against errors than applying no control at all. However, in some instances (notably on ibmq\_kolkata), the difference is slight. We also highlight the optimised 128 gate sequence on ibmq\_mumbai, which is able to better protect the information but is essentially the same as the $XY4$ sequence repeated 32 times. 
The dominant correlated error channel here is correlated dephasing along the $Z$ axis. On ibm\_auckland, there is a clear separation between the three sets of states. The upshot here is that the noise is more complex, and hence one benefits from a tailored profiling.

These results warrant further structured investigation into finding optimal DD sequences in different scenarios, and comparing with standard protocols to determine if a priori characterisation is required.
It would also be interesting to see the performance in actual quantum algorithms, or on hardware platforms with different noise characteristics to the fixed-frequency transmons of IBM Quantum devices. In the simulated instance, the clear disparity between sequences indicates the potential power of bespoke DD sequences for multi-axis noise. In IBM systems, this appears not to always be the case. In some instances (ibm\_auckland and ibmq\_mumbai) there is an error reduction between 20 and 60\%. In other cases, we were not able to substantially improve coherence in these instances with tailored DD sequences. Our analysis reveals if and when a device might benefit from optimised control.

Nevertheless, we have established a cheap and straightforward method through which bespoke DD sequences can be applied to any given noisy setup, and are hence optimal in the number of pulses for that given setup. However, suppose that correlations persisted across different time-scales (i.e. short time and long time). Then an MPO estimation may not appropriately suffice to determine these memory structures; the resulting optimal DD sequence would only cancel out the short-time correlations rather than the long-time ones. There are two immediate approaches that one could consider to solve this problem. First, the short-time dynamics could be characterised and an optimal DD sequence found. Then, this sequence could be fixed in the process, such that all fast errors were cancelled. Next, characterise a longer time-scale process tensor around the fixed sequence. After optimising for this DD, one would then find the concatenation of an inner and an outer DD sequence which respectively cancelled the short time and long time interactions. Alternatively, one could use more sophisticated tensor networks -- such as tree tensor networks or the multiscale entanglement renormalisation ansatz~\cite{PhysRevLett.101.110501} -- which are designed to capture different scales of correlation~\cite{dowling2023process}, estimate these and then optimise for a DD sequence. 

\section{Technical Methods}
\label{sec:technical-methods}
Although Section~\ref{sec:background} provided a high level summary of the technical methods employed in this work, here we will discuss the details of our model and implementation. In particular, the results discussed in this section are essential for constructing the models used to perform the various characterisation demonstrations in Section~\ref{sec:applied-results}. We will first elucidate our proposed anatomy of quantum noise in Section~\ref{sec:categories}, followed by a derivation of self-consistent PTT in Section~\ref{app:nmgst-derivation}---which places these concepts on firm mathematical footing. Lastly, in Sections~\ref{sec:PT-LPDOs} and~\ref{sec:self-consistency} we construct tensor network representations of the above models and develop the requisite methods for estimating them in this compact form.

\subsection{Categorising and Modelling Correlated Quantum Noise}\label{sec:categories}

This work presents an operational viewpoint which is designed to be expressive, capturing and controlling dynamical effects not treated elsewhere in the literature. Moreover, we endeavour to make the approach modular so that one can straightforwardly make changes based on expected physics.
Before we get into the actual methodology of characterisation, we first discuss in this section our partitioning of quantum dynamical effects that might reasonably be construed as `noise'. The classification is, of course, subjective -- 
and the intent to is not be prescriptive -- but we aim to meaningfully motivate our choices according to the philosophy of what an experimenter can and cannot control.

Let us emphasise that from the perspective of quantum stochastic processes with idealised control, there is no ambiguity on the matter. Non-Markovianity refers to a process whose dynamics do not factor into the product of dynamical maps. 
In the context of quantum computing, however, gate error cannot be taken as idealised, and processes can change in the midst of their characterisation. Temporal correlations, then, are not only witnessed by control operations but can indeed by the source of them. 
A self-consistent model must therefore make the distinction between control and process error, which is not often done in practice. 
We base our discussion around a separation between process, control, and the interplay between the two. All noise can be broadly categorised in these camps. Our framework thus models each of these in the language of process tensors, superprocesses, and testers as depicted in Figure~\ref{fig:superprocess}.

\subsubsection{Delineating Process and Control}

\begin{figure}[!b]
	\centering
	\includegraphics[width=0.95\linewidth]{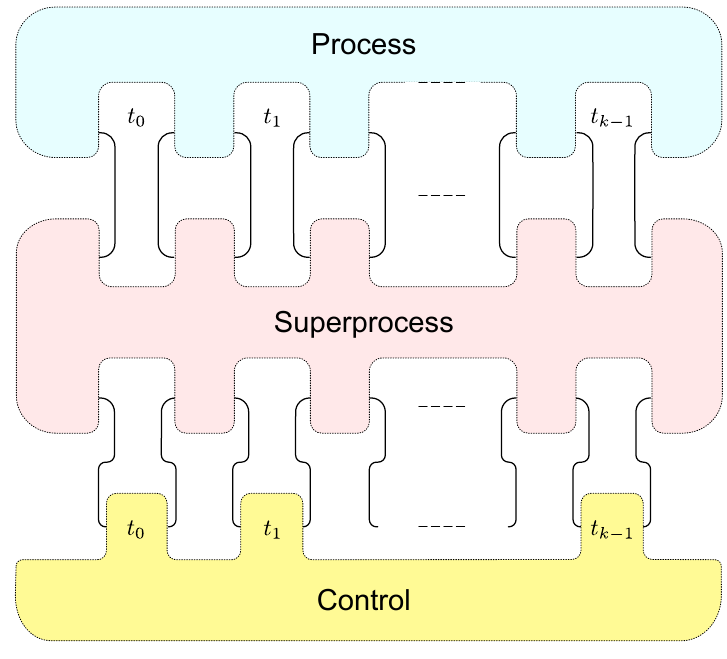}
	\caption[General model of driven quantum dynamics with arbitrary errors ]{General model of driven quantum dynamics with arbitrary errors. We conceptually partition process, control, and spillage effects from one another. These are respectively represented by process tensors, testers, and superprocess objects. This formalism captures a general setting of system-environment interaction, faulty correlated control, and active overlap between control and environment.}
	\label{fig:superprocess}
\end{figure}

The historical issue with defining quantum stochastic processes stems from the non-commutativity of observables at different times~\cite{Milz2021PRXQ}. This was resolved with the insight that the underlying process and experimenter-implemented control need to be formally separated to account for the non-commutativity of observables at different times. Consequently, we resolve all driven dynamics into two objects: process tensors and their dual, testers -- respectively process and control in Figure~\ref{fig:superprocess}. Testers are multi-time instruments: they probe the system with the possibility of carrying memory within the control operations~\cite{Milz2021PRXQ}.

One might argue that there is an epistemic difference between process tensors and testers, but not an ontological distinction. In this context, testers can be viewed as the process that we, as experimenters, have control over. This perspective assumes a certain state of knowledge about the process being implemented as a control operation.
However, when we relax these assumptions, the distinction between the two objects becomes less clear. Intuitively, the process is the piece of dynamics that occurs in each experiment, while control is what allows one experiment to be differentiated from another at the behest of the experimenter. 
That is to say, some elements are under direct experimenter control, some are under indirect experimenter control, and others are not at all controllable.
We start with this formal separation of processes as a foundation to categorise, model, and estimate non-Markovian open quantum systems in full generality. We consider registers of qubits in this discussion but note that it is also valid with respect to $d$-dimensional systems.

To start with, suppose we have a quantum device represented by a series of qubits $\mathbf{Q}_N :=\{q_1,q_2,\cdots, q_N\}$.
We define the system $S\subseteq \mathbf{Q}_N$ whose state space is $\mathcal{H}_S = \prod_{q_i\in S} \mathcal{H}_{q_i}$.
Everything else, \emph{including the device qubits not in $S$}, is considered to be the environment $E$. This distinction is important for clarifying whether crosstalk-induced dynamics are Markovian or non-Markovian. The system is permitted to evolve across a window $[t_0,t_k]$, divided into $k$ steps to define the set $\mathbf{T}_k:=\{t_0,t_1,\cdots, t_k\}$. Across $\mathbf{T}_k$, the system is manipulated by a multi-time instrument $\mathbf{A}_{k-1:0}$, followed by a POVM $\{\Pi_x\}$. The instrument may be time local (factor into a tensor product of operations), but it need not be. 

We start with our first definition, covering the underlying dynamics.
\begin{definition}[Process Error]
	A process error is when, without experimenter intervention, we have $S$ evolving according to some Hamiltonian $H$ which differs from the ideal dynamics prescribed by $H_{\text{ideal}}$. $H$ may contain interaction terms with the nearby bath. That is, it is an always-on interaction between an experimentally accessible system and its experimentally inaccessible environment. 
\end{definition}
Process errors are exactly stochastic noise as we have discussed so far in this paper. It is stochastic noise that occurs no matter what the actions of an experimenter are. These effects are entirely encoded in the idealised model of a process tensor $\Upsilon_{k:0}$.
We next have perhaps the most conventional notion of an error in a quantum device, errors due to instruments themselves. 
\begin{definition}[Control Error]
	A control error is an operation, or sequence of operations, designed to manipulate a system whose manipulation of the system deviates from the intended effect. The dynamics should be (a) \underline{controllable:} the effect should be able to be switched on and off by the experimenter, and (b) \underline{not always-on:} the effect should not be present in all experiments, otherwise it is part of the process. This latter condition can equivalently be read as \emph{gate-dependent} noise.
\end{definition}
Control errors are dynamical effects on the system whose origin stems from control equipment. Thus, even if the control itself cannot be perfect, these effects can always be switched on or off at will be an experimenter. 

Finally, we consider another unintended, but distinct, consequence of imperfect control operations. 
\begin{definition}[Superprocess Error]
A superprocess error is a \emph{control operation} (may be turned on or off by the experimenter) which, (a) acts on an extended Hilbert space such that not only the system is manipulated, but part of the environment, and (b) does so in a way that modifies the part of the environment responsible for any future process error.
\end{definition}
See Fig.~\ref{fig:superprocess} for a depiction of a superprocess transforming a process. This terminology will become clear momentarily when we introduce spillage errors. Superprocess errors overlap with process and control errors. They are a subset of control errors, in that the physical origin is control equipment and the effect may be switched on or off by the experimenter. But since they also manipulate the broader environment with which the system interacts, these can modify the always-on interaction between $S$ and $E$, and hence change the nature of the process. These are not describable under the present process tensor framework.

\subsubsection{Informal Definitions}

We will now further subdivide these categories and offer some physical motivation for each case. Note that these categories have overlapping features, and the distinction we offer are to properties we believe are worth highlighting in particular. We introduce these notions informally here, and expand on the definitions as well as offering some concrete examples in Appendix~\ref{app:definitions}.

\emph{Fully Markovian.} Widely accepted in the quantum characterisation community as a definition of complete Markovianity is the case where a series of control operations $\{\mathcal{A}_i\}$ and time steps $\{\mathcal{E}_{j:j-1}\}$ factorise, i.e., they can be composed: $\mathcal{A}_{j} \circ \mathcal{E}_{j:j-1}\circ \cdots \circ \mathcal{A}_1 \circ \mathcal{E}_{1:0}\circ\rho_0^S$. 
These have been well studied in the literature~\cite{noise-coherence-2015,postler2022demonstration}, see especially Ref.~\cite{blume2022taxonomy}. This is consistent with our definition of both process tensor and tester represented as product states, and includes both gate dependent and independent noise. 

\emph{Active and passive crosstalk.} Crosstalk is a major source of correlated noise in almost all current hardware platforms. Crosstalk initiates either an undesired interaction between different qubits or an unwanted effect on a separate qubit as a result of controlling a different system. Under this umbrella, the separate qubits must be considered as the system under study.
This definition coincides with that of Ref.~\cite{Sarovar2020detectingcrosstalk}, with the added dichotomy that we distinguish between \emph{active} and \emph{passive} crosstalk. By active, we mean that the error on one qubit is a result of the application of control on a different qubit. This could be a weight-one error, such as a Stark shift~\cite{mckay2020correlated} from a nearby drive tone, or a higher weight error such as overlap in M\o lmer-S\o rensen gates. 
By passive, we mean that the undesired interaction between qubits is governed by some always-on Hamiltonian with non-local terms between the qubits.

\underline{\emph{Process non-Markovianity.}} The definition of non-Markovianity is well-established in the quantum stochastic literature~\cite{Pollock2018}. As a consequence of system-environment interactions, a process is non-Markovian if its process tensor does not factorise into a tensor product of dynamical maps. This might stem, for example, from correlations generated by a classical field, or interactions with a nearby coherent quantum system. We add the \emph{process} description here to once more indicate passivity: interactions which occur regardless of which gate is applied by the experimenter. 

\underline{\emph{Control non-Markovianity.}} The above class of non-Markovianity is not broad enough to capture all temporally correlated dynamics. The chief problem with this notion -- as it has been studied previously -- is that control operations are assumed to be fully understood and fully intentional. In reality, however, the systems used for control of a set of qubits might themselves be capable of mediating temporal correlations. We distinguish this from process non-Markovianity because these correlations might not manifest in every single experiment, they are dependent on the applied gate sequence. In this work, we use the framework of quantum testers (yellow comb at the bottom of Fig.~\ref{fig:superprocess}) to characterise control non-Markovianity. Testers, the dual to process tensors, can be time-non-local inputs to a process. As a consequence, if a sequence of gates is described by a tester whose Choi state does not factorise, then there is control non-Markovianity in the dynamics.

\underline{\emph{Time-dependent Markovianity, and drift.}}
In this work, we treat separately the notions of time-dependent, time-independent, and non-Markovian dynamics based on relevance to characterisation. Ideally, one knows the time at which dynamics occur exactly and can always condition on that time, rendering time-dependent processes perfectly describable under a Markov model. The fact that characterisation takes place over a non-trivial interval means this is no longer true in a practical setting. Suppose the Hamiltonian governing the dynamics of a system and its surrounds changes over a characteristic timescale $\tau$. Now, suppose we have two clocks: a circuit clock $\{t_0,t_1,\cdots,t_k\}$ at which gates are applied; and a wall clock $\{s_1, s_2, \cdots, s_N\}$ across which each shot of data is collected for a single circuit, and $\{c_1, c_2, \cdots, c_K\}$ at which each bucket of data is collected for a given circuit. There are three broad scenarios to consider:
\begin{enumerate}
	\item $\tau \sim \Delta t$: will admit a \emph{time-dependent} Markov model, where the dynamics factorise into dynamical maps indexed by time -- so long as those dynamics are periodic on the scale of the circuit, repeating when the circuit is re-initialised.
	\item $\tau \sim \Delta s$: is \emph{quastistatic} noise, where, for each shot of a circuit, a single stochastic variable is drawn from some stationary distribution and has the ability to affect all steps in the circuit. For example, the value of the magnetic field at that time. Because any characterisation will collect many shots for each circuit, and hence marginalise over the relevant timescales, the resulting model will look non-Markovian.
	\item $\tau \sim \Delta c$: is \emph{drift}, where dynamics change much slower than the time-scale of the device repetition rate. From bucket to bucket, the model of the dynamics may be inconsistent with one another. Once more, when marginalised, this looks non-Markovian.
\end{enumerate}

The main distinction between quasistatic noise and drift from a practical standpoint is that a model can capture quasistatic noise and be validated on future runs of that experiment because a future run will also marginalise the time-dependence across many runs of the same circuit in $\{s_i\}$. Whereas if the experiment drifts, then the $\{c_i\}$ are marginalised to produce a non-Markovian model, but the model may not be valid for any specific $c_i$. If the data is rasterised~\cite{su2023characterizing,proctor2020detecting} (each circuit is run once before repeating shots) then a drift model will be transformed into a quasistatic model, and PTT will appropriately characterise any drift as non-Markovianity. However, little can be done to suppress its effect in software.
We expand on each of these points in Appendix~\ref{app:definitions}. It is worth also emphasising that in the case where time-dependent effects manifest as time-correlated, the resulting memory will always look classical because the situation is a convex mixture of Markovian processes. That is, the resulting process tensor will be separable. Any detection of temporal entanglement will hence rule out marginalised time-dependent effects.

\underline{\emph{Non-linear spillage.}} We have so far considered two primary mechanisms for noise in quantum processors: (possibly correlated) background dynamics, and (possibly correlated) control operations. We lastly introduce a third possible mechanism, which we formally refer to as a \emph{superprocess} error, and colloquially as a \emph{non-linear spillage} error. We will usually refer to this only spillage for brevity. However, its non-linear nature sets it apart from usual quantum noise varieties. It is also its non-linearity that makes spillage very difficult to characterise and control.

This mechanism describes the situation where interventions manipulate not only the system, but also inadvertently the environment with which it interacts.
For example, a control operation implemented by an experimenter drives a range of frequencies, including some which are resonant with environment transitions. 
Since we do not have any reliable access to the environment, we cannot simply dilate the dynamics to include it in the model. Instead, we continue with our open systems philosophy and model how such effects can modify the dynamics with respect to our system. Because we are introducing a conceptually different framework to describe these types of dynamics, we will devote some more attention to the task. Specifically, we define spillage effects in the discussion as follows.

Spillage effects change the environment as a result of a control operation, which then affects the system non-trivially in the future.
At first glance, it might seem that spillage errors are the same as non-Markovian environment back-action. In both cases, we could have two instruments $\mathcal{A},\mathcal{B}$ applied at time $t_j$. Then, conditioned on these instruments, we have two different future dynamics at some time $t_{j'}$: $\mathcal{E}_{j':j'-1}^{(\mathcal{A})}$ and $\mathcal{E}_{j':j'-1}^{(\mathcal{B})}$. The distinction here is that spillage errors need not be linear in the control. That is, for control operation $\mathcal{C} = \alpha\mathcal{A} + \beta\mathcal{B}$ we may have 
\begin{equation}
	\mathcal{E}_{j':j'-1}^{(\mathcal{C})}  \neq \alpha\mathcal{E}_{j':j'-1}^{(\mathcal{A})} + \beta \mathcal{E}_{j':j'-1}^{(\mathcal{B})}.
\end{equation}
Hence, the above scenario cannot be described with process tensors alone. 

Physically, spillage is similar to the case of active crosstalk, but with environment degrees of freedom. Spillage incorporates well-studied active \emph{leakage} errors, where the state of a qubit leaves the computational subspace as a consequence of active driving~\cite{varbanov2020leakage,strikis2019quantum}. In short, spillage describes the implementation of control whose action affects the environment. This means that there is an important distinction to note here: with non-Markovian quantum processes, an instrument modifies a system at some time $t_i$ and then, by virtue of a naturally occurring system-environment interaction, this affects the environment before $t_{i+1}$. In the case of spillage, however, the control operation could taken to directly map the state of the environment. That is,
\begin{equation}
	\mathcal{A}_j\in \mathcal{B}(\mathcal{H}_S\otimes \mathcal{H}_E).
\end{equation}
In practice, we expect this to only be an incredibly small subspace of $E$. Nevertheless, we wish to consider descriptions that reference only the system in its model.

To discuss spillage in the language of process tensors we employ \emph{superprocesses}. A superprocess~\cite{PhysRevA.81.062348} is a way of describing transformations on a particular open quantum evolution, and are typically used in the context of resource theories~\cite{berk,berk2021extracting}. A higher order map $\mathbf{Z}_{k:0}$ is a mapping from process tensors to process tensors, or equivalently its dual action is from control sequences to control sequences. To ease the subsequent discussion, we will employ a vectorised notation $\Upsilon_{k:0}\rightarrow |\Upsilon_{k:0}\rangle\!\rangle$. Written in this superoperator representation, the multi-time expectation value of some sequence of control operations $\mathbf{A}_{k:0}$ with respect to a process \pt{} is 
\begin{equation}
	\langle\!\langle \hat{\mathbf{A}}_{k:0} | \Upsilon_{k:0}\rangle\!\rangle.
\end{equation}

A superprocess extends this notion to 
\begin{equation}
	\langle\!\langle \hat{\mathbf{A}}_{k:0} |\mathbf{Z}_{k:0}| \Upsilon_{k:0}\rangle\!\rangle,
\end{equation}
where, implicitly, $\mathbf{Z}_{k:0}$ retains causal order of control operations on input and output spaces. In other words, superprocesses are completely positive and causality-preserving transformations of process tensors: $|\Upsilon_{k:0}\rangle\!\rangle \mapsto \mathbf{Z}_{k:0} |\Upsilon_{k:0}\rangle\!\rangle  = |\Upsilon_{k:0}'\rangle\!\rangle$. Spillage noise, as we have defined it, is a control action on both system and environment. We can view this equivalently then as a type of transformation on the process tensor. By definition, the control operation $\mathcal{A}_0$ is applied at time $t_0$. If $\mathcal{A}_0$ acts on the system only, then the future dynamics are well described by $\Upsilon_{k:1}^{(\mathcal{A}_0)} = \Tr_{\mathfrak{i}_1\mathfrak{o}_0}[\hat{\mathcal{A}}_0^{\text{T}} \Upsilon_{k:0}]$. But if $\mathcal{A}_0$ also acts non-trivially on the environment, then we still obtain a valid future process, but it is no longer defined by contraction of $\hat{\mathcal{A}}_0$ into $\Upsilon_{k:0}$. Instead, we denote this by $\Upsilon_{k:1}'^{(\mathcal{A}_0)}$, where 
\begin{equation}
	\label{eq:superprocess-expectation}
	|\Upsilon_{k:1}'^{(\mathcal{A}_0)}\rangle\!\rangle = \mathbf{Z}_{k:0}^{(\mathcal{A}_0)} |\Upsilon_{k:0}\rangle\!\rangle.
\end{equation}

Here, we have introduced $\mathbf{Z}_{k:0}^{(\mathcal{A}_0)}$ as the superprocess which (i) implements (some attempt at) $\mathcal{A}_0$, and (ii) transforms $\Upsilon_{k:1}$.
This is the situation described earlier in Figure~\ref{fig:superprocess}. 
The transformation is still extremely general. After all, the new process could be \emph{anything}. In principle, we could throw away our previous environment and replace it with something entirely different. 
To ensure that the transformation is not too significant, we once more incorporate the language of tensor networks to model a sparse set of spillage errors only. 

Note, however, that we do not actually \emph{require} the superprocess representation. Since the spillage effect is intrinsically tied to the control operation, we can always absorb $\mathbf{Z}_{k:0}^{(\mathcal{A}_0)}$ into $\mathbf{A}_{k:0}$ to create $|\mathbf{A}_{k:0}'\rangle\!\rangle := \mathbf{Z}_{k:0}^{(\mathcal{A}_0)} |\mathbf{A}_{k:0}\rangle\!\rangle $. That is to say, the superprocess cannot be varied independent of $\mathcal{A}_0$. The two models are physically indistinguishable, and so we opt for the simpler one.
We described the situation with superprocesses for conceptual reasoning, but in practice this shows that we do not need to explicitly model the superprocess, as it is always contained in the tester.
In full generality, the bond dimension of $\mathbf{A}_{k:0}'$ would be the dimension of the future process tensor $\Upsilon_{k:j+1}$ itself, but in practice we can significantly compress things so long as the control only modifies a small part of the environment. 
This concludes the discussion regarding the different facets of non-Markovian noise; we can now turn to its determination.

\subsection{Derivation of Self-Consistent Process Tensor Tomography}
\label{app:nmgst-derivation}

Now that we have formally categorised various categories of noise processes, we wish to develop methods to characterise each of these noise processes in practice. This poses the challenge of separating control noise from process noise, and spillage from each of these.

To do so, we now derive a method to incorporate both experiment design and reconstruction of these general process and control dynamics. This is intended to be in the spirit of GST and guided by the derivation of Ref.~\cite{greenbaum2015introduction}: the gate set is not taken for granted, and instead we aim to self-consistently estimate the underlying dynamics. With this formalism in hand, we will then turn again to tensor networks to aid our practical compression and estimation in an experimental setting.

The tenets of GST are highly relevant to near-term quantum devices: SPAM errors are significant, and gates not perfect. Thus, it is important to account for each of these noisy mechanisms individually in order to paint a holistic picture of device performance. But GST has its drawbacks. Most notably, it adopts a totally Markovian model. Probabilities are obtained from the model by matrix multiplication, which is only an accurate representation if all dynamics (both process and control) are completely time local, an inadequate assumption in practice. Moreover, gates themselves as well as the surrounding process are combined into the one object, and no distinction is made between the two.
We have seen the ability of process tensors to encode temporal correlations, as well as to be estimated in practice~\cite{White-NM-2020,White-MLPT,white2021many,white2023filtering}. 
We wish now to further develop our formalism to incorporate the philosophy of GST -- to assume only a model and not the parameters of our control operations -- but to also characterise non-Markovian processes as we have so far done. The result is a robust estimate of the process tensor, as well as the possibly-noise interventions used to probe that process.

The language of GST is to start with a gate set $\mathcal{G} = \{\kket{\rho},\bbra{E}, G_0, G_1,G_2,\cdots,G_K\}$ which includes a reliably preparable initial state $\rho$, a POVM $E$, and the set of control operations that an experimenter may implement $\{G_j\}$. 
The set of gates contains two restrictions: it must include the special case of $G_0 = \{\}$, i.e. the do-nothing-for-no-time gate~\footnote{The $G_0$ gate must not simply have its target be the identity, it must be a perfect implementation of the identity superoperator. Experimentally, this amounts to doing literally nothing.}; and compositions of the gates must be able to be able to generate an IC set of state preparations and measurements.
The experiment design seeks to simultaneously estimate the entire gate set self-consistently. 

\begin{figure*}[!t]
	\centering
	\includegraphics[width = 0.85\linewidth]{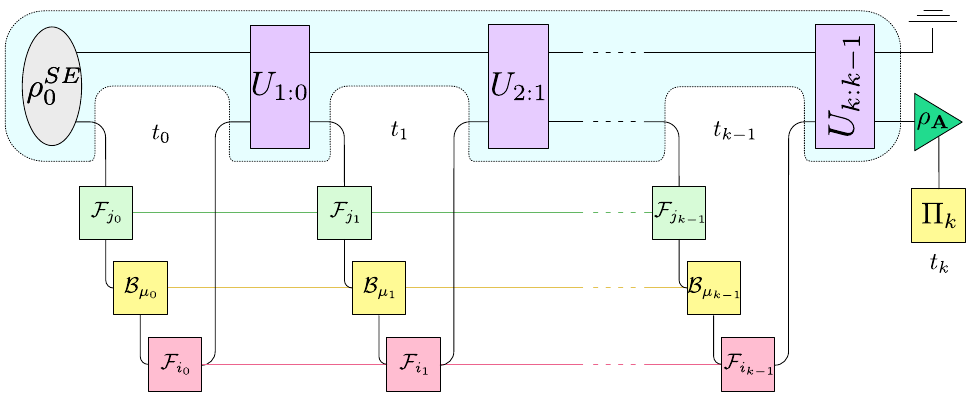}
	\caption{Circuit diagram for performing self-consistent non-Markovian process characterisation. Given a fixed, unknown process $\Upsilon_{k:0}$, and fixed, unknown set of testers $\{\mathbf{B}_{k-1:0}^{\vec{\mu}}\}$, the underlying objects can be self-consistently determined (up to a gauge freedom) by forming combinations of IC testers at each time step. Each fiducial $\mathbf{F}$ is drawn from $\mathscr{B}$. }
	\label{fig:nm-gst}
\end{figure*}

Let us clearly lay out our goals. In PTT, we use a known basis of multi-time instruments $\{\mathbf{B}_{k-1:0}^\mu\}$ and terminating POVM $\mathcal{J}$ to determine an underlying quantum stochastic process $\Upsilon_{k:0}$. Now, suppose that both instruments and measurements are taken to be unknown. We have an equivalent `gate set': $\{\Upsilon_{k:0}, \mathcal{J}, \mathbf{B}_{k-1:0}^{(1)}, \mathbf{B}_{k-1:0}^{(2)},\cdots,\mathbf{B}_{k-1:0}^{(K)}\}$ which we wish to estimate. This encodes both Markovian and non-Markovian \emph{process} errors into \pt{}, \emph{control} errors into $\{\mathbf{B}_{k-1:0}\}$, and measurement errors into $\mathcal{J}$. Our intent for this section is to first formally construct a fully general non-Markovian equivalent to GST. This establishes a firm foundation on which we can consider self-consistent PTT. However, this general formalism is simply too large to treat in practice, so we will chip away at the model to apply simplifications before eventually constructing a computationally efficient method to characterise such processes using tensor networks as our tool.

To set the intuition, we can suppose that the resulting Choi state from a CJI circuit could be used as an input state to a GST experiment on $2k+1$ qubits. Before, we had a single fiducial state $|\rho\rangle\!\rangle$, which we could reliably prepare but not necessarily know; a two-outcome dual effect $\langle\!\langle E|$ that allows us to make observations; and a set of linear transformations $\{G_i\}$. We now make the identification of a $k$-step process whose dynamics we can reliably prepare -- but not necessarily know. As we have stated, the dual to a process is a multi-time instrument known as a tester. 

Self-consistently determining multi-time control operations will require an IC basis of process tensors $\{\Upsilon_{k:0}^\nu\}$. One could take this to mean the independent generation of a set of process tensors by manipulating the environment, which will be difficult in general. Alternatively, we can take a single fiducial process \pt{} and use control operations to transform it at the system level to generate an IC set. This is equivalent to having a single preparable state $\rho$ and generating an IC set of states by transforming the one state with gates $\{G_i\rho G_i^\dagger\}$. We also require an IC set of testers. In keeping with the GST analogy, we fix a single fiducial tester which shall be the `do-nothing' operation at each time $t_j$, followed by measurement with POVM $\mathcal{J} := \{\Pi_x\}$. This single tester can then be transformed by composing the do-nothing operation with some form of control.
The reason for these distinctions will become clear momentarily.

Both process and instruments can be transformed with superprocess transformations. This is, of course, a very general statement. Any change in the dynamics or the control instruments is indeed a superprocess mechanism. Typically, we use an instrument $\mathcal{A}_j$ to extract the state from a process at some $\mathfrak{o}_j$, transform it, and then feed into $\mathfrak{i}_{j+1}$. But suppose we fixed the instrument as a part of our process to define new process tensor $\Upsilon_{k:0}'$. The new output leg $\mathfrak{o}_j'$ is now the old process but with $\mathcal{A}_j$ applied. What we have done is rather than contract an operation into the process, we have \emph{composed} an operation with the process. For a Choi state, this is performed through the link product:  
\begin{equation}
	\Upsilon_{k:0}' = \Tr_{\mathfrak{o}_j}[\hat{\mathcal{A}}_{\mathfrak{o}_j'\mathfrak{o}_j}^{\text{T}_{\mathfrak{o}_j}}\Upsilon_{k:0}].
\end{equation}
As above in the superoperator representation, this is simply matrix multiplication:
\begin{equation}
	|\Upsilon_{k:0}'\rangle\!\rangle = \mathbf{Z}_{k:0}^{(\mathcal{A}_j)} |\Upsilon_{k:0}\rangle\!\rangle,
\end{equation}
and similarly for input legs. Note that if $\mathcal{A}_j$ is a trace-decreasing map, then the resulting process tensor must be trace-normalised in order to remain causal. This amounts to post-selecting on instrument outcomes in practice. 
From the states, gates, and measurements, we make the identification:
\begin{equation}
	\begin{split}
		|\rho\rangle\!\rangle &\rightarrow |\Upsilon_{k:0}\rangle\!\rangle, \\
		\langle\!\langle E | &\rightarrow \langle\!\langle \mathcal{J} \otimes \mathcal{I}\otimes\cdots\otimes \mathcal{I}|,\\
		\mathcal{G} = \{G_l\} & \rightarrow \mathscr{B} = \{\mathbf{B}_{k-1:0}^{\mu}\}.
	\end{split}
\end{equation}

We will drop the time subscripts temporarily to make indexing clearer. Also, let $\bbra{\mathcal{J}}$ be a proxy for $\langle\!\langle \mathcal{J} \otimes \hat{\mathcal{I}}\otimes\cdots\otimes \hat{\mathcal{I}}|$.
We define a fiducial set $\mathscr{F} = \{\mathbf{F}_1,\mathbf{F}_2,\cdots, \mathbf{F}_N\}$ to be drawn from $\mathscr{B}$ which is also to be estimated. For simplicity, this is best chosen to be a deterministic subset such as sequences of unitary operations. We denote the single-time components of each tester without the bold face, and with a subscript denoting the time.
Iterating over each of these sets generates the probabilities
\begin{equation}
	p_{ij\mu} = \langle \!\langle \mathcal{J}|\mathbf{F}_i\mathbf{B}_\mu \mathbf{F}_j| \Upsilon\rangle\!\rangle.
\end{equation}
The circuit diagram for the above described scenario is depicted in Figure~\ref{fig:nm-gst}. Absorbing $\kket{\Upsilon}$ and $\bbra{\mathcal{J}}$ into $\mathbf{F}_j$ and $\mathbf{F}_i$ respectively, we can define new matrices $A$ and $C$ such that $A_{ir} = \langle\!\langle \mathcal{J} | \mathbf{F}_i|r\rangle\!\rangle$ and $C_{sj} = \langle\!\langle s | \mathbf{F}_j | \Upsilon\rangle\!\rangle$. Here, the $\langle\!\langle s|$ and $|r \rangle\!\rangle$ are vectorised elements of some orthonormal basis, such the set of $2k+1$-qubit Pauli matrices. This implies that $p_{ij\mu} = (A\mathbf{B}_\mu C)_{ij}$. That is, varying the fiducials and measuring the outcomes for a given $\mu$ is the measurement of the elements of the matrix $\tilde{\mathbf{B}}_\mu = A\mathbf{B}_\mu C$. In the special case where $\mathbf{B}_0$ is chosen to be the null gate at each time, we fix 
\begin{equation}
	g = \tilde{\mathbf{B}}_1 = AC,
\end{equation}
which is the Gram matrix of the fiducials $\bbra{\mathcal{J}}\mathbf{F}_i\mathbf{F}_j\kket{\Upsilon}$. For each other $\tilde{\mathbf{B}}_\mu$, left multiplying by the inverse of $g$ obtains 
\begin{equation}
	g^{-1} \tilde{\mathbf{B}}_\mu = C^{-1} A^{-1} A \mathbf{B}_\mu C = C^{-1} \mathbf{B}_\mu C.
\end{equation}
Note that although this assumes the existence of $g^{-1}$, this can be taken as a given from the informational completeness of $\mathscr{F}$ (and the extremely mild assumption that the experimental implementation is also informationally complete).
Hence, 
\begin{equation}
	\bar{\mathbf{B}}_\mu := g^{-1}\tilde{\mathbf{B}}_\mu
\end{equation}
is an estimate of each control operation up to similarity transformation by matrix $C$. Note that, like in GST, this gauge freedom is an unavoidable consequence of taking no preferred reference frame. If we set 
\begin{equation}
	\label{eq:ptt-gauge}
	\begin{split}
		|\Upsilon\rangle\!\rangle &\mapsto C'|\Upsilon\rangle\!\rangle\\
		\mathbf{B}_\mu &\mapsto C'^{-1} \mathbf{B}_\mu C,\\
		\langle\!\langle \mathcal{J} | &\mapsto \langle\!\langle \mathcal{J} | C,
	\end{split}
\end{equation}
then the transformed probabilities 
\begin{equation}
	p_{ij\mu}' = \langle \!\langle \mathcal{J}|CC^{-1}\mathbf{F}_iCC^{-1}\mathbf{B}_\mu CC^{-1}\mathbf{F}_jCC^{-1}| \Upsilon\rangle\!\rangle,
\end{equation}
are identical to the $p_{ij\mu}$. This means that any reconstruction will \emph{always} be up to an unobservable gauge freedom across the set. 
The gauge freedom manifests itself as a superprocess.
In practice, gauges are often set by taking the transformation (as in Equation~\eqref{eq:ptt-gauge}) and performing a gauge optimisation. One has the freedom to choose gauges that obey certain properties -- such as maintaining CPTP -- and also to select the gauge that takes the estimated gate set as close to the target gate set as possible. This is completely without loss of generality, and allows our estimates to look as familiar as possible with what we expect.

To obtain the two SPAM vectors, we note that we have the measurable estimate $g_{i0} = (AC)_{i0} = \langle\!\langle \mathcal{J}|\mathbf{F}_i|\Upsilon\rangle\!\rangle$. From this, we get the two equivalent relations 
\begin{equation}
	\begin{split}
		\langle\!\langle \mathcal{J} | \mathbf{F}_i |\Upsilon\rangle\!\rangle &= \sum_r\langle\!\langle \mathcal{J}|\mathbf{F}_i|r\rangle\!\rangle\!\langle\!\langle r|\Upsilon\rangle\!\rangle = \sum_rA_{ir}|\Upsilon\rangle\!\rangle_r := |\tilde{\Upsilon}\rangle\!\rangle_i, \\
		\langle\!\langle \mathcal{J} | \mathbf{F}_j|\Upsilon\rangle\!\rangle &= \sum_s\langle\!\langle \mathcal{J}|s\rangle\!\rangle\!\langle\!\langle s|\mathbf{F}_j|\Upsilon\rangle\!\rangle = \sum_s\langle\!\langle \mathcal{J}|_sC_{sj} := \langle\!\langle \tilde{\mathcal{J}}|_j.
	\end{split}
\end{equation}
and hence, the final estimate (up to gauge transformation) of each vector is:
\begin{equation}
	\begin{split}
		|\bar{\Upsilon}\rangle\!\rangle &= g^{-1}|\tilde{\Upsilon}\rangle\!\rangle = C^{-1}|\Upsilon\rangle\!\rangle,\\
		\langle\!\langle \bar{\mathcal{J}}| &= \langle\!\langle \tilde{\mathcal{J}}| = \langle\!\langle \mathcal{J}|C.
	\end{split}
\end{equation}

One might (rightly) point out that this formalism neglects the possibility of short-time correlations between the $\mathbf{F}_i$, $\mathbf{B}_\mu$, and $\mathbf{F}_j$, since we have assumed that they compose like tensor products within a given leg of the process tensor. On this issue, we make several remarks.
\begin{enumerate}[label=(\roman*)]
	\item The effect of ignoring these correlations in the formalism is equivalent to marginalising over them. This means, for example, that a given $\mathbf{B}_\mu$ is conditioned on the average case $\mathbf{F}_j$ which precedes it, rather than the specific one. In general, this practice can be dangerous, since if the two objects are strongly correlated then the conditional case can vary greatly from the average case. However, we do not expect pulse-level correlations to be the dominating problem here. Instead, it appears that the typical frequency of control fluctuations is on the time scale of dozens of gates, as evidenced by the large range of GST experiments across different devices in the literature~\cite{white-POST,RBK2017, Dehollain_2016, kim_microwave-driven_2015}.
	\item One could not, in general, bootstrap correlated instruments to determine all information about each other. This can be seen with a simple parameter-counting argument: suppose at a single time we have the composition of three gates $\mathcal{F}_i \circ \mathcal{B}_\mu \circ \mathcal{F}_j$, except now they are generically correlated with one another. The collective sequence is thus described by a set of testers $\{\mathbf{A}_{2:0}^{ij\mu}\}$, whose IC set size is $16^3$ with a total of $16^6$ unique parameters. Note that the action of composition here is now replaced by the projection onto two Bell states, so that the output of $\mathcal{F}_j$ feeds to the input of $\mathcal{B}_\mu$, and the output of $\mathcal{B}_\mu$ into the input of $\mathcal{F}_i$. 
	Let us presume that we have more capabilities than our present GST experiment, we have the ability to perform QPT on the input of $\mathcal{F}_j$ and the output of $\mathcal{F}_i$. Even in this idealised scenario, we have only have $(4\times 4) \times 16^3 = 16^4$ linearly independent experiments we can perform. Hence, we lack full information to determine the entire object. 
	\item With undercomplete information, we \emph{could} always construct a plausible $\{\mathbf{A}_{2:0}^{ij\mu}\}$ to fit the data, but we would gain little extra insight about the instruments due to the large gauge freedoms. 
	\item Nevertheless, the discussion is rendered somewhat moot, since it is too onerous to characterise in complete generality. Hence, as we shall see in the following section, we can resolve this issue by both constructing and estimating compressed models using tensor networks; these \emph{do} adequately account for both short and long time correlations in the instruments. 
\end{enumerate}

\subsubsection{Time Local Bases and Time Local Processes}

We have derived a formulation of non-Markovian GST in full generality. This accounts for the full swathe of noisy processes as we have introduced them in the previous section. In particular, the use of multi-time instruments allows one to effectively model the possibility of control non-Markovianity, time-dependent Markovianity, and control spillage, as well as process non-Markovianity. If one were to expect -- or one wished to test a model of -- either process non-Markovian and control Markovian, or process Markovian and control non-Markovian, then this amounts respectively to treating $\{\mathbf{B}_{k-1:0}^\mu\}$ or \pt{} as a tensor product. Let us consider the former case, i.e. that $\{\hat{\mathbf{B}}_{k-1:0}^{\vec{\mu}}\} = \left\{\bigotimes_{i=0}^{k-1} \hat{\mathcal{B}}_{\mu_i}\right\}$. That is, all of the sequences of operations are generated from the same set of $d_S^4$ time-local instruments, and the \emph{effects of an instrument is the same} no matter at which time it is applied. This substantially reduces the problem size. First, it allows the matrix $A$ to take a tensor product structure: 
\begin{equation}
	A_{\vec{i}\vec{r}} = \bigotimes_{j=0}^{k-1}\bigotimes_{l=0}^{k-1}\langle\!\langle \mathcal{J}_{j_l}|F_{j_l}|r_l\rangle\!\rangle.
\end{equation}
Hence, rather than measuring $i$ and $\mu$ from $1$ to $d_S^{4k}$, one only needs to measure each from 1 to $d_S^4$. Note that for full determination of $|\Upsilon\rangle\!\rangle$, one still has $j$ scaling like $d_S^{4k}$, since it is generally encodes arbitrary temporal correlations.
Another consequence of this is that it allows us to take the gauge freedom to also maintain a tensor product structure, since right multiplying by $g^{-1}$ selects $A$ as the gauge matrix rather than $C$, and we have already fixed this to be a tensor product. This simplifies the problem somewhat, but it is not sufficient to reduce the problem from being exponential in the system size, number of gates, and number of steps. For this, we must turn to tensor network methods. 

\subsection{Process Tensors as Locally Purified Density Operators}
\label{sec:PT-LPDOs}

\begin{figure*}[t]
	\centering
	\includegraphics[width=0.95\linewidth]{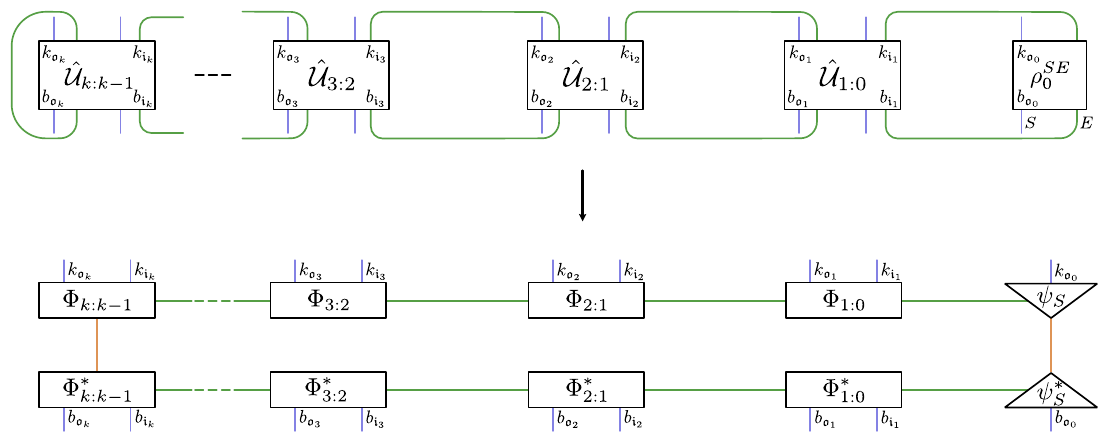}
	\caption[The link product representation of the process tensors shows how it can readily be expressed as an locally purified density operators (LPDO), where each step is pure except for the start and end.]{The link product representation of the process tensors shows how it can readily be expressed as an LPDO, where each step is pure except for the start and end. The size of the non-Markovian memory hence corresponds to the horizontal bond dimension, and the vertical bonds correspond to the total size of the environment, as sensed by the process.
	}
	\label{fig:mpo-ring}
\end{figure*}

In many physically relevant processes, the non-Markovian memory -- or the size of the effective environment required to carry that memory -- ought not to grow too large. Indeed, the bond dimension of the MPO representation of a process tensor has been shown to be a measure of non-Markovianity~\cite{Pollock2018a}.
However, most analyses of process tensor MPOs have been in the form of numerical experiments. 
One of the chief difficulties in translating this to an experimental setting is actually estimating this object in a robust and physically sensible way. 
In this section, we focus on this problem. Given an experimental dataset, how can we perform a sparse version of PTT and estimate a tensor network representation of our process? In contrast with finite Markov order models, which are approximate~\cite{White-MLPT,taranto1}, we will see many examples where tensor networks can \emph{exactly} represent a process. This comes at the expense of greater classical computational cost, and a lack of convergence guarantee in the estimation. However, we will present a method which details a powerful estimation procedure that we find effective in characterising multi-qubit, multi-time non-Markovian processes. Our approach is modular: we construct a tensor network ansatz for the process, employ a log-likelihood objective function, use \mbox{autodifferentiation} to obtain the gradients of the individual tensors, and finally perform a variant of stochastic gradient descent to find the maximum-likelihood model. The flexibility here is that the tensor network model for the process is fully generic and can be user-chosen. 

This section and the next lay out detailed breakdowns of how we model multi-qubit multi-time processes and instruments as positive tensor networks which may then be estimated. Readers who are predominantly interested in results and control applications may skip forward to Section~\ref{sec:synth-demonstrations} and beyond.

\subsubsection{Representation} 

Process tensors are naturally equipped to be represented with locally purified density operators (LPDO)s, the central object used in Ref.~\cite{torlai2020quantum}. Apart from an initial (possibly) mixed state and the trace over the environment at the end, dynamics are unitary. Hence, only the start and end of a process require operator representations. We expand on the theoretical prescription given in Ref.~\cite{Pollock2018a}. A process tensor Choi state, being a density operator, has an MPO representation which we can write as

\begin{widetext}
\begin{equation}
	\label{eq:pt-mpo}
	\Upsilon_{k:0} = \sum (\Gamma_{k:k-1})_{b_{\mathfrak{o}_k}b_{\mathfrak{i}_k}}^{k_{\mathfrak{o}_k}k_{\mathfrak{i}_k}}
	\cdots (\Gamma_{1:0})_{b_{\mathfrak{o}_1}b_{\mathfrak{i}_1}}^{k_{\mathfrak{o}_1}k_{\mathfrak{i}_1}}(\Gamma_0)_{b_{\mathfrak{o}_0}}^{k_{\mathfrak{o}_0}} |k_{\mathfrak{o}_k}k_{\mathfrak{i}_k}\cdots k_{\mathfrak{o}_1}k_{\mathfrak{i}_1}k_{\mathfrak{o}_0}\rangle \! \langle b_{\mathfrak{o}_k}b_{\mathfrak{i}_k}\cdots b_{\mathfrak{o}_1}b_{\mathfrak{i}_1}b_{\mathfrak{o}_0}|,
\end{equation}
\end{widetext}
where, for $j\neq k,0$, the following are $d_E^2\times d_E^2$ matrices representing $SE$ evolution:
\begin{equation}
	(\Gamma_{j:j-1})_{b_{\mathfrak{o}_j}b_{\mathfrak{i}_j}}^{k_{\mathfrak{o}_j}k_{\mathfrak{i}_j}} = \langle b_{\mathfrak{o}_j} | U_{j:j-1}|b_{\mathfrak{i}_j}\rangle \otimes \langle k_{\mathfrak{o}_j}| U_{j:j-1}^\ast | k_{\mathfrak{i}_{j}}\rangle.
\end{equation}
The final step, then, is a length $d_E^2$ row vector $(\Gamma_{k:k-1})_{b_{\mathfrak{o}_k}b_{\mathfrak{i}_k}}^{k_{\mathfrak{o}_k}k_{\mathfrak{i}_k}}$ which can be expressed as
\begin{equation}
	 \sum_{\gamma}\langle b_{\mathfrak{o}_k} \gamma_E| U_{k:k-1}|b_{\mathfrak{i}_j}\rangle \otimes \langle k_{\mathfrak{o}_k} \gamma_E| U_{k:k-1}^\ast | k_{\mathfrak{i}_{k}}\rangle,
\end{equation}
accounting for a final trace over the environment. That is, $d_E$ is the dimension of the environment and $\gamma_E$ is the index for its state. Lastly, supposing the initial state has eigendecomposition
\begin{equation}
	\rho_0^{SE} = \sum_i p_i |\psi_i\rangle \! \langle \psi_i|,
\end{equation}
then we have a length $d_E^2$ column vector
\begin{equation}
  \label{eq:initial-state}
  (\Gamma_0)_{b_{\mathfrak{o}_0}}^{k_{\mathfrak{o}_0}} = \sum_i p_i \langle b_{\mathfrak{o}_0}|\psi_i\rangle \otimes \langle k_{\mathfrak{o}_0}|\psi_i\rangle^\ast.
\end{equation}
The intention of writing these decompositions is to show that the non-Markovian environment is the natural local purification of process tensors. 
This relationship is more easily seen graphically if we write the process in its the link product form~\cite{Milz2021PRXQ}
\begin{equation}
	\Upsilon_{k:0} =  \Tr_E[\hat{\mathcal{U}}_{k:k-1}\star \hat{\mathcal{U}}_{k-1:k-2}\star \cdots \star \hat{\mathcal{U}}_{1:0}\star \rho_0^{SE}],
\end{equation}
sketched in Figure~\ref{fig:mpo-ring}. $(\star)$ here denotes a Choi-map composition on $E$ and tensor product on $S$.
We notice several features. First, only $\Gamma_0$ and $\Gamma_{k:k-1}$ are non-pure, as these respectively represent a generically mixed initial state, and the trace over the environment at the end of the dynamics. Each other intermediate step is pure. Second, the bond dimension grows with the size of the environment.
We can introduce two ancilla systems $A$ as proxies for the environment, one to purify $\Gamma_0$ and one to purify $\Gamma_{k:k-1}$. In the former case, we have 
\begin{equation}
	\begin{split}
		|\Psi_0^{SAE}\rangle &= \sum_{\mu_0=1}^{\chi_{\mu_0}}\sqrt{p_{\mu_0}}|\psi_{\mu_0}\rangle\otimes |a_{\mu_0}^{(1)}\rangle,\\
	\end{split}
\end{equation}
which we can contract with the ancilla bond of its complex conjugate to re-express the same local tensor from Equation~\eqref{eq:initial-state}, $(\Gamma_0)^{k_{\mathfrak{o}_0}}_{b_{\mathfrak{o}_0}}$, as
\begin{equation}
	\begin{split}
	 &\sum_{\mu_0=1}^{\chi_{\mu_0}} [\Psi_0]^{k_{\mathfrak{o}_0}}_{\mu_0,\nu_1}[\Psi_0^\ast]^{\mu_0,\nu_1'}_{b_{\mathfrak{o}_0}}, \\
	&= \sum_{\mu_0=1}^{\chi_{\mu_0}}p_{\mu_0} \langle k_{\mathfrak{o}_0}|\psi_{\mu_0}\rangle\otimes \langle a_{\mu_0}^{(1)}|a_{\mu_0}^{(1)}\rangle \langle\psi_{\mu_0}|b_{\mathfrak{o}_0}\rangle^\ast\otimes \langle a_{\mu_0}^{(1)}|a_{\mu_0}^{(1)}\rangle,\\
	&= \sum_{\mu_0}p_{\mu_0}\langle b_{\mathfrak{o}_0}|\psi_{\mu_0}\rangle \otimes \langle k_{\mathfrak{o}_0}|\psi_{\mu_0}\rangle^\ast.
	\end{split}
\end{equation}
Explicitly, $\Gamma_0$ is a reshaping of $|\Psi_0^{SAE}\rangle\!\langle \Psi_0^{SAE}|$, with $\{k_{\mathfrak{o}_0}, b_{\mathfrak{o}_0}\}$ the physical indices of $S$, $\{\nu_1,\nu_1'\}$ the physical indices of $E$, and $\mu_0$ the convex sum over a complete basis on $A$. This is an expliclty positive form.
We say then that $\Gamma_0$ is locally purified in the same sense as Ref.~\cite{torlai2020quantum}, because the addition of the bond $\mu_0$ represents an ancilla system that encodes the mixedness of the reduced state $\Gamma_0$. The same procedure is repeated for $\Gamma_{k:k-1}$. This LPDO process representation immediately gives rise to a ring-like structure, as in Figure~\ref{fig:mpo-ring}b. One great advantage to using a locally purified state rather than a generic MPO as in Equation~\eqref{eq:pt-mpo} is that this form naturally encodes positivity of the state. As well as being a generally desirable physical property, this parametrisation produces only positive, real probabilities when evaluated. Hence, it is well-behaved when considering the log-likelihood to find our optimal model. 
For a single-system, we can therefore construct an LPDO parametrisation of a quantum stochastic process by casting its process tensor representation into this form:

\begin{widetext}
\begin{equation}
\label{eq:PT-TN}
[\Upsilon_{k:0}^{\vec{\theta}}]^{\vec{k_{\mathfrak{o}}}\vec{k_{\mathfrak{i}}}}_{\vec{b_{\mathfrak{o}}}\vec{b_{\mathfrak{i}}}} = \sum_{\mu_k,\mu_0} \sum_{\vec{\nu},\vec{\nu}'} [\Phi_{k:k-1}]_{\mu_k,\nu_k}^{k_{\mathfrak{o}_k}k_{\mathfrak{i}_k}} [\Phi_{k:k-1}^\ast]^{\mu_k,\nu'_k}_{b_{\mathfrak{o}_k}b_{\mathfrak{i}_k}}\left(\prod_{j=1}^{k-1}
	[\Phi_{j:j-1}]_{\nu_{j+1}\nu_j}^{k_{\mathfrak{o}_j}k_{\mathfrak{i}_j}}[\Phi_{j:j-1}^\ast]^{\nu'_{j+1}\nu'_j}_{b_{\mathfrak{o}_j}b_{\mathfrak{i}_j}}\right) [\Psi_{0}]_{\mu_0,\nu_1}^{k_{\mathfrak{o}_0}} [\Psi_{0}^\ast]^{\mu_0,\nu'_1}_{b_{\mathfrak{o}_0}},
\end{equation}
\end{widetext}
where $\vec{\theta} = \{\Phi_{k:k-1},\{\Phi_{j:j-1}\},\Psi_0\}$. In this variational form, bond dimensions must be chosen \emph{a priori}. We have Kraus rank $\chi_{\mu_0},\chi_{\mu_k}$ reflective respectively of the rank of the initial $SE$ state and number of environment degrees of freedom traced at the end of the circuit. 
The $\chi_{\nu_j}$ are measures of non-Markovianity: they decree the effective size of the non-Markovian environment relevant to possible memory effects. Implicitly, we will always absorb eigenvalues and singular values into the left and right singular vectors of each site. This is simply a gauge choice that eases notational overload.

\begin{figure}[t!]
    \centering
    \includegraphics[width=\linewidth]{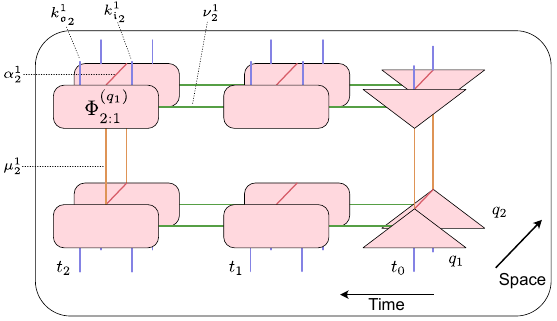}
	\caption[An indicative block of the 3D tensor network used to represent multi-qubit, multi-time quantum stochastic processes processes ]{An indicative block of the 3D tensor network we employ to represent multi-qubit, multi-time quantum stochastic processes processes. Green horizontal bonds represent non-Markovian memory; orange vertical bonds indicate local purification (Kraus bonds); red bonds into the page indicate qubit interactions, or crosstalk.}
	\label{fig:tn-3d-exp}
\end{figure}

Before proceeding to fit this model, we make one final adjustment. Each of these variational tensors are dense representations with local system sites of size $d_S^2$. But if we wish to choose a system whose Hilbert space structure is composite -- for example a chain of spins -- then we can further decompose this to take advantage of the sparseness of spatial correlations. For a system $S$ defined as a register of qubits $\{q_1,\cdots,q_n\}$, we can perform a series of singular value decompositions across the subsystems such that 
\begin{equation}
	\Phi_{j:j-1,\mu_j\nu_j}^{k_{\mathfrak{o}_j}k_{\mathfrak{i}_j}} = \sum_{\vec{\alpha}}\prod_{i=1}^n\Phi_{j:j-1, \alpha_j^{i-1}\alpha_j^i, \mu_j^i\nu_j^i}^{(q_i)k^i_{\mathfrak{o}_j}k^i_{\mathfrak{i}_j}}.
\end{equation}
This rather messy piece of index notation represents a 3D tensor network representation of our process, summarised both in the following list and in Figure~\ref{fig:tn-3d-exp}.

\begin{itemize}
	\item A single site $\Phi_{j:j-1}^{(q_i)}$ is the local purification of the $j$th dynamical map on the $i$th qubit. $k_{\mathfrak{o}_j}^i$ and $k_{\mathfrak{i}_j}^i$ are the local site indices for the input and output spaces of the dynamical map represented for that particular qubit at that particular time step.
	\item The bonds in the $x$-direction are indexed by $\nu_j^i$ and encode temporal correlations on that qubit.
	\item The $y$-direction bonds are indexed by $\mu_j^i$, these are Kraus bonds and are only present for $j=0,k$.
	\item The $z$-direction bonds are indexed by $\alpha_j^i$: these encode spatial correlations between the qubits generated across a given time step.
\end{itemize}

The efficiency of this ansatz relies on the process obeying an area law growth. 
That is, the maximum bond dimension of the whole network should be bounded by a constant. 
Incidentally, although we have three bond dimensions here to make this a geometrically three-dimensional network because of the ring-like structure naturally encoded by process tensors, this is to be interpreted as a layered 2D tensor network. 
Nevertheless, we must be careful. Two-dimensional tensor networks might encode a state efficiently -- hence requiring fewer quantum resources, but they cannot be contracted efficiently~\cite{Orus2014} -- and so the classical computation grows exponentially in the area of the network. For this reason, we limit our approach to either only few time steps and many qubits, or many qubits and few time steps. To rectify this, one could combine the above approach with a Markov order ansatz so that this were efficient in 1+1 spacetime.

\begin{figure*}[t!]
	\centering
	\includegraphics*[width=0.95\linewidth]{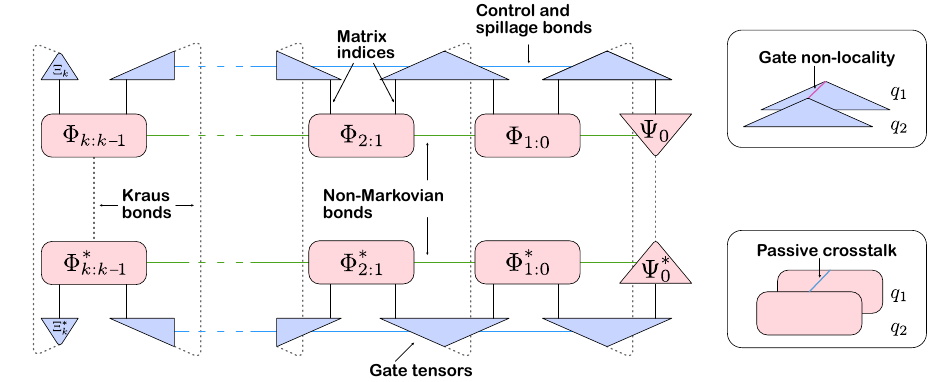}
	\caption{A tensor network diagram of locally purified process tensors represented alongside correlated control operations. This models all the different dynamical effects discussed in the present work.}
	\label{fig:self-consistent-tn}
\end{figure*}

\subsubsection{Self-Consistency}

In Section~\ref{app:nmgst-derivation} we derived a formalism to simultaneously capture various notions of control and process non-Markovianity, as well as the interplay between the two. For fully general processes, the complexity of the problem scales exponentially in the number of time steps $k$, the number of qubits, and the number of gates in the gate set. In Ref.~\cite{White-MLPT}, this regular PTT was already limited to a single qubit at three time steps for the fully general case before being overcome by practical considerations. Although full PTT is somewhat practical, implementing general self-consistency is not. 
We hence consider a tensor network method to be necessary to implement this characterisation.
Not only does this compress the representation significantly, but allows for modularity based on the expected physics of the system. We build on the approach from the previous section and develop a self-consistent method to estimate any quantum stochastic process as well as the noisy instrument used to probe it. In Sections~\ref{sec:synth-demonstrations} and~\ref{ssec:nisq-tn} we demonstrate the characterisation on a variety of synthetic and real data, showing how one may accurately capture a wide range of noisy quantum dynamics.

For a system $S$ composed of $N$ qubits characterised across times $\mathbf{T}_k$, we start with \pt{} representing the process, and $\{\hat{\mathbf{A}}_{k:0}^{(i)}\}_{i=1}^M$ representing the Choi states for the series of the $M$ multi-time control operations. Note that these testers include measurement outcomes in the index $i$, and are hence trace non-increasing. First, we cast these in variational tensor network form as LPDOs, as depicted in Figure~\ref{fig:self-consistent-tn}. \pt{}, once again, can be written as in Equation~\eqref{eq:PT-TN}.
The control operations may similarly be locally purified, but since they do not stem from a continuous unitary evolution from a dilated environment, they do not inherit the same ring-like structure. These are given as 
\begin{widetext}
\begin{equation}
	[\hat{\mathbf{A}}_{k:0}^{\vec{\phi}_{i}}]^{\vec{b_{\mathfrak{o}}}\vec{b_{\mathfrak{i}}}}_{\vec{k_{\mathfrak{o}}}\vec{k_{\mathfrak{i}}}} = \sum_{\vec{\delta}} \sum_{\vec{\gamma},\vec{\gamma}'} 
	[\Xi_{k}]^{\delta_k,\gamma_k}_{k_{\mathfrak{o}_k}} [\Xi_{k}^\ast]_{\delta_k,\gamma'_k}^{b_{\mathfrak{o}_k}}
	\left(\prod_{j=1}^{k-1}
	[\Gamma_{j:j-1}]^{\gamma_{j+1}\gamma_j}_{k_{\mathfrak{i}_j}k_{\mathfrak{o}_{j-1}}}[\Gamma_{j:j-1}^\ast]_{\gamma'_{j+1}\gamma'_j}^{b_{\mathfrak{i}_j}b_{\mathfrak{o}_{j-1}}}\right),
\end{equation}
\end{widetext}
where $\vec{\phi}_{(i)} =  \{\Xi_{k}^{(i)},\{\Gamma_{j:j-1}^{(i)}\}\}$. Let us further subdivide the control index $i$ into $(j,x)$, where $j$ indicates the deterministically chosen multi-time instrument, and $x$ is its corresponding measurement outcome (or sequence of measurement outcomes). Then, when a quantum circuit is run with instrument $\mathbf{A}_{k:0}^{(j)}$, the probability of obtaining outcome $x$ is
\begin{equation}
	p_{x\mid j} = \Tr[\hat{\mathbf{A}}_{k:0}^{(j)\text{T}}\Upsilon_{k:0}],
\end{equation}
modelled by our parametrised tensor networks as
\begin{equation}
	\label{eq:self-consistent-prob}
	p_{x\mid j}^{\vec{\theta}\vec{\phi_{j}}} = \sum_{\vec{k_{\mathfrak{o}}}\vec{k_{\mathfrak{i}}}\vec{b_{\mathfrak{o}}}\vec{b_{\mathfrak{i}}}} [\Upsilon_{k:0}^{\vec{\theta}}]^{\vec{k_{\mathfrak{o}}}\vec{k_{\mathfrak{i}}}}_{\vec{b_{\mathfrak{o}}}\vec{b_{\mathfrak{i}}}}[\hat{\mathbf{A}}_{k:0}^{\vec{\phi}_{j}}]^{\vec{b_{\mathfrak{o}}}\vec{b_{\mathfrak{i}}}}_{\vec{k_{\mathfrak{o}}}\vec{k_{\mathfrak{i}}}}.
\end{equation}
That is to say, it is a single tensor network contraction of the two representations. We are now in a position to estimate these objects.

\subsection{Implementation of Efficient and Self-Consistent Process Tensor Tomography}
\label{sec:self-consistency}

We now introduce our tensor network estimation procedure for arbitrary non-Markovian quantum stochastic processes. We have made some specific choices about the form of the gate operations here, but these do not affect the generality of the procedure. Specifically, for the purposes of demonstration and simplicity of exposition, we restrict ourselves to learning unitary operations only, in addition to the process tensor and final measurement operation. 
However, the procedure is readily generalisable to any (possibly time-non-local) IC basis if available. 
To perform PTT with LPDOs, we continue with the same structure. Our multi-time instrument will consist of sequences of unitary operations followed by a final terminating measurement. We take a time-local basis of operations $\{\mathcal{B}_j\}$, moreover for concreteness we take this basis to be the set of (for the time being) single-qubit Clifford operations. 
We gain a significant reduction in required classical and quantum computational resources to understand complex correlated quantum noise.

\begin{figure}
	\centering
	\includegraphics[width=\linewidth]{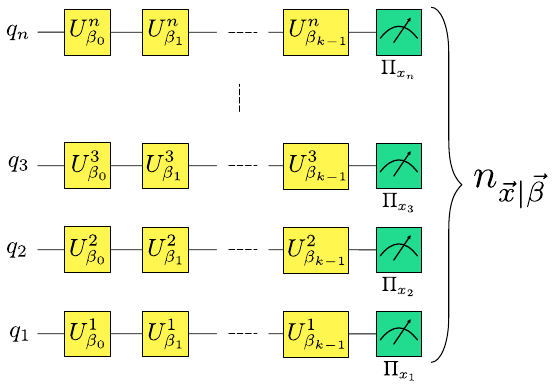}
	\caption{Circuit representation of the data collection procedure for tensor network characterisation of process tensors. }
	\label{fig:tn-circuit-char}
\end{figure}

As in fully general PTT, one should decide on the stochastic process they wish to estimate, defined across a system $S$ and a series of times $\mathbf{T}_k$. At each time $t_j$, a random Clifford operation $\mathcal{U}_j^i$ is applied to qubit $i$, followed by a final projective measurement in a random Clifford basis, obtaining outcome $x_i$. Let us index the Clifford sequence by $\vec{\beta}$. The circuit for this is shown in Figure~\ref{fig:tn-circuit-char}.
The sequence of unitaries is represented by 
\begin{equation}
	\hat{\mathcal{B}}_{\vec{\beta}} = \bigotimes_{j=0}^{k-1} \bigotimes_{i=1}^n \hat{\mathcal{U}}_{\beta_j}^i,
\end{equation}
with outcome 
\begin{equation}
	\Pi_{\vec{x}} = \bigotimes \Pi_{x_i}.
\end{equation}
Both of these control maps are part of the model to be estimated.

A single experiment run with $N$ shots returns a series of at most $N$ non-zero frequencies $\{n_{\vec{x} \mid \vec{\beta}_0}\}_{\vec{x} = (0,0,\cdots,0)}^{(1,1,\cdots,1)}$ for each of the observed outcomes. 
Correspondingly, the tensor network ansatz can make a prediction for each of the frequencies which depends on the local tensor parameters $\vec{\theta}$ of the process, and $\{\vec{\phi}\}$ of the gates:
\begin{equation}
	p_{\vec{x} \mid \vec{\beta}}^{\vec{\theta},\vec{\phi_j}} = \Tr[\Upsilon_{k:0}^{\theta}\cdot (\Pi_{\vec{x}}\otimes \hat{\mathcal{B}}_{\vec{\beta}})].
\end{equation}
The experimental procedure runs $M$ experiments and collect $N$ shots per experiment, forming the set of runs $\mathcal{R} = \{\vec{\beta}\}_{\vec{\beta} = \vec{\beta}_1}^{\vec{\beta_M}}$ with corresponding dataset $\mathcal{D} = \{{n_{\vec{x}\mid\vec{\beta}}}\}$. Our objective function $f$, which quantifies the goodness-of-fit of our model, is the log-likelihood:
\begin{equation}
	\label{eq:tn-ll}
	f(\vec{\theta},\{\vec{\phi_j}\}) = \sum_{\vec{x}}\sum_{\vec{\beta}} -n_{\vec{x}\mid\vec{\beta}}\log p_{\vec{x} \mid \vec{\beta}}^{\vec{\theta},\vec{\phi_j}}.
\end{equation}
In practice, evaluating Equation~\eqref{eq:tn-ll} can be computationally arduous, and so at each iteration we randomly select a subset of $\mathcal{D}$ of size $M_{\text{batch}}$. This has the additional benefit of rendering the objective function to be stochastic, which can be useful for overcoming local minima in the optimisation space. 

\subsubsection{Regularisation}

Our tensor network ansatz is defined to be positive, but it has no constraints to ensure it is causal (and therefore physical). In Ref.~\cite{White-MLPT}, we rigorously encoded causality into our process tensors by projecting onto the linear space of causal quantum states at each step. This projected gradient descent came equipped with performance guarantees, namely that the final output of the maximum likelihood estimation algorithm would be guaranteed to lie on the intersection of the cone of positive semidefinite matrices with linear space defined by causality conditions. As we turn to larger problems, this approach becomes infeasible due to the exponential scaling of the projection. We encode positivity of the process into our LPDO parametrisation, but we propose a more heuristic approach to maintain causality by regularising it in the objective function.

The set of causality conditions is captured by the set of conditions $\Tr_{\mathfrak{o}_k}[\Upsilon_{k:0}] =\mathbb{I}_{\mathfrak{i}_k}\otimes \Upsilon_{k-1:0}\:\:\forall \:\:k$. We can map this onto the requirement that a specific set of local Pauli expectation values (on the process) must be zero. If we adopt the perspective that these causality requirements constitute data that we can feed our model, then we can combine these into the experimentally observed data at only slight extra computational expense. There are exponentially many Pauli expectation values to rule out, and so we must randomise over these as described below. Remarkably, we find this to be an extremely effective approach to ruling out acausal estimates. 
Let $\mathbf{P}^{(n)}$ be the full set of $n$-qubit Pauli matrices, and $\tilde{\mathbf{P}}^{(n)}$ the set of traceless $n$-qubit Pauli matrices. 

We generate some number $M_{\text{causal}}$ of Pauli tensor products $P_{\text{c}} = P_{\mathfrak{o}_k}\otimes P_{\mathfrak{i}_k}\otimes \cdots\otimes P_{\mathfrak{o}_1}\otimes P_{\mathfrak{i}_1}\otimes P_{\mathfrak{o}_0}$ for which causality demands that $\Tr[P_{\text{c}}\cdot \Upsilon_{k:0}^{\vec{\theta}}] = 0$~\cite{White-MLPT}. 
To generate a set of $M_{\text{causal}}$ Pauli causality constraints $\mathcal{C}:=\{P_{\text{c}}\}$, we repeat the following procedure:
\begin{enumerate}
	\item Select a number $j$ between 1 and $k$ (inclusive) at random.
	\item Set $P_{\mathfrak{o}_j} = \mathbb{I}_{\mathfrak{o}_j}$.
	\item Set $\tilde{P}_{\mathfrak{i}_j}$ to be a random member of $\tilde{\mathbf{P}}^{(n)}$.
	\item Set all $P_{\mathfrak{o}_m}$ and $P_{\mathfrak{i}_m}$ to be equal to $\mathbb{I}$, for $j < m \leq k$. 
	\item Set $P_{\mathfrak{o}_m}$ and $P_{\mathfrak{i}_m}$ to be (independently) random members of $\mathbf{P}^{(n)}$ for $0 < m < j$ .
    \item Add $P_{\mathfrak{o}_k}\otimes P_{\mathfrak{i}_k}\otimes \cdots\otimes P_{\mathfrak{o}_1}\otimes P_{\mathfrak{i}_1}\otimes P_{\mathfrak{o}_0}$ to $\mathcal{C}$.
\end{enumerate}

Evaluating $\langle P_{\text{c}}\rangle$ with respect to a valid process tensor will always result in zero. Hence, we can regularise our tensor network model by evaluating this expectation value across randomly generated sets $\mathcal{C}$. This will bias our optimisation towards valid process tensors (which we may then later verify). 

\emph{Control trace-preservation.} We can derive an equivalent procedure to enforce the physicality of control operations.  
The Choi states of multi-time instruments have the same positivity requirement (and imposition) but need not be causal. The natural intuition for this follows from the two-time case: quantum channels must be TP, but quantum instruments may be trace non-increasing since the classical outcomes may be stochastically obtained. A sequence of measurements then, for example, will be an acausal multi-time intrument and trace will not be preserved. We might also consider the use of an ancilla tester through a simple example: suppose an ancilla qubit is placed into equal superposition and interacts with the system through only diagonal interactions. The ancilla is projectively measured at some early time $t_i$ and some later time $t_f$ in the same basis. This gives rise to four testers, one for each of the outcomes $\{0_i0_f,0_i1_f,1_i0_f,1_i1_f\}$. But, since the tester is defined by interactions that do not change the population of the ancilla qubit, the probability of obtaining outcomes $0_i1_f$ or $1_i0_f$ is zero. Hence, if we post-select on the later measurement outcome, we update the statistics of the earlier measurement.

The condition satisfied by testers must be that for a given $j$, if we sum over outcomes $x$, the result is causal. This is necessary to preserve probabilities across the whole range of multi-time instruments. Much like the causality of the process tensor, we can regularise this by sampling the Pauli expectation values of 
\begin{equation}
	\hat{\mathbf{A}}_{k:0}^{(j)} = \sum_x \hat{\mathbf{A}}_{k:0}^{(j,x)}.
\end{equation}

Let $\mathbf{P}_c$ be the total set of $2k+1$-qubit Pauli operators for which causality conditions dictate the expectation values must be zero for the respective gate set objects. That is to say, 
\begin{equation}
	\Tr[P\cdot \Upsilon_{k:0}] \overset{!}{=} 0 \quad \text{and} \quad \Tr[P \cdot \hat{\mathbf{A}}_{k:0}^{(j)}] \overset{!}{=} 0 \quad \forall \:j,P\in \mathbf{P}_c.
\end{equation}
At each iteration, we randomly generate $|D|+1$ subsets of $\mathbf{P}_c$ -- denoted by $\{\mathcal{C}_i\}_{i=0}^{|D|}$ -- corresponding to a subset $D\subset \mathcal{D}$ of the total dataset. $\mathcal{C}_0$ is a causality regularisation of the process, and each $\mathcal{C}_i$ is a TP regularisation of each unique tester involved in producing the data. The size of $|D|$ is a hyperparameter $M_{\text{batch}}$, and the size of each $\mathcal{C}_i$ is some $M_{\text{causal}}$. In this work, we typically take the former to be 1000 and the later to be 200. 

\subsubsection{Model Fitting}

We are now in a position to set up the problem of fitting our tensor network model for a multi-time, multi-qubit quantum stochastic process to some experimental data. The objective function we choose is a sum of the (average) log-likelihood, and the above causal regularisation, and the TP regularisation:
\begin{widetext}
\begin{equation}
\label{eq:tn-objective}
	f(\vec{\theta},\{\vec{\phi_j}\}) = \overset{\text{Model log-likelihood}}{\overbrace{\sum_{d\in D}-n_d\ln p_d^{\vec{\theta}\vec{\phi_{j}}} }}
	+ \overset{\text{Process causality}}{\overbrace{\sum_{P\in \mathcal{C}_0} \Tr\left[P\cdot [\Upsilon_{k:0}^{\vec{\theta}}]\right]}}
	+ \overset{\text{Control trace preservation}}{\overbrace{\sum_{d\in D}\sum_{P\in\mathcal{C}_d} \Tr\left[P\cdot \sum_x \hat{\mathbf{A}}_{k:0}^{\vec{\phi}_{(j,x)}}\right]}}.
\end{equation}
\end{widetext}
emphasising again that the respective sums are of size $M_{\text{batch}}$ and $M_{\text{causal}}$, and are chosen randomly at each evaluation. $\kappa$ here is a meta-parameter of the optimisation governing the strength of the regularisation. If it is too small, the optimal model may not be causal; if it is too large then it may slow down convergence. The first term in Equation~\eqref{eq:tn-objective} is called the cross entropy (equivalent to log-likelihood) and attains a minimum value at the data entropy. That is, when $p_{\vec{x}\mid \vec{\beta}} = n_{\vec{x}\mid \vec{\beta}} $ for each $\vec{x}$, $\vec{\beta}$.

Now our aim is to minimise Equation~\eqref{eq:tn-objective} with respect to $\theta$. We do this using the Adam optimiser, which has found remarkable success in optimising stochastic objective functions~\cite{kingma2014adam}. For tensor network semantics, we use the Python library \texttt{quimb}~\cite{quimb}. To obtain the gradients of the objective function with respect to the local tensors, $\nabla_{\vec{\theta},\vec{\phi}}f$, we use the library JAX for numerical autodifferentiation~\cite{jax2018github}. We additionally collect a smaller validation dataset $\mathcal{D}_v$ on which Equation~\eqref{eq:tn-objective} is evaluated. By computing the likelihood of the model with respect to this validation dataset, we can cross-validate the model to ensure that no-overfitting has occurred.

\subsubsection{Avoiding Premature Convergence} 

A consequence of adopting a completely positive parametrisation of our process tensor in its tensor network form is that the objective function is now non-linear, and hence the optimisation problem is no longer convex. This is a problem broadly applicable to machine learning and is often tackled by using stochastic optimisation methods such as stochastic gradient descent or Adam. Randomising the data partitions fed into the objective function can also be a powerful approach to escape the effects of local minima. Moreover, a common supplementary technique is to begin the optimisation from randomly generated seeds to avoid getting stuck in the same local minima, and ideally find the globally optimal solution.
Although we do adopt these approaches, through a series of trial and error we still find that the problem of tensor network learning can be prohibitively slow, requiring a large number of iterations or cold restarts to converge. This is in general unideal, but it is particularly a problem if the purpose of the characterisation is to feed forward into calibration of a device. Further, one cannot say in general whether failure to converge is a deficiency of the chosen bond dimensions of the model, or simply whether not enough trial seeds have been examined. 
Consider instead that often a process is in the neighbourhood of something expected. By this we mean close to
\begin{equation}
	\Upsilon_{k:0}^{\text{(ideal)}} = \bigotimes_{j=1}^k|\Phi^+\rangle\!\langle \Phi^+| \otimes |0\rangle\!\langle 0|.
\end{equation}
This is the identity process, representing an initial state of zero and a series of do-nothing evolutions. Rather than starting from a random seed, we let each 
\begin{equation}
  [\Phi_{j:j-1}]_{\nu_{j+1}\nu_j}^{k_{\mathfrak{o}_j}k_{\mathfrak{i}_j}}\mapsto [|\Phi^+\rangle]_{\nu_{j+1}\nu_j}^{k_{\mathfrak{o}_j}k_{\mathfrak{i}_j}} + [\tilde{R}_{j:j-1}]_{\nu_{j+1}\nu_j}^{k_{\mathfrak{o}_j}k_{\mathfrak{i}_j}}
\end{equation}
where $|\Phi^+\rangle$ is a Bell state made into the appropriate tensor shape by padding out the extra axes with zeros, and $\tilde{R}$ is an equivalently shaped tensor with its entries chosen from a complex Gaussian distribution $\mathcal{N}(0,0.1)$. \par

In all simulated and real demonstrations, we find that this starting point is not only useful, it is essential to obtaining a solution in a reasonable time. Moreover, we find convergence from this starting point in all cases. It is, however, not entirely reasonable to expect that one would always know the process around which to perturb. But we anticipate that the only scenario in which post-processing time is a significant factor is in the event that an experimenter wished to tune up their device to apply, for example, error suppression or correction protocols. In this event, the experimenter will always know their target channel -- the device is aimed at producing a clean, identity set of dynamics. If, however, the purpose of characterisation is to learn a completely unknown quantum stochastic process (for example, from a quantum sensor in an unknown environment), then the post-processing time is not such a significant factor and an eventual solution could be found using the myriad approaches in machine learning.

Another meta-parameter relevant to the problem of obtaining convergence is the number of shots per experiment. This, too, requires some fine-tuning. Too few shots-per-experiment and the landscape may be extremely noisy, preventing the fit to an adequate model. Too many shots-per-experiment, and we tend to find that the model overfits with respect to its data. The tuning here is not particularly fine, and in practice we often find that a large range of values results in an excellent model fit. Nevertheless, we examine this problem heuristically for different problem sizes to find a good practical guide. Appendix~\ref{app:benchmarking} presents a series of simulated benchmarks to determine both estimation time and accuracy of estimation across a wide range of processes.

\subsubsection{Basis Gateset}
In general, we will not simultaneously employ each feature of the model as outlined here for reasons of computational convenience and simplicity of exposition. Although accounting for the entire set of possible effects is not intractable, we find it more practical to design a model around what is reasonable for the expected physics. This is why the modularity of our approach is so useful. If we did not expect correlations in the control instrument, for example, we would simply set the bond dimension to be equal to 1, and the multi-time instrument would be completely time-local. 

\begin{figure}[!t]
	\centering
	\includegraphics[width=\linewidth]{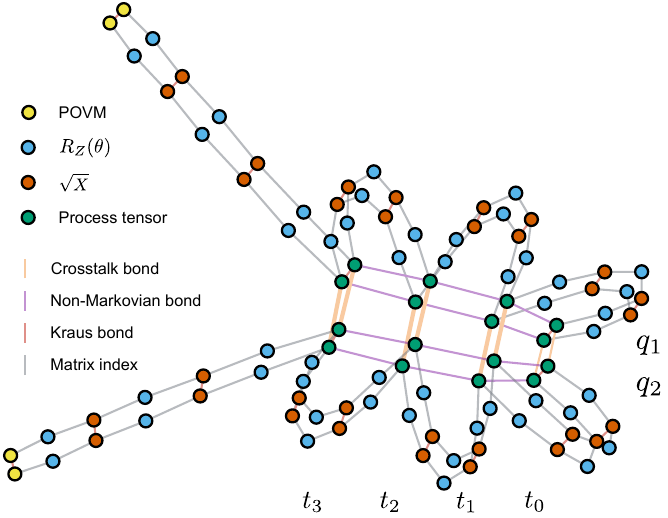}
	\caption[Tensor network decomposition of a three step, two qubit process partitioned into process and control models ]{Tensor network decomposition of a three step, two qubit process partitioned into process and control -- specific to IBM Quantum gate decompositions -- both of which are estimated. The background process partitions into non-Markovian (horizontal) and crosstalk (vertical) bonds as before. The single qubit control operations decompose into virtual $Z$-rotations and physical $X$-pulses, followed by final POVM.}
	\label{fig:dragonfly_fig}
\end{figure}

We can also design hardware-specific simplifications of the model. In IBM Quantum devices (and many other hardware platforms), for example, single-qubit unitary gates decompose into a sequence of physical $X$ pulses and virtual $Z$ rotations, such that an arbitrary single-qubit unitary can be taken as a function of three parameters, $u(\theta,\phi,\lambda)$: 
\begin{equation}
	\label{eq:u3-decomp}
	R_Z(\phi + 3\pi)\cdot R_X(\pi/2)\cdot R_Z(\theta + \pi)\cdot R_X(\pi/2)\cdot R_Z(\lambda).
\end{equation}
This way, only a single physical gate needs tuning up. Consequently, only a single physical gate needs to be estimated. It suffices to incorporate only this single physical gate into our model, taking the remainder to be perfect $Z-$rotations. The single-qubit rotations hence decompose into the tensor network shown in Figure~\ref{fig:dragonfly_fig}.
Assuming no control correlations then, the gate set is completely defined by $\{\Upsilon_{k:0}, \hat{R}_{X}(\pi/2)^{(q_i)}, \Pi_0^{(q_i)}, \Pi_1^{(q_i)}\}$. In fact, for many of our simulations and our demonstrations on IBM Quantum devices, this minimal model is the one we shall estimate, and we find it to be extremely accurate.

In accordance with the availabilities of current technology, we consider predominantly the estimation of restricted process tensors here. By this we mean we have a sequence of unitaries operations followed by a final terminating measurement. This means that each tester $\mathbf{A}_{k:0}^{(j,x)}$ is only a two-outcome probe. The procedure is hence designed not to uniquely learn the properties of the process tensor, but to capture extrinsic behaviours of the dynamics. We emphasise, however, that the formalism is wholly compatible with the estimation of multi-time statistics, as per the methods in Ref.~\cite{white2021many}.

\section{Discussion}
\label{sec:discussion}

In this work, we address the challenge of correlated noise on two fronts: unifying diverse notions of non-Markovian processes into a cohesive framework, and developing efficient methods to characterise such noise even in adversarial settings. We validate our methods both with extensive simulations and real noise on IBM Quantum devices, which highlights the efficacy of sparse models in capturing spatiotemporally relevant effects without information loss. Our contributions advance (i) conceptual understanding of quantum noise, (ii) state-of-the-art noise characterisation, (iii) practical analysis of exotic correlated noise, and (iv) novel methods for optimal control of quantum devices.

Our approach enables systematic circuit compilation tailored to a device's noise fingerprint. Discussions on optimal control are usually fine-grained to the level of pulse shaping~\cite{werschnik2007quantum, PRXQuantum.2.030333}. One difficulty with this approach is that in practice, whenever a signal is sent to the device, there is always distortion. Pulse samples are not true step functions, and the physical realisation can vary from qubit to qubit. Hence, even though theoretical prescriptions can tailor control to the physics of hardware, some data collection and feed-forward are always necessary. Further, designated pulse shaping cancels out only fast-acting correlated errors and do not account, for example, for context dependence in sequences of gates~\cite{veitia2020testing, PhysRevX.9.021045}. Our methods, in contrast, take the gate pulses as they are, but learn about them as much as possible. The result is more compatible with the established control software of the system, and designed to stop errors from propagating in an algorithmic setting. Robust characterisation of this form also allows the user to determine where there is room left for active control to help. If non-Markovianity is present at the timescale in question, then a judicious choice of unitary operations can always be used to extend the processor's performance.

The pervasiveness of non-Markovian noise is already apparent in quantum computing hardware~\cite{wei2022hamiltonian,muller2019towards,wilen2021correlated,RevModPhys.87.1419,ParradoRodriguez2021crosstalk,kuhlmann2013charge,yoneda2022noise,rojas2023spatial,wu2021concise,Terhal_2020,PhysRevB.97.054508,Nielsen2021gatesettomography,RBK2017,PhysRevX.9.021045}.
Less appreciated, however, is the potential impact of pathological noise on future devices and experiments. In particular, the adversarial nature of correlated noise can stunt the progression of fault-tolerant quantum computing~\cite{correlated-qec,Clader2021,PhysRevResearch.5.043161,wang2023dgr}. 
In an area where hardware is both expensive and scarce, software tools to control and compensate for these dynamics in detail will be valuable in achieving more with less. The upsides to scrutinising real device noise are myriad. In the context of error suppression, we have seen that deterministic control may be chosen to optimise the performance of circuit compilation and dynamical decoupling sequences. But correlated noise characterisation will also be valuable for error mitigation~\cite{cai2023quantum,PRXQuantum.2.040330}, where real noise models could circumvent pessimistic analyses~\cite{quek2023exponentially}. Further, characterisation can naturally integrate into tensor network decoders~\cite{chubb2021statistical,piveteau2023tensor} for augmented decoding practices and even allow for the construction of hardware-tailored error correcting codes~\cite{farrelly2021tensor,su2023discovery,mauron2023optimization,nautrup2019optimizing}.
Our methods -- both in concept and in execution -- push the forefront of this important problem, and we anticipate their application across the broad church of quantum computing software.

Our presented results achieve pseudo-scalability for characterisation methods, reducing sampling overheads to polynomial and enabling arbitrary numbers of times. However, real scalability to benchmark dozens or thousands of qubits requires further advancements. There, the 2D tensor networks employed here become intractible, and will necessitate the introduction of spacetime Markov order concepts to causally disconnect regions~\cite{taranto1,White-MLPT}. One might also consider the exploration of other geometries. In systems with polynomially decaying correlations, exploring tree-like structures may capture essential physics more effectively~\cite{dowling2023process}. This introduces a connection between fine and coarse-grained process memory and quantum error correction protocols, where noise can vary in scale at the pulse, gate, and logical levels. Establishing a hierarchical notion of correlated noise in theoretical settings is valuable for understanding its impact on quantum processors.

Our paper builds on the program to operationally understand non-Markoian quantum noise. A key aim of this program is to build tools to characterise and control complex noise. This paper provides a powerful tool for quantum engineers to add to their toolkit. Namely, it combines the desired features of GST -- a widely used tool for quantum benchmarking -- with methods for handling non-Markovian noise, which is a gap in the current benchmarking toolkit. Importantly, the tools we develop here are designed to readily employ approximation methods for scalability, such as those of quantum Markov order~\cite{taranto1, White-MLPT}, standard tensor network compression tools, or Pauli-twirled Clifford circuits~\cite{chen2024efficientselfconsistentlearninggate, hockings2024scalablenoisecharacterisationsyndrome} to enable large-scale implementations. Another future avenue will be to integrate the current tools with fast Bayesian tomography~\cite{PhysRevApplied.17.024068,su2023characterizing} methods to yield autonomous real-time characterisation of noise. The compact description of non-Markovian noise will aid the development of quantum error correction resilient to complex noise models.

We envision further work on precision estimation within our models, drawing inspiration from GST~\cite{Nielsen2021gatesettomography}. We have already noted that the $\chi=1$ case is a Markov model, and hence can be interpreted as a scalable implementation of GST. Advances on this front have already been made in the context of compressive data using tensor networks, but not yet explicitly in the system dimension~\cite{Brieger_2023}. Essential features for robust and comprehensive characterisation of near-term quantum devices include coherent parameter amplification for enhanced detection, optimisation of the unobservable gauge, and to identification of non-Markovian noise generators~\cite{blume2022taxonomy}. Further, our characterisation model is currently constructed ad-hoc, involving user-made choices like selecting bond dimensions. We could benefit from integrating automated model selection processes, such as using genetic algorithms~\cite{gentile2021learning}.

Precision estimation of parameters extends beyond noise detection in quantum computers to any open quantum system. We emphasise therefore that the results of this work are not solely applicable to the study of noise. These findings may interest the metrology community, particularly in quantum sensing of non-Markovian environments. PTT serves as a form of learning about an environment through structured control, translating the problem of optimal estimation into adaptive metrology using quantum combs, especially in practical, finite-sampling scenarios~\cite{meyer2023quantum}. The challenge of learning quantum environments is broadly applicable, and this paper marks a crucial step in understanding and addressing complex quantum noise in quantum hardware.

\begin{acknowledgments}
	We thank Robin Blume-Kohout, Aidan Dang, Michael Gullans, Lloyd Hollenberg, Corey Ostrove, and Kevin Young for valuable discussions. 
    G.A.L.W. was supported by an Australian Government Research Training Program Scholarship for the duration of this work. 
	K.M. is supported through Australian Research Council Discovery Project DP220101793.
	K.M. and C.D.H. acknowledge the support of Australian Research Council's Discovery Project DP210100597.
	K.M. and C.D.H. were recipients of the International Quantum U Tech Accelerator award by the US Air Force Research Laboratory.\\
	
\end{acknowledgments}

\section*{Data availability}
The code and data necessary to replicate all of the results in this work may be found in our public GitHub repository~\cite{PTT_github_repo}. 

\section*{References}

\clearpage
\appendix
\onecolumngrid

\section{Definitions and examples in the anatomy of quantum noise}
\label{app:definitions}

\subsubsection*{Markovian Error Channels}
We have seen a great many conditions (both operational and not) for which quantum dynamics might be said to be Markovian. Very commonly accepted in the literature is the case where a series of control operations $\{\mathcal{A}_i\}$ and time steps $\{\mathcal{E}_{j:j-1}\}$ factorise, i.e., they can be composed: $\mathcal{A}_{j} \circ \mathcal{E}_{j:j-1}\circ \cdots \circ \mathcal{A}_1 \circ \mathcal{E}_{1:0}\circ\rho_0^S$. 
These have been well characterised in the literature~\cite{resch2021benchmarking,noise-coherence-2015,postler2022demonstration}, see especially Ref.~\cite{blume2022taxonomy}. This is consistent with our definition of both process tensor and tester represented as product states, and includes both gate dependent and independent noise. 
	
\textbf{Process Example:} Relaxation processes, quantified by $T_1$ decay, are irreversible. A qubit couples weakly to a cold, large, environment and exchanges energy 
with the environment. But qubits are engineered to have resonant frequencies $\omega \gg k_B T$ for ambient temperature $T$, meaning that energy gain by the qubit is exponentially suppressed~\cite{krantz2019quantum}. Practically, then, energy leaks in one direction  until the qubit has relaxed into its lowest energy state. The coupling is fixed, and hence stationary.

\textbf{Control Example:} A qubit is driven around its $x$-axis with pulse envelope $e(t)$ and coupling strength $h_d$ between the qubit and the drive field. The resulting time evolution operator is given by 
\begin{equation}
	U(t)=\exp\left(\frac{i}{2}h_d \int_0^{t}e(t')\text{d}t' X\right).
\end{equation}
This equation can be solved to rotate about the $x$-axis by angle
\begin{equation}
	\Theta(t) = -h_d\int_0^t e(t')\text{d}t'.
\end{equation}
If the coupling strength $h_d$ is not accurately known, the pulse envelope will not correspond to the desired $\Theta(t)$ and will be a coherent over or under rotation compared to the ideal gate.

\subsubsection*{Active and Passive Crosstalk}
Crosstalk denotes the presence of undesirable interactions between different qubits on a device. We first make the distinction that crosstalk only refers to the case where the qubits involved are entirely contained in $S$. Otherwise, if not accounted for, the qubits constitute an environment and will be treated differently. We make the further distinction between \emph{passive} and \emph{active} crosstalk. 

	\begin{definition}[Passive Crosstalk]
		For qubits $\{q_i\}\in S$, if the uncontrolled dynamics feature at least a weight-2 Pauli term in the Hamiltonian, this has the capacity to generate entanglement between different qubits, irrespective of the control operations. This is said to be {passive crosstalk}.
	\end{definition}
	\textbf{Example:} Fixed-frequency transmon qubits with fixed couplings enables entangling interactions between qubits connected to the same coupling bus~\cite{krantz2019quantum}. But the coupling leads to state-dependent frequency shifts of coupled qubits in the form of a static $ZZ$ term in the interaction Hamiltonian~\cite{PhysRevLett.129.060501}. The coupling dresses the energy levels of each qubit: if either qubit is in a $Z$-eigenstate, this resonance shift causes coherent $Z$-rotations in the other. If both qubits have non-zero transverse Bloch sphere components, this will lead to an entangling interaction.

	\begin{definition}[Active Crosstalk]
		If a control operation $\mathcal{A}$ is intended to act on some qubit $q_j\in S$, but instead also manipulates other qubits within the system, this is said to be {active crosstalk}.
	\end{definition}
	\textbf{Example:} In optically-addressed trapped ion devices, a widely employed method to implement entangling gates is known as the M\o lmer-S\o rensen (MS) gate~\cite{PhysRevLett.82.1835}. For a given string of qubits $q_1,q_2,\cdots,q_n$, the gate uses a common vibrational mode to mediate entanglement, provided that the laser illuminates each qubit in the string uniformly. 
	If the beam of the laser is not tightly controlled, then residual light can illuminate unintended neighbouring ions. This allows them to participate in the same MS interaction, driving unwanted entangling interactions whenever the MS gate is used~\cite{ParradoRodriguez2021crosstalk}.

\subsubsection*{Process non-Markovianity}

\begin{definition}[Process non-Markovianity]
	The always-on interaction between a system $S$ and its environment $E$ defines a quantum stochastic process, whose representation is given by a process tensor \pt{}. These dynamics are said to be {process non-Markovian} if \pt{} cannot be expressed as a product of its marginals. That is:
	\begin{equation}
		S(\Upsilon_{k:0}\mid\mid \bigotimes_{j=1}^k \hat{\mathcal{E}}_{j:j-1}\otimes \rho_0) > 0,
	\end{equation}
\end{definition}
for relative entropy $S[\rho\mid\mid\sigma]:=\text{Tr}[\rho(\log\rho - \log\sigma)]$.
This is exactly the definition of non-Markovianity employed in the context of quantum stochastic processes, introduced in Ref.~\cite{Pollock2018}. That is, we have temporal quantum correlations generated by an environment with which a system interacts. The reason we have included the `process' modifier in the label is that now we are including control operations in the world view, we wish to be precise about the physical origin of the correlations. 

\textbf{Example:} The electron and nuclear spins of a $^{31}$P donor in isotopically pure $^{28}$Si (which has nuclear spin 0) can be used as qubits. However, if the silicon is not isotopically pure, then both electron and nuclear spin may coherently interact with the extraneous silicon nuclei. This always-on interaction will generate temporal correlations in the dynamics of the systems. 

\subsubsection*{Control non-Markovianity}
\begin{definition}[Control non-Markovianity]
	For a system probed across multiple times by a multi-time instrument $\mathbf{A}_{k:0}$, we say that the hardware is {control non-Markovian} if the (appropriately normalised) instrument does not factor into a product of its marginals. That is:
	\begin{equation}
		S(\hat{\mathbf{A}}_{k:0}\mid\mid \Pi_k\otimes \bigotimes_{j=0}^{k-1} \hat{\mathcal{A}}_j) > 0,
	\end{equation}
\end{definition}
for trace-normalised $\hat{\mathbf{A}}_{k:0}$ and marginals. Note that this is whether or not the control correlations are deliberately created.

\textbf{Example:} Ion trap qubits are optically addressed by lasers for their single qubit gates. The control lasers may have quasistatically fluctuating power spectra. An unaccounted for power fluctuation (i.e., if the gate time is not altered) has the effect that the unitary will coherently over or under rotate about its rotation axis. If the power fluctuates on the timescale of the circuit, then each $\mathcal{A}_i$ may over/under rotate by the same amount, some random variable $\epsilon$ drawn from the power spectrum of the laser. The error is then coherently correlated across different gates.

\subsubsection*{Time-dependent Markovianity}
The next category we consider is that of time-dependent Markovianity. Some authors in the literature consider this to be non-Markovian, because it breaks the tenets of stationarity, even if not the tenets of temporal locality. 
We start with a circuit clock $\{t_0,t_1,\cdots,t_k\}$ to designate the spacing between steps in a process. This clock is reset with every run of the circuit.
\begin{definition}[Time-dependent process]
	A $k$-step process whose process tensor Choi state \pt{} can be expressed as 
	\begin{equation}
		\Upsilon_{k:0} = \bigotimes_{j=1}^{k} \hat{\mathcal{E}}_{j:j-1}\otimes \rho_0,\:\text{if } \exists\: j\neq l\text{ s.t. }\hat{\mathcal{E}}_{j:j-1} \neq \hat{\mathcal{E}}_{l:l-1}
	\end{equation}
	is said to have {time-dependent process Markovianity}.
\end{definition}

Additionally, we have instruments which do not behave as expected:
\begin{definition}[Time-dependent control]
	A gate $\mathcal{A}$ is said to be {time-dependent control Markovian} if, for $k$ applications of $\mathcal{A}$, its multi-time instrument is
	\begin{equation}
		\begin{split}
		\hat{\mathbf{A}}_{k-1:0}  &= \bigotimes_{i=0}^{k-1}\hat{\mathcal{A}}_i,\:\text{but}\\
		\hat{\mathbf{A}}_{k-1:0}&\neq \bigotimes_{i=0}^{k-1}\hat{\mathcal{A}},
		\end{split}
	\end{equation}
	where $\hat{\mathbf{A}}_{k-1:0}$ is the repeated application of $\mathcal{A}$. That is, the tester is a product state (and hence carries no correlations) but the effect of the gate takes on a time dependence.
\end{definition}
A key distinction with control here is intent. Of course, if one were to apply a series of different perfect gates, then this constitutes time-dependent control Markovianity in a sense. To tighten this, we consider it to be the case that one \emph{intends} to repeat a gate, and the action of the gate changes over time, even if the effects are still time-local. 
	
\textbf{Example:} We have a faulty $X$ gate whose properties we wish to determine. We perform a GST experiment with $k$ sequential applications of the gate for various values of $k$. But the gate degrades over time due to the laser heating up: at $t_0$, a perfect $R_X(\pi)$ rotation is performed. At $t_1$, $R_X(\pi + \theta_1)$ occurs, where $\theta_1$ is drawn randomly from some distribution. At $t_2$, $R_X(\pi+\theta_2)$ occurs where $\theta_2$ is drawn from a different distribution that has no covariance with $\theta_1$ -- and so on. Attempts, then, to fit this the one gate to different $(X)^k$ expressions will fail.
The situation is rectified by treating each circuit time as an independent gate, collected in $\{X_0, X_1, \cdots, X_{k-1}\}$. The dynamics is then accurately modelled by the composition of these.

\subsubsection{Process and Control Drift}
	
\begin{definition}[Drift]
	{Drift} is if: with respect to a wall clock, either the underlying model for control, or the underlying model for process/control, changes with time.
	Suppose we have $M$ experiments (shots), each taking place at some wall clock time $\{s_1,\cdots s_M\}$. The wall clock time distinguishes between different runs of an identical circuit. Both process \pt{} and control $\mathbf{A}_{k:0}$ may be indexed by the times at which experiments are run to collect data. If there is any variation within the sets $\{\Upsilon_{k:0}^{(s_j)}\}$ and $\{\mathbf{A}_{k:0}^{(s_j)}\}$, then either the process or the control is said to have drifted, respectively. 
\end{definition}
The difficulty in integrating drift into the process tensor framework is because drift is a practical effect that arises as a consequence of the time it takes to characterise. The process tensor is sensitive only to a circuit clock, rather than a wall clock.

By ``look'' we mean that any characterisation designed to detect non-Markovianity would flag it as such. In theoretical treatments of quantum processes, time is typically taken for granted as a parameter known to arbitrary precision. The distinction then between time-dependent Markovian and non-Markovian processes in this case is always clear. However, in practice where this is not true, time can act as an effective environment.
The conflation with time-dependent processes is that whether they look Markovian or non-Markovian depends on whether one has fast enough control over the system that different times may be \emph{conditioned on} at the time scales over which change is effected. If times are marginalised over then these can be indistinguishable from a non-Markovian environment. One way to understand this is to note that the set of Markovian processes is non-convex; convex combinations of different Markovian processes can result in something non-Markovian. 

Effects of drift are non-Markovian in the sense that we can reconstruct process tensors that have correlations between different points in time. An important distinction to make is that we are marginalising over a time variable rather than a space variable (a physical system). We can evaluate a perfectly good process tensor model for the time-averaged characterisation. But as soon as we attempt to apply our model to predict dynamics, we will be conditioning on a single value for time, at which point the model will break down. 
The introduction of clocks somewhat obfuscates the ontological existence of processes in a single-shot setting. Treating drift and time-dependent Markovianity introduces an epistemic ignorance into the matter, we cannot in practice always condition on the parameter values of a clock. Here, then, time is a basic form of memory. 
The following example emphasises how marginalising over a wall clock can induce an identical non-Markovian ensemble as a classical environment. \par 
\textbf{Example:} Consider the reconstruction of two separate two step processes on a single qubit.
		\begin{enumerate}
			
			\item The system $S$ is coupled to a single qubit environment $E$, which is initialised in a $\ket{+}$ state. The process then undergoes a series of two CNOT gates controlled by $E$ and targetted on $S$. This is non-Markovian because $E$ is a random variable controlling the correlated evolution of the system. If $E=\ket{0}$, the system undergoes a series of identity evolutions. If $E=\ket{1}$, it will be subject to a series of $X$ gates. The dynamics are the same across the entire period that an experimenter reconstructs the process tensor.
			\item The system is isolated from its environment but driven with faulty control electronics. An experimenter aims to reconstruct a process tensor by collecting statistics across a ten-minute period. Unbeknownst to the experimenter, for the first five minutes, the dynamics are perfect. But after five minutes, the control electronics randomly switch to applying sequential $X$ gates.
		\end{enumerate}
		In this scenario, the two reconstructed process tensors will be identical:
		\begin{equation}
			\Upsilon_{3:0} = \left(0.5|\Phi^+\rangle\!\langle \Phi^+|\otimes |\Phi^+\rangle\!\langle \Phi^+| + 0.5 |\Psi^+\rangle\!\langle \Psi^+|\otimes |\Psi^+\rangle\!\langle \Psi^+|\right) \otimes \rho_0.
		\end{equation}
		That is, the process is (classically) non-Markovian. In the first instance, a nearby qubit is responsible for the correlations. In the second, although it might be described nominally as Markovian, once we marginalise over the wall clock time variable we obtain the same non-Markovian process. The problem here is that the process tensor framework does not account for the wall clock, we have assumed an identical and repeated process in our experiments.

Let us comment further on time-dependent effects. As discussed in the main text, one could consider three different time scales with respect to the circuit repetition rate at which time-dependent dynamics behave differently. By circuit repetition rate, we mean the time from initialisation of a single shot to the next.
		\begin{enumerate}[label=(\roman*)]
			\item \underline{Faster than circuit:} Dynamics which change faster than the time-scale of the circuit admit a \emph{time-dependent Markovian} description because the relevant timescales are not marginalised. For example, a gate that degrades upon each application due to heating effects (and resets upon reinitialisation) can be well-described by the composition of dynamical maps, each with progressively more noise.
			\item \underline{Quasistatic:} Dynamics which change at the time-scale of the circuit will only admit \emph{non-Markovian} process characterisation, because -- by definition -- the time-scale cannot by probed by available control, and is hence marginalised. For example, suppose qubit's resonant frequency shifts depending on the state of a nearby fluctator. Upon each initialisation, the frequency will take a value according to the fluctuator state and the value stays fixed for the duration of that circuit, causing some level of coherent error. Across the ensemble of shots, this amounts to time-correlated noise where the error $\epsilon_i$ at time $t_i$ is the same as $\epsilon_j$ at $t_j$, and each $\epsilon$ is drawn from a probability distribution at the start of each circuit. Here, the distribution itself will be stationary.
			\item \underline{Drift:} Dynamics which change much slower than at the time-scale of the device repetition rate are said to have drifted, but the Markovian/non-Markovian distinction will depend on whether the drift happens faster or slower than the characterisation process. If it is slower, then the dynamics will naturally look static. A process tensor $\Upsilon_{k:0}^{\text{day 1}}$ may differ from $\Upsilon_{k:0}^{\text{day 7}}$, but both can be Markovian across the time taken to characterise. If the dynamics are faster, then we obtain the same non-Markovian effect as quasistatic processes. 
			So long as the data collection is \emph{rasterised} -- that is, if each individual circuit is iterated for one shot before any circuit is repeated -- then a process tensor can always be validly fit. This ensures proper marginalisation.
			However, one key difference is that -- in contrast to quasistatic noise -- the effects of drift can at least be witnessed, given that a round of collecting validation data (a few circuits) can be completed in a timeframe much shorter than characterisation takes. If characterisation takes place across a series of wall clock times $\{s_1,\cdots, s_M\}$, then the model will be averaged over all $s_i$. This can then be used to make predictions about a sequence of validation circuits at an effectively single time. If the predictions differ from what is actually observed, then it can be seen that the marginal wall clock value differs from a specific wall clock value.
		\end{enumerate}

\subsubsection{Environment Spillage}
	
\begin{definition}[Spillage Error]
	If an intended control operation $\mathcal{A}_j\in \mathcal{B}(\mathcal{H}_S)$ at time $t_j$ excites some transition \emph{outside} $\mathcal{H}_S$, \emph{and} there exists a QPT experiment for a future dynamics $\mathcal{E}_{j':j'j-1},\ t_{j'} > t_j$, which can detect the effects of $\mathcal{A}_j$, then we term this either a superprocess or a spillage error.
\end{definition}
	
\textbf{Example:} A quantum system passively interacts with some two-level system (TLS) defect in the environment. An experimenter applies a control operation to $S$ by switching on a drive field with pulse envelope $e(t)$. The Fourier transform of the envelope reveals a spectrum with frequencies resonant with the TLS. Some population of the TLS is then driven from the ground $|g\rangle$ to the excited $|e\rangle$ state. The system is then evolving with respect to a \emph{different} environment, and hence no longer described by the same quantum stochastic process.

\section{Numerical Benchmarking}
\label{app:benchmarking}
\subsection{Demonstrations}
We have constructed both an efficient ansatz for representing multi-qubit quantum  stochastic processes and an algorithm to fit the ansatz to experiment data. We are now in a position to robustly test it. Some fine-tuning is required in the fitting process, and so we point out several best-usage policies which were discovered by trial and error. 

\subsubsection{Benchmarking Time Taken}

To start, we consider the time taken to contract expectation values of various process tensor networks. Linearly connected tensor networks -- such as MPOs -- have efficient algorithms to evaluate, and hence the time taken grows approximately linearly in the number of steps for fixed qubit number. However, it is in general a \texttt{\#P-complete} problem to find the optimal contraction~\cite{PhysRevResearch.2.013010}. \texttt{opt\_einsum}~\cite{daniel2018opt} is a Python package designed for heuristically finding hypercontraction paths for tensor networks. After some exploration, we found that the \texttt{auto-hq} algorithm performed best with our specific topology. Although contraction costs grow exponentially for 2D tensor networks, using this approach we find it is feasible to fit tensor network ans\"atze up to around 4-5 qubits on a personal computer. 

\begin{figure}[htbp]
	\centering
	\includegraphics[width=\linewidth]{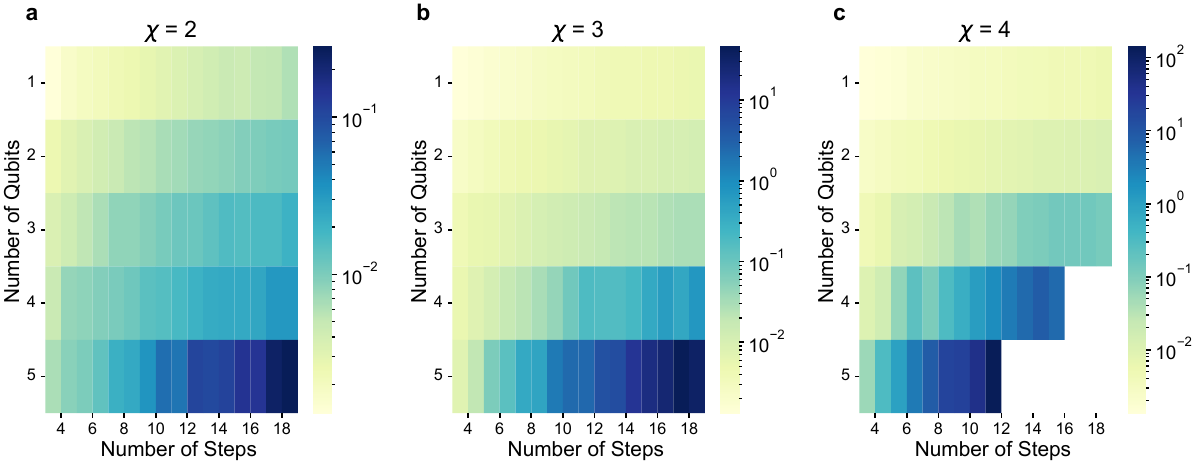}
	\caption[Benchmarking contraction times for different spatiotemporal tensor network expectation values.]{Benchmarking contraction times for a single expectation value evaluation across different maximal bond dimensions $\chi$, number of qubits $n$, and number of steps $k$. The contraction time grows exponentially in the area of the network. Not all cells are filled in the $\chi=4$ instance due to memory limitations. }
	\label{fig:contraction-times}
\end{figure}

In Figure~\ref{fig:contraction-times} we benchmark the time it takes to perform a single contraction across different values of $k$, $n$, and $\chi$ on an 3.2 GHz 6-Core Intel Core i7 processor. This is performed using the Python library \texttt{quimb} for tensor network semantics, which deploys to the \texttt{numpy.einsum} function for computation. We can clearly identify the scaling behaviour between different network sizes and bond dimensions. Small numbers of times, qubits, and bond dimensions, are feasible on a personal computer, but estimating larger than this may take the use of GPUs or cluster computing.

In practice, while performing the optimisation. we use a compiled version of this code using the \texttt{JAX} library~\cite{jax2018github}, for which the individual contractions are approximately two orders of magnitude faster. Using the ADAM~\cite{kingma2014adam} optimiser, we randomise over the data at each iteration of stochastic gradient descent. A good trade-off between memory requirements and convergence rates seems to be to use 1000 data points per iteration of the fit. 

\subsubsection{Benchmarking Experimental Requirements}

We now demonstrate this on synthetic data, scaling both number of qubits and number of time steps. To simulate non-Markovian behaviour, we couple a chain of qubits via a Heisenberg interaction with random couplings to a randomly initialised single-qubit environment. We also include random $ZZ$ couplings between each of the nearest-neighbour qubits in the chain. We are predominantly interested in controlling such non-Markovian systems, and so we will consider only restricted process tensors here. Note, however, that no modifications need to be included to estimate full process tensors. 

As part of our investigation, we look extensively into the effects of different metaparameters on both the quality of the optimisation and the rate of convergence. As we have already mentioned, we use the \texttt{auto-hq} path-finding algorithm to find appropriate contraction paths. 
A good choice of the regularisation metaparameter $\kappa$ seems to be $\kappa=1$: if this is too low, it can sometimes lead to unphysical process tensor estimates; too high, and it can slow down the convergence unreasonably. In our numerical experiments, this choice reliably produces process tensors with acausal Pauli expectation values suppressed to $10^{-7}$ or lower.

\begin{figure}[!htbp]
	\centering
	\includegraphics[width=0.9\linewidth]{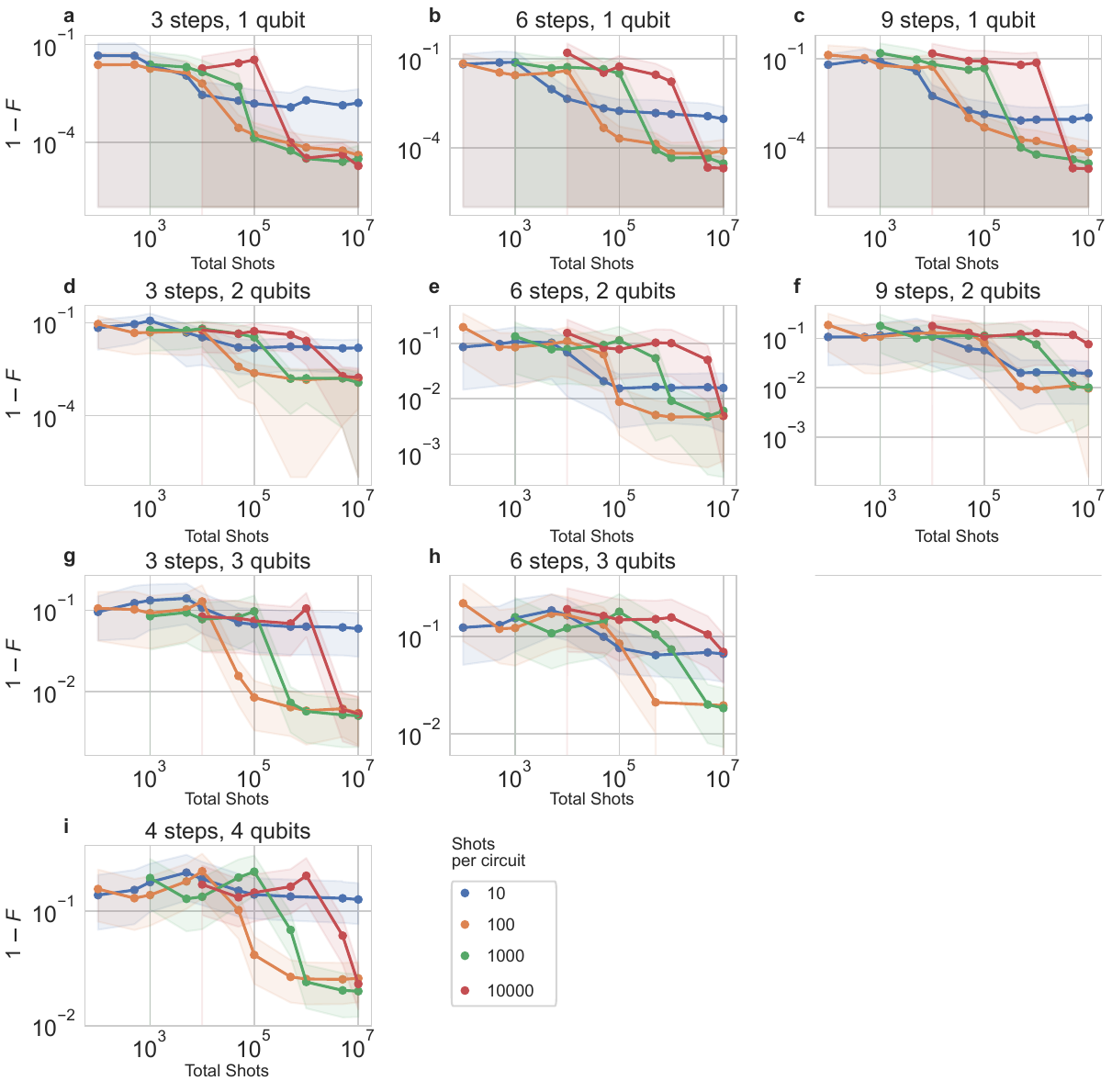}
	\caption[Benchmarking 2D tensor network fits from synthetic data]{Benchmarking 2D tensor network fits from synthetic data. For a variety of step numbers and qubit numbers, we fit tensor network models to qubits coupled via exchange interactions to a non-Markovian bath. We observe convergence properties as dependent on both the total number of shots, and the number of shots per circuit. We see a trade-off between quality and quantity of data, observing that for small numbers of shots, more circuits is preferable and vice versa. In all cases, we see orders of magnitudes fewer circuits required. }
	\label{fig:tn-convergence-graph}
\end{figure}

The next metaparameter we experimented with was the total number of shots per circuit. Given a total number of experiments run, one has the broad choice of collecting either numerous statistics on a few circuits, or few repetitions of many circuits. The trade-off here is quality of the data versus number of points in the space of parameter values. 
For similar problems applied to QPT, Ref.~\cite{torlai2020quantum} suggest using single shot measurements across many random circuits. This is in the spirit of unsupervised machine learning, where it is envisaged that a rough optimisation landscape can help to avoid local minima, as well as avoid overfitting to produce an approximately generalisable estimate. We find two issues with this: the first is that for extremely stochastic data, the time taken to converge can be unreasonably long. Secondly, although for very few numbers of shots the final estimate can be marginally better than a tensor network which has overfit, it severely underperforms for even modest amounts of data.

We explore this key point in Figure~\ref{fig:tn-convergence-graph}, where we numerically demonstrate our process tensor network fits for the above dynamics. For a variety of different numbers of qubits and numbers of steps, we fit a process tensor estimate and then compute the reconstruction (in)fidelities for 300 randomly generated validation sequences at $16\ 384$ shots. We perform this for differently sized pools of data, ranging from $10^2$ up to $10^7$.
We see in Figure~\ref{fig:tn-convergence-graph} the effects of changing the shots-per-circuit metaparameter in data collection. Note that the \emph{total} amount of data is fixed entirely by the $x$-axis, and is the same for each numbers of shots-per-circuit. That is, a single circuit run $10\ 000$ times will be equivalently compared to 100 circuits each run 100 times. For lower numbers of total data, it appears preferable to choose a lower number of shots-per-circuit. These fits appear to converge to a much more generalisable estimate of the underlying process. 
Likely, this is because the landscape for higher quality data is much smoother, causing the optimisation to either get stuck in a local minimum or simply to overfit. However, at larger numbers of total shots $10^6$--$10^7$, it appears better to smooth that data, since the case of 10 shots-per-circuit tends to hit a plateau and fails to improve substantially. 
The upshot is that 100--1000 shots-per-circuit appears to be a quasi-optimal choice for moderate amounts of data.

Note the dependence of fit quality on number of qubits or number of times: for a given pool of data, the quality of the fit is relatively insensitive to the number of steps. 
This is less true for increasing the number of qubits, since central tensor blocks are not spatially repeated in the same way that they are temporally repeated. Further, increased size of the probability distribution -- as measured by number of measurement outcomes -- will naturally increase variance of the fit. Nevertheless, we see that even for a four qubit process, we still achieve an average reconstruction fidelity close to $0.99$, despite having a comparable number of data points supplied when compared with the three-step process tensors in Ref.~\cite{White-MLPT}. This reduces estimation from the previously infeasible $\mathcal{O}(10^{19})$ experiments, to the much more palatable $10^7$--$10^8$. 
Since we are estimating fewer parameters, we also achieve a greater accuracy than in the fully general version of PTT, even for three-step processes, as benchmarked in Ref.~\cite{White-MLPT}.

\begin{thebibliography}{137}%
\makeatletter
\providecommand \@ifxundefined [1]{%
 \@ifx{#1\undefined}
}%
\providecommand \@ifnum [1]{%
 \ifnum #1\expandafter \@firstoftwo
 \else \expandafter \@secondoftwo
 \fi
}%
\providecommand \@ifx [1]{%
 \ifx #1\expandafter \@firstoftwo
 \else \expandafter \@secondoftwo
 \fi
}%
\providecommand \natexlab [1]{#1}%
\providecommand \enquote  [1]{``#1''}%
\providecommand \bibnamefont  [1]{#1}%
\providecommand \bibfnamefont [1]{#1}%
\providecommand \citenamefont [1]{#1}%
\providecommand \href@noop [0]{\@secondoftwo}%
\providecommand \href [0]{\begingroup \@sanitize@url \@href}%
\providecommand \@href[1]{\@@startlink{#1}\@@href}%
\providecommand \@@href[1]{\endgroup#1\@@endlink}%
\providecommand \@sanitize@url [0]{\catcode `\\12\catcode `\$12\catcode
  `\&12\catcode `\#12\catcode `\^12\catcode `\_12\catcode `\%12\relax}%
\providecommand \@@startlink[1]{}%
\providecommand \@@endlink[0]{}%
\providecommand \url  [0]{\begingroup\@sanitize@url \@url }%
\providecommand \@url [1]{\endgroup\@href {#1}{\urlprefix }}%
\providecommand \urlprefix  [0]{URL }%
\providecommand \Eprint [0]{\href }%
\providecommand \doibase [0]{http://dx.doi.org/}%
\providecommand \selectlanguage [0]{\@gobble}%
\providecommand \bibinfo  [0]{\@secondoftwo}%
\providecommand \bibfield  [0]{\@secondoftwo}%
\providecommand \translation [1]{[#1]}%
\providecommand \BibitemOpen [0]{}%
\providecommand \bibitemStop [0]{}%
\providecommand \bibitemNoStop [0]{.\EOS\space}%
\providecommand \EOS [0]{\spacefactor3000\relax}%
\providecommand \BibitemShut  [1]{\csname bibitem#1\endcsname}%
\let\auto@bib@innerbib\@empty
\bibitem [{\citenamefont {Postler}\ \emph {et~al.}(2022)\citenamefont
  {Postler}, \citenamefont {Heussen}, \citenamefont {Pogorelov}, \citenamefont
  {Rispler}, \citenamefont {Feldker}, \citenamefont {Meth}, \citenamefont
  {Marciniak}, \citenamefont {Stricker}, \citenamefont {Ringbauer},
  \citenamefont {Blatt}, \citenamefont {Schindler}, \citenamefont {M\"uller},\
  and\ \citenamefont {Monz}}]{postler2022demonstration}%
  \BibitemOpen
  \bibfield  {author} {\bibinfo {author} {\bibfnamefont {Lukas}\ \bibnamefont
  {Postler}}, \bibinfo {author} {\bibfnamefont {Sascha}\ \bibnamefont
  {Heussen}}, \bibinfo {author} {\bibfnamefont {Ivan}\ \bibnamefont
  {Pogorelov}}, \bibinfo {author} {\bibfnamefont {Manuel}\ \bibnamefont
  {Rispler}}, \bibinfo {author} {\bibfnamefont {Thomas}\ \bibnamefont
  {Feldker}}, \bibinfo {author} {\bibfnamefont {Michael}\ \bibnamefont {Meth}},
  \bibinfo {author} {\bibfnamefont {Christian~D.}\ \bibnamefont {Marciniak}},
  \bibinfo {author} {\bibfnamefont {Roman}\ \bibnamefont {Stricker}}, \bibinfo
  {author} {\bibfnamefont {Martin}\ \bibnamefont {Ringbauer}}, \bibinfo
  {author} {\bibfnamefont {Rainer}\ \bibnamefont {Blatt}}, \bibinfo {author}
  {\bibfnamefont {Philipp}\ \bibnamefont {Schindler}}, \bibinfo {author}
  {\bibfnamefont {Markus}\ \bibnamefont {M\"uller}}, \ and\ \bibinfo {author}
  {\bibfnamefont {Thomas}\ \bibnamefont {Monz}},\ }\bibfield  {title} {\enquote
  {\bibinfo {title} {Demonstration of fault-tolerant universal quantum gate
  operations},}\ }\href {\doibase 10.1038/s41586-022-04721-1} {\bibfield
  {journal} {\bibinfo  {journal} {Nature}\ }\textbf {\bibinfo {volume} {605}},\
  \bibinfo {pages} {675--680} (\bibinfo {year} {2022})}\BibitemShut {NoStop}%
\bibitem [{\citenamefont {Kim}\ \emph {et~al.}(2023)\citenamefont {Kim},
  \citenamefont {Eddins}, \citenamefont {Anand}, \citenamefont {Wei},
  \citenamefont {Van Den~Berg}, \citenamefont {Rosenblatt}, \citenamefont
  {Nayfeh}, \citenamefont {Wu}, \citenamefont {Zaletel}, \citenamefont {Temme}
  \emph {et~al.}}]{kim2023evidence}%
  \BibitemOpen
  \bibfield  {author} {\bibinfo {author} {\bibfnamefont {Youngseok}\
  \bibnamefont {Kim}}, \bibinfo {author} {\bibfnamefont {Andrew}\ \bibnamefont
  {Eddins}}, \bibinfo {author} {\bibfnamefont {Sajant}\ \bibnamefont {Anand}},
  \bibinfo {author} {\bibfnamefont {Ken~Xuan}\ \bibnamefont {Wei}}, \bibinfo
  {author} {\bibfnamefont {Ewout}\ \bibnamefont {Van Den~Berg}}, \bibinfo
  {author} {\bibfnamefont {Sami}\ \bibnamefont {Rosenblatt}}, \bibinfo {author}
  {\bibfnamefont {Hasan}\ \bibnamefont {Nayfeh}}, \bibinfo {author}
  {\bibfnamefont {Yantao}\ \bibnamefont {Wu}}, \bibinfo {author} {\bibfnamefont
  {Michael}\ \bibnamefont {Zaletel}}, \bibinfo {author} {\bibfnamefont
  {Kristan}\ \bibnamefont {Temme}},  \emph {et~al.},\ }\bibfield  {title}
  {\enquote {\bibinfo {title} {Evidence for the utility of quantum computing
  before fault tolerance},}\ }\href {\doibase
  https://doi.org/10.1038/s41586-023-06096-3} {\bibfield  {journal} {\bibinfo
  {journal} {Nature}\ }\textbf {\bibinfo {volume} {618}},\ \bibinfo {pages}
  {500--505} (\bibinfo {year} {2023})}\BibitemShut {NoStop}%
\bibitem [{\citenamefont {Abobeih}\ \emph {et~al.}(2022)\citenamefont
  {Abobeih}, \citenamefont {Wang}, \citenamefont {Randall}, \citenamefont
  {Loenen}, \citenamefont {Bradley}, \citenamefont {Markham}, \citenamefont
  {Twitchen}, \citenamefont {Terhal},\ and\ \citenamefont
  {Taminiau}}]{abobeih2022fault}%
  \BibitemOpen
  \bibfield  {author} {\bibinfo {author} {\bibfnamefont {MH}~\bibnamefont
  {Abobeih}}, \bibinfo {author} {\bibfnamefont {Y}~\bibnamefont {Wang}},
  \bibinfo {author} {\bibfnamefont {J}~\bibnamefont {Randall}}, \bibinfo
  {author} {\bibfnamefont {SJH}\ \bibnamefont {Loenen}}, \bibinfo {author}
  {\bibfnamefont {CE}~\bibnamefont {Bradley}}, \bibinfo {author} {\bibfnamefont
  {M}~\bibnamefont {Markham}}, \bibinfo {author} {\bibfnamefont
  {DJ}~\bibnamefont {Twitchen}}, \bibinfo {author} {\bibfnamefont
  {BM}~\bibnamefont {Terhal}}, \ and\ \bibinfo {author} {\bibfnamefont
  {TH}~\bibnamefont {Taminiau}},\ }\bibfield  {title} {\enquote {\bibinfo
  {title} {Fault-tolerant operation of a logical qubit in a diamond quantum
  processor},}\ }\href {\doibase https://doi.org/10.1038/s41586-022-04819-6}
  {\bibfield  {journal} {\bibinfo  {journal} {Nature}\ }\textbf {\bibinfo
  {volume} {606}},\ \bibinfo {pages} {884--889} (\bibinfo {year}
  {2022})}\BibitemShut {NoStop}%
\bibitem [{\citenamefont {Noiri}\ \emph {et~al.}(2022)\citenamefont {Noiri},
  \citenamefont {Takeda}, \citenamefont {Nakajima}, \citenamefont {Kobayashi},
  \citenamefont {Sammak}, \citenamefont {Scappucci},\ and\ \citenamefont
  {Tarucha}}]{noiri2022fast}%
  \BibitemOpen
  \bibfield  {author} {\bibinfo {author} {\bibfnamefont {Akito}\ \bibnamefont
  {Noiri}}, \bibinfo {author} {\bibfnamefont {Kenta}\ \bibnamefont {Takeda}},
  \bibinfo {author} {\bibfnamefont {Takashi}\ \bibnamefont {Nakajima}},
  \bibinfo {author} {\bibfnamefont {Takashi}\ \bibnamefont {Kobayashi}},
  \bibinfo {author} {\bibfnamefont {Amir}\ \bibnamefont {Sammak}}, \bibinfo
  {author} {\bibfnamefont {Giordano}\ \bibnamefont {Scappucci}}, \ and\
  \bibinfo {author} {\bibfnamefont {Seigo}\ \bibnamefont {Tarucha}},\
  }\bibfield  {title} {\enquote {\bibinfo {title} {Fast universal quantum gate
  above the fault-tolerance threshold in silicon},}\ }\href {\doibase
  https://doi.org/10.1038/s41586-021-04182-y} {\bibfield  {journal} {\bibinfo
  {journal} {Nature}\ }\textbf {\bibinfo {volume} {601}},\ \bibinfo {pages}
  {338--342} (\bibinfo {year} {2022})}\BibitemShut {NoStop}%
\bibitem [{\citenamefont {{Google~Quantum~AI}}(2023)}]{google2023suppressing}%
  \BibitemOpen
  \bibfield  {author} {\bibinfo {author} {\bibnamefont {{Google~Quantum~AI}}},\
  }\bibfield  {title} {\enquote {\bibinfo {title} {Suppressing quantum errors
  by scaling a surface code logical qubit},}\ }\href {\doibase
  https://doi.org/10.5281/zenodo.6804040} {\bibfield  {journal} {\bibinfo
  {journal} {Nature}\ }\textbf {\bibinfo {volume} {614}},\ \bibinfo {pages}
  {676--681} (\bibinfo {year} {2023})}\BibitemShut {NoStop}%
\bibitem [{\citenamefont {Bluvstein}\ \emph {et~al.}(2023)\citenamefont
  {Bluvstein}, \citenamefont {Evered}, \citenamefont {Geim}, \citenamefont
  {Li}, \citenamefont {Zhou}, \citenamefont {Manovitz}, \citenamefont {Ebadi},
  \citenamefont {Cain}, \citenamefont {Kalinowski}, \citenamefont {Hangleiter}
  \emph {et~al.}}]{bluvstein2023logical}%
  \BibitemOpen
  \bibfield  {author} {\bibinfo {author} {\bibfnamefont {Dolev}\ \bibnamefont
  {Bluvstein}}, \bibinfo {author} {\bibfnamefont {Simon~J}\ \bibnamefont
  {Evered}}, \bibinfo {author} {\bibfnamefont {Alexandra~A}\ \bibnamefont
  {Geim}}, \bibinfo {author} {\bibfnamefont {Sophie~H}\ \bibnamefont {Li}},
  \bibinfo {author} {\bibfnamefont {Hengyun}\ \bibnamefont {Zhou}}, \bibinfo
  {author} {\bibfnamefont {Tom}\ \bibnamefont {Manovitz}}, \bibinfo {author}
  {\bibfnamefont {Sepehr}\ \bibnamefont {Ebadi}}, \bibinfo {author}
  {\bibfnamefont {Madelyn}\ \bibnamefont {Cain}}, \bibinfo {author}
  {\bibfnamefont {Marcin}\ \bibnamefont {Kalinowski}}, \bibinfo {author}
  {\bibfnamefont {Dominik}\ \bibnamefont {Hangleiter}},  \emph {et~al.},\
  }\bibfield  {title} {\enquote {\bibinfo {title} {Logical quantum processor
  based on reconfigurable atom arrays},}\ }\href {\doibase
  https://doi.org/10.1038/s41586-022-04592-6} {\bibfield  {journal} {\bibinfo
  {journal} {Nature}\ }\textbf {\bibinfo {volume} {604}},\ \bibinfo {pages}
  {451–456} (\bibinfo {year} {2023})}\BibitemShut {NoStop}%
\bibitem [{\citenamefont {Proctor}\ \emph {et~al.}(2022)\citenamefont
  {Proctor}, \citenamefont {Rudinger}, \citenamefont {Young}, \citenamefont
  {Nielsen},\ and\ \citenamefont {Blume-Kohout}}]{proctor2022measuring}%
  \BibitemOpen
  \bibfield  {author} {\bibinfo {author} {\bibfnamefont {Timothy}\ \bibnamefont
  {Proctor}}, \bibinfo {author} {\bibfnamefont {Kenneth}\ \bibnamefont
  {Rudinger}}, \bibinfo {author} {\bibfnamefont {Kevin}\ \bibnamefont {Young}},
  \bibinfo {author} {\bibfnamefont {Erik}\ \bibnamefont {Nielsen}}, \ and\
  \bibinfo {author} {\bibfnamefont {Robin}\ \bibnamefont {Blume-Kohout}},\
  }\bibfield  {title} {\enquote {\bibinfo {title} {Measuring the capabilities
  of quantum computers},}\ }\href {\doibase
  https://doi.org/10.5281/zenodo.5197499. T} {\bibfield  {journal} {\bibinfo
  {journal} {Nature Physics}\ }\textbf {\bibinfo {volume} {18}},\ \bibinfo
  {pages} {75--79} (\bibinfo {year} {2022})}\BibitemShut {NoStop}%
\bibitem [{\citenamefont {Aharonov}\ and\ \citenamefont
  {Ben-Or}(1997)}]{aharonov1997fault}%
  \BibitemOpen
  \bibfield  {author} {\bibinfo {author} {\bibfnamefont {Dorit}\ \bibnamefont
  {Aharonov}}\ and\ \bibinfo {author} {\bibfnamefont {Michael}\ \bibnamefont
  {Ben-Or}},\ }\bibfield  {title} {\enquote {\bibinfo {title} {Fault-tolerant
  quantum computation with constant error},}\ }in\ \href {\doibase
  https://doi.org/10.48550/arXiv.quant-ph/9906129} {\emph {\bibinfo {booktitle}
  {Proceedings of the twenty-ninth annual ACM symposium on Theory of
  computing}}}\ (\bibinfo {year} {1997})\ pp.\ \bibinfo {pages}
  {176--188}\BibitemShut {NoStop}%
\bibitem [{\citenamefont {Huang}\ \emph {et~al.}(2020)\citenamefont {Huang},
  \citenamefont {Zhang}, \citenamefont {Newman}, \citenamefont {Cai},
  \citenamefont {Gao}, \citenamefont {Tian}, \citenamefont {Wu}, \citenamefont
  {Xu}, \citenamefont {Yu}, \citenamefont {Yuan}, \citenamefont {Szegedy},
  \citenamefont {Shi},\ and\ \citenamefont {Chen}}]{huang2020classical}%
  \BibitemOpen
  \bibfield  {author} {\bibinfo {author} {\bibfnamefont {Cupjin}\ \bibnamefont
  {Huang}}, \bibinfo {author} {\bibfnamefont {Fang}\ \bibnamefont {Zhang}},
  \bibinfo {author} {\bibfnamefont {Michael}\ \bibnamefont {Newman}}, \bibinfo
  {author} {\bibfnamefont {Junjie}\ \bibnamefont {Cai}}, \bibinfo {author}
  {\bibfnamefont {Xun}\ \bibnamefont {Gao}}, \bibinfo {author} {\bibfnamefont
  {Zhengxiong}\ \bibnamefont {Tian}}, \bibinfo {author} {\bibfnamefont
  {Junyin}\ \bibnamefont {Wu}}, \bibinfo {author} {\bibfnamefont {Haihong}\
  \bibnamefont {Xu}}, \bibinfo {author} {\bibfnamefont {Huanjun}\ \bibnamefont
  {Yu}}, \bibinfo {author} {\bibfnamefont {Bo}~\bibnamefont {Yuan}}, \bibinfo
  {author} {\bibfnamefont {Mario}\ \bibnamefont {Szegedy}}, \bibinfo {author}
  {\bibfnamefont {Yaoyun}\ \bibnamefont {Shi}}, \ and\ \bibinfo {author}
  {\bibfnamefont {Jianxin}\ \bibnamefont {Chen}},\ }\href@noop {} {\enquote
  {\bibinfo {title} {Classical simulation of quantum supremacy circuits},}\ }
  (\bibinfo {year} {2020}),\ \Eprint {http://arxiv.org/abs/2005.06787}
  {arXiv:2005.06787 [quant-ph]} \BibitemShut {NoStop}%
\bibitem [{\citenamefont {Clader}\ \emph {et~al.}(2021)\citenamefont {Clader},
  \citenamefont {Trout}, \citenamefont {Barnes}, \citenamefont {Schultz},
  \citenamefont {Quiroz},\ and\ \citenamefont {Titum}}]{Clader2021}%
  \BibitemOpen
  \bibfield  {author} {\bibinfo {author} {\bibfnamefont {B.~D.}\ \bibnamefont
  {Clader}}, \bibinfo {author} {\bibfnamefont {Colin~J.}\ \bibnamefont
  {Trout}}, \bibinfo {author} {\bibfnamefont {Jeff~P.}\ \bibnamefont {Barnes}},
  \bibinfo {author} {\bibfnamefont {Kevin}\ \bibnamefont {Schultz}}, \bibinfo
  {author} {\bibfnamefont {Gregory}\ \bibnamefont {Quiroz}}, \ and\ \bibinfo
  {author} {\bibfnamefont {Paraj}\ \bibnamefont {Titum}},\ }\bibfield  {title}
  {\enquote {\bibinfo {title} {{Impact of correlations and heavy tails on
  quantum error correction}},}\ }\href {\doibase 10.1103/PhysRevA.103.052428}
  {\bibfield  {journal} {\bibinfo  {journal} {Physical Review A}\ }\textbf
  {\bibinfo {volume} {103}},\ \bibinfo {pages} {052428} (\bibinfo {year}
  {2021})},\ \Eprint {http://arxiv.org/abs/2101.11631} {arXiv:2101.11631}
  \BibitemShut {NoStop}%
\bibitem [{\citenamefont {Nickerson}\ and\ \citenamefont
  {Brown}(2019)}]{correlated-qec}%
  \BibitemOpen
  \bibfield  {author} {\bibinfo {author} {\bibfnamefont {Naomi~H.}\
  \bibnamefont {Nickerson}}\ and\ \bibinfo {author} {\bibfnamefont
  {Benjamin~J.}\ \bibnamefont {Brown}},\ }\bibfield  {title} {\enquote
  {\bibinfo {title} {{Analysing correlated noise on the surface code using
  adaptive decoding algorithms}},}\ }\href {\doibase 10.22331/q-2019-04-08-131}
  {\bibfield  {journal} {\bibinfo  {journal} {Quantum}\ }\textbf {\bibinfo
  {volume} {3}},\ \bibinfo {pages} {131} (\bibinfo {year} {2019})},\ \Eprint
  {http://arxiv.org/abs/1712.00502} {arXiv:1712.00502} \BibitemShut {NoStop}%
\bibitem [{\citenamefont {Bonilla~Ataides}\ \emph {et~al.}(2021)\citenamefont
  {Bonilla~Ataides}, \citenamefont {Tuckett}, \citenamefont {Bartlett},
  \citenamefont {Flammia},\ and\ \citenamefont {Brown}}]{bonilla2021xzzx}%
  \BibitemOpen
  \bibfield  {author} {\bibinfo {author} {\bibfnamefont {J~Pablo}\ \bibnamefont
  {Bonilla~Ataides}}, \bibinfo {author} {\bibfnamefont {David~K}\ \bibnamefont
  {Tuckett}}, \bibinfo {author} {\bibfnamefont {Stephen~D}\ \bibnamefont
  {Bartlett}}, \bibinfo {author} {\bibfnamefont {Steven~T}\ \bibnamefont
  {Flammia}}, \ and\ \bibinfo {author} {\bibfnamefont {Benjamin~J}\
  \bibnamefont {Brown}},\ }\bibfield  {title} {\enquote {\bibinfo {title} {The
  {XZZX} surface code},}\ }\href {\doibase
  https://doi.org/10.1038/s41467-021-22274-1} {\bibfield  {journal} {\bibinfo
  {journal} {Nature Communications}\ }\textbf {\bibinfo {volume} {12}},\
  \bibinfo {pages} {2172} (\bibinfo {year} {2021})}\BibitemShut {NoStop}%
\bibitem [{\citenamefont {Tuckett}\ \emph {et~al.}(2018)\citenamefont
  {Tuckett}, \citenamefont {Bartlett},\ and\ \citenamefont
  {Flammia}}]{tuckett2018ultrahigh}%
  \BibitemOpen
  \bibfield  {author} {\bibinfo {author} {\bibfnamefont {David~K}\ \bibnamefont
  {Tuckett}}, \bibinfo {author} {\bibfnamefont {Stephen~D}\ \bibnamefont
  {Bartlett}}, \ and\ \bibinfo {author} {\bibfnamefont {Steven~T}\ \bibnamefont
  {Flammia}},\ }\bibfield  {title} {\enquote {\bibinfo {title} {Ultrahigh error
  threshold for surface codes with biased noise},}\ }\href {\doibase
  10.1103/PhysRevLett.120.050505} {\bibfield  {journal} {\bibinfo  {journal}
  {Physical review letters}\ }\textbf {\bibinfo {volume} {120}},\ \bibinfo
  {pages} {050505} (\bibinfo {year} {2018})}\BibitemShut {NoStop}%
\bibitem [{\citenamefont {Harper}\ and\ \citenamefont
  {Flammia}(2023)}]{PRXQuantum.4.040311}%
  \BibitemOpen
  \bibfield  {author} {\bibinfo {author} {\bibfnamefont {Robin}\ \bibnamefont
  {Harper}}\ and\ \bibinfo {author} {\bibfnamefont {Steven~T.}\ \bibnamefont
  {Flammia}},\ }\bibfield  {title} {\enquote {\bibinfo {title} {Learning
  correlated noise in a 39-qubit quantum processor},}\ }\href {\doibase
  10.1103/PRXQuantum.4.040311} {\bibfield  {journal} {\bibinfo  {journal} {PRX
  Quantum}\ }\textbf {\bibinfo {volume} {4}},\ \bibinfo {pages} {040311}
  (\bibinfo {year} {2023})}\BibitemShut {NoStop}%
\bibitem [{\citenamefont {Nautrup}\ \emph {et~al.}(2019)\citenamefont
  {Nautrup}, \citenamefont {Delfosse}, \citenamefont {Dunjko}, \citenamefont
  {Briegel},\ and\ \citenamefont {Friis}}]{nautrup2019optimizing}%
  \BibitemOpen
  \bibfield  {author} {\bibinfo {author} {\bibfnamefont {Hendrik~Poulsen}\
  \bibnamefont {Nautrup}}, \bibinfo {author} {\bibfnamefont {Nicolas}\
  \bibnamefont {Delfosse}}, \bibinfo {author} {\bibfnamefont {Vedran}\
  \bibnamefont {Dunjko}}, \bibinfo {author} {\bibfnamefont {Hans~J}\
  \bibnamefont {Briegel}}, \ and\ \bibinfo {author} {\bibfnamefont {Nicolai}\
  \bibnamefont {Friis}},\ }\bibfield  {title} {\enquote {\bibinfo {title}
  {Optimizing quantum error correction codes with reinforcement learning},}\
  }\href {\doibase https://doi.org/10.22331/q-2019-12-16-215} {\bibfield
  {journal} {\bibinfo  {journal} {Quantum}\ }\textbf {\bibinfo {volume} {3}},\
  \bibinfo {pages} {215} (\bibinfo {year} {2019})}\BibitemShut {NoStop}%
\bibitem [{\citenamefont {Farrelly}\ \emph {et~al.}(2021)\citenamefont
  {Farrelly}, \citenamefont {Harris}, \citenamefont {McMahon},\ and\
  \citenamefont {Stace}}]{farrelly2021tensor}%
  \BibitemOpen
  \bibfield  {author} {\bibinfo {author} {\bibfnamefont {Terry}\ \bibnamefont
  {Farrelly}}, \bibinfo {author} {\bibfnamefont {Robert~J}\ \bibnamefont
  {Harris}}, \bibinfo {author} {\bibfnamefont {Nathan~A}\ \bibnamefont
  {McMahon}}, \ and\ \bibinfo {author} {\bibfnamefont {Thomas~M}\ \bibnamefont
  {Stace}},\ }\bibfield  {title} {\enquote {\bibinfo {title} {Tensor-network
  codes},}\ }\href {\doibase 10.1103/PhysRevLett.127.040507} {\bibfield
  {journal} {\bibinfo  {journal} {Physical Review Letters}\ }\textbf {\bibinfo
  {volume} {127}},\ \bibinfo {pages} {040507} (\bibinfo {year}
  {2021})}\BibitemShut {NoStop}%
\bibitem [{\citenamefont {Eisert}\ \emph {et~al.}(2020)\citenamefont {Eisert},
  \citenamefont {Hangleiter}, \citenamefont {Walk}, \citenamefont {Roth},
  \citenamefont {Markham}, \citenamefont {Parekh}, \citenamefont {Chabaud},\
  and\ \citenamefont {Kashefi}}]{eisert2020quantum}%
  \BibitemOpen
  \bibfield  {author} {\bibinfo {author} {\bibfnamefont {Jens}\ \bibnamefont
  {Eisert}}, \bibinfo {author} {\bibfnamefont {Dominik}\ \bibnamefont
  {Hangleiter}}, \bibinfo {author} {\bibfnamefont {Nathan}\ \bibnamefont
  {Walk}}, \bibinfo {author} {\bibfnamefont {Ingo}\ \bibnamefont {Roth}},
  \bibinfo {author} {\bibfnamefont {Damian}\ \bibnamefont {Markham}}, \bibinfo
  {author} {\bibfnamefont {Rhea}\ \bibnamefont {Parekh}}, \bibinfo {author}
  {\bibfnamefont {Ulysse}\ \bibnamefont {Chabaud}}, \ and\ \bibinfo {author}
  {\bibfnamefont {Elham}\ \bibnamefont {Kashefi}},\ }\bibfield  {title}
  {\enquote {\bibinfo {title} {Quantum certification and benchmarking},}\
  }\href {\doibase 10.1038/s42254-020-0186-4} {\bibfield  {journal} {\bibinfo
  {journal} {Nature Reviews Physics}\ }\textbf {\bibinfo {volume} {2}},\
  \bibinfo {pages} {382--390} (\bibinfo {year} {2020})}\BibitemShut {NoStop}%
\bibitem [{\citenamefont {White}\ \emph {et~al.}(2023)\citenamefont {White},
  \citenamefont {Modi},\ and\ \citenamefont {Hill}}]{white2023filtering}%
  \BibitemOpen
  \bibfield  {author} {\bibinfo {author} {\bibfnamefont {G.~A.~L.}\
  \bibnamefont {White}}, \bibinfo {author} {\bibfnamefont {K.}~\bibnamefont
  {Modi}}, \ and\ \bibinfo {author} {\bibfnamefont {C.~D.}\ \bibnamefont
  {Hill}},\ }\bibfield  {title} {\enquote {\bibinfo {title} {{Filtering
  Crosstalk from Bath Non-Markovianity via Spacetime Classical Shadows}},}\
  }\href {\doibase 10.1103/PhysRevLett.130.160401} {\bibfield  {journal}
  {\bibinfo  {journal} {Phys. Rev. Lett.}\ }\textbf {\bibinfo {volume} {130}},\
  \bibinfo {pages} {160401} (\bibinfo {year} {2023})}\BibitemShut {NoStop}%
\bibitem [{\citenamefont {Fux}\ \emph {et~al.}(2021)\citenamefont {Fux},
  \citenamefont {Butler}, \citenamefont {Eastham}, \citenamefont {Lovett},\
  and\ \citenamefont {Keeling}}]{PhysRevLett.126.200401}%
  \BibitemOpen
  \bibfield  {author} {\bibinfo {author} {\bibfnamefont {Gerald~E.}\
  \bibnamefont {Fux}}, \bibinfo {author} {\bibfnamefont {Eoin~P.}\ \bibnamefont
  {Butler}}, \bibinfo {author} {\bibfnamefont {Paul~R.}\ \bibnamefont
  {Eastham}}, \bibinfo {author} {\bibfnamefont {Brendon~W.}\ \bibnamefont
  {Lovett}}, \ and\ \bibinfo {author} {\bibfnamefont {Jonathan}\ \bibnamefont
  {Keeling}},\ }\bibfield  {title} {\enquote {\bibinfo {title} {{Efficient
  Exploration of Hamiltonian Parameter Space for Optimal Control of
  Non-Markovian Open Quantum Systems}},}\ }\href {\doibase
  10.1103/PhysRevLett.126.200401} {\bibfield  {journal} {\bibinfo  {journal}
  {Physical Review Letters}\ }\textbf {\bibinfo {volume} {126}},\ \bibinfo
  {pages} {200401} (\bibinfo {year} {2021})}\BibitemShut {NoStop}%
\bibitem [{\citenamefont {Butler}\ \emph {et~al.}(2023)\citenamefont {Butler},
  \citenamefont {Fux}, \citenamefont {Ortega-Taberner}, \citenamefont {Lovett},
  \citenamefont {Keeling},\ and\ \citenamefont
  {Eastham}}]{butler2023optimizing}%
  \BibitemOpen
  \bibfield  {author} {\bibinfo {author} {\bibfnamefont {Eoin~P.}\ \bibnamefont
  {Butler}}, \bibinfo {author} {\bibfnamefont {Gerald~E.}\ \bibnamefont {Fux}},
  \bibinfo {author} {\bibfnamefont {Carlos}\ \bibnamefont {Ortega-Taberner}},
  \bibinfo {author} {\bibfnamefont {Brendon~W.}\ \bibnamefont {Lovett}},
  \bibinfo {author} {\bibfnamefont {Jonathan}\ \bibnamefont {Keeling}}, \ and\
  \bibinfo {author} {\bibfnamefont {Paul~R.}\ \bibnamefont {Eastham}},\
  }\href@noop {} {\enquote {\bibinfo {title} {{Optimizing performance of
  quantum operations with non-Markovian decoherence: the tortoise or the
  hare?}}}\ } (\bibinfo {year} {2023}),\ \Eprint
  {http://arxiv.org/abs/2303.16002} {arXiv:2303.16002 [quant-ph]} \BibitemShut
  {NoStop}%
\bibitem [{\citenamefont {Ball}\ \emph {et~al.}(2021)\citenamefont {Ball},
  \citenamefont {Biercuk}, \citenamefont {Carvalho}, \citenamefont {Chen},
  \citenamefont {Hush}, \citenamefont {Castro}, \citenamefont {Li},
  \citenamefont {Liebermann}, \citenamefont {Slatyer}, \citenamefont {Edmunds},
  \citenamefont {Frey}, \citenamefont {Hempel},\ and\ \citenamefont
  {Milne}}]{Ball_2021}%
  \BibitemOpen
  \bibfield  {author} {\bibinfo {author} {\bibfnamefont {Harrison}\
  \bibnamefont {Ball}}, \bibinfo {author} {\bibfnamefont {Michael~J}\
  \bibnamefont {Biercuk}}, \bibinfo {author} {\bibfnamefont {Andre R~R}\
  \bibnamefont {Carvalho}}, \bibinfo {author} {\bibfnamefont {Jiayin}\
  \bibnamefont {Chen}}, \bibinfo {author} {\bibfnamefont {Michael}\
  \bibnamefont {Hush}}, \bibinfo {author} {\bibfnamefont {Leonardo A~De}\
  \bibnamefont {Castro}}, \bibinfo {author} {\bibfnamefont {Li}~\bibnamefont
  {Li}}, \bibinfo {author} {\bibfnamefont {Per~J}\ \bibnamefont {Liebermann}},
  \bibinfo {author} {\bibfnamefont {Harry~J}\ \bibnamefont {Slatyer}}, \bibinfo
  {author} {\bibfnamefont {Claire}\ \bibnamefont {Edmunds}}, \bibinfo {author}
  {\bibfnamefont {Virginia}\ \bibnamefont {Frey}}, \bibinfo {author}
  {\bibfnamefont {Cornelius}\ \bibnamefont {Hempel}}, \ and\ \bibinfo {author}
  {\bibfnamefont {Alistair}\ \bibnamefont {Milne}},\ }\bibfield  {title}
  {\enquote {\bibinfo {title} {Software tools for quantum control: improving
  quantum computer performance through noise and error suppression},}\ }\href
  {\doibase 10.1088/2058-9565/abdca6} {\bibfield  {journal} {\bibinfo
  {journal} {Quantum Science and Technology}\ }\textbf {\bibinfo {volume}
  {6}},\ \bibinfo {pages} {044011} (\bibinfo {year} {2021})}\BibitemShut
  {NoStop}%
\bibitem [{\citenamefont {Chalermpusitarak}\ \emph {et~al.}(2021)\citenamefont
  {Chalermpusitarak}, \citenamefont {Tonekaboni}, \citenamefont {Wang},
  \citenamefont {Norris}, \citenamefont {Viola},\ and\ \citenamefont
  {Paz-Silva}}]{PRXQuantum.2.030315}%
  \BibitemOpen
  \bibfield  {author} {\bibinfo {author} {\bibfnamefont {Teerawat}\
  \bibnamefont {Chalermpusitarak}}, \bibinfo {author} {\bibfnamefont {Behnam}\
  \bibnamefont {Tonekaboni}}, \bibinfo {author} {\bibfnamefont {Yuanlong}\
  \bibnamefont {Wang}}, \bibinfo {author} {\bibfnamefont {Leigh~M.}\
  \bibnamefont {Norris}}, \bibinfo {author} {\bibfnamefont {Lorenza}\
  \bibnamefont {Viola}}, \ and\ \bibinfo {author} {\bibfnamefont {Gerardo~A.}\
  \bibnamefont {Paz-Silva}},\ }\bibfield  {title} {\enquote {\bibinfo {title}
  {Frame-based filter-function formalism for quantum characterization and
  control},}\ }\href {\doibase 10.1103/PRXQuantum.2.030315} {\bibfield
  {journal} {\bibinfo  {journal} {PRX Quantum}\ }\textbf {\bibinfo {volume}
  {2}},\ \bibinfo {pages} {030315} (\bibinfo {year} {2021})}\BibitemShut
  {NoStop}%
\bibitem [{\citenamefont {Suzuki}\ \emph {et~al.}(2022)\citenamefont {Suzuki},
  \citenamefont {Endo}, \citenamefont {Fujii},\ and\ \citenamefont
  {Tokunaga}}]{PRXQuantum.3.010345}%
  \BibitemOpen
  \bibfield  {author} {\bibinfo {author} {\bibfnamefont {Yasunari}\
  \bibnamefont {Suzuki}}, \bibinfo {author} {\bibfnamefont {Suguru}\
  \bibnamefont {Endo}}, \bibinfo {author} {\bibfnamefont {Keisuke}\
  \bibnamefont {Fujii}}, \ and\ \bibinfo {author} {\bibfnamefont {Yuuki}\
  \bibnamefont {Tokunaga}},\ }\bibfield  {title} {\enquote {\bibinfo {title}
  {Quantum error mitigation as a universal error reduction technique:
  Applications from the nisq to the fault-tolerant quantum computing eras},}\
  }\href {\doibase 10.1103/PRXQuantum.3.010345} {\bibfield  {journal} {\bibinfo
   {journal} {PRX Quantum}\ }\textbf {\bibinfo {volume} {3}},\ \bibinfo {pages}
  {010345} (\bibinfo {year} {2022})}\BibitemShut {NoStop}%
\bibitem [{\citenamefont {Wang}\ \emph {et~al.}(2023)\citenamefont {Wang},
  \citenamefont {Liu}, \citenamefont {Liu}, \citenamefont {Gu}, \citenamefont
  {Baker}, \citenamefont {Chong},\ and\ \citenamefont {Han}}]{wang2023dgr}%
  \BibitemOpen
  \bibfield  {author} {\bibinfo {author} {\bibfnamefont {Hanrui}\ \bibnamefont
  {Wang}}, \bibinfo {author} {\bibfnamefont {Pengyu}\ \bibnamefont {Liu}},
  \bibinfo {author} {\bibfnamefont {Yilian}\ \bibnamefont {Liu}}, \bibinfo
  {author} {\bibfnamefont {Jiaqi}\ \bibnamefont {Gu}}, \bibinfo {author}
  {\bibfnamefont {Jonathan}\ \bibnamefont {Baker}}, \bibinfo {author}
  {\bibfnamefont {Frederic~T.}\ \bibnamefont {Chong}}, \ and\ \bibinfo {author}
  {\bibfnamefont {Song}\ \bibnamefont {Han}},\ }\href@noop {} {\enquote
  {\bibinfo {title} {Dgr: Tackling drifted and correlated noise in quantum
  error correction via decoding graph re-weighting},}\ } (\bibinfo {year}
  {2023}),\ \Eprint {http://arxiv.org/abs/2311.16214} {arXiv:2311.16214
  [quant-ph]} \BibitemShut {NoStop}%
\bibitem [{\citenamefont {Wei}\ \emph {et~al.}(2022{\natexlab{a}})\citenamefont
  {Wei}, \citenamefont {Magesan}, \citenamefont {Lauer}, \citenamefont
  {Srinivasan}, \citenamefont {Bogorin}, \citenamefont {Carnevale},
  \citenamefont {Keefe}, \citenamefont {Kim} \emph
  {et~al.}}]{wei2022hamiltonian}%
  \BibitemOpen
  \bibfield  {author} {\bibinfo {author} {\bibfnamefont {K.~X.}\ \bibnamefont
  {Wei}}, \bibinfo {author} {\bibfnamefont {E.}~\bibnamefont {Magesan}},
  \bibinfo {author} {\bibfnamefont {I.}~\bibnamefont {Lauer}}, \bibinfo
  {author} {\bibfnamefont {S.}~\bibnamefont {Srinivasan}}, \bibinfo {author}
  {\bibfnamefont {D.~F.}\ \bibnamefont {Bogorin}}, \bibinfo {author}
  {\bibfnamefont {S.}~\bibnamefont {Carnevale}}, \bibinfo {author}
  {\bibfnamefont {G.~A.}\ \bibnamefont {Keefe}}, \bibinfo {author}
  {\bibfnamefont {Y.}~\bibnamefont {Kim}},  \emph {et~al.},\ }\bibfield
  {title} {\enquote {\bibinfo {title} {Hamiltonian engineering with multicolor
  drives for fast entangling gates and quantum crosstalk cancellation},}\
  }\href {\doibase 10.1103/PhysRevLett.129.060501} {\bibfield  {journal}
  {\bibinfo  {journal} {Phys. Rev. Lett.}\ }\textbf {\bibinfo {volume} {129}},\
  \bibinfo {pages} {060501} (\bibinfo {year} {2022}{\natexlab{a}})}\BibitemShut
  {NoStop}%
\bibitem [{\citenamefont {M{\"u}ller}\ \emph {et~al.}(2019)\citenamefont
  {M{\"u}ller}, \citenamefont {Cole},\ and\ \citenamefont
  {Lisenfeld}}]{muller2019towards}%
  \BibitemOpen
  \bibfield  {author} {\bibinfo {author} {\bibfnamefont {Clemens}\ \bibnamefont
  {M{\"u}ller}}, \bibinfo {author} {\bibfnamefont {Jared~H}\ \bibnamefont
  {Cole}}, \ and\ \bibinfo {author} {\bibfnamefont {J{\"u}rgen}\ \bibnamefont
  {Lisenfeld}},\ }\bibfield  {title} {\enquote {\bibinfo {title} {Towards
  understanding two-level-systems in amorphous solids: insights from quantum
  circuits},}\ }\href {\doibase https://doi.org/10.1088/1361-6633/ab3a7e}
  {\bibfield  {journal} {\bibinfo  {journal} {Reports on Progress in Physics}\
  }\textbf {\bibinfo {volume} {82}},\ \bibinfo {pages} {124501} (\bibinfo
  {year} {2019})}\BibitemShut {NoStop}%
\bibitem [{\citenamefont {Wilen}\ \emph {et~al.}(2021)\citenamefont {Wilen},
  \citenamefont {Abdullah}, \citenamefont {Kurinsky}, \citenamefont {Stanford},
  \citenamefont {Cardani}, \citenamefont {d’Imperio}, \citenamefont {Tomei},
  \citenamefont {Faoro}, \citenamefont {Ioffe}, \citenamefont {Liu} \emph
  {et~al.}}]{wilen2021correlated}%
  \BibitemOpen
  \bibfield  {author} {\bibinfo {author} {\bibfnamefont {Christopher~D}\
  \bibnamefont {Wilen}}, \bibinfo {author} {\bibfnamefont {S}~\bibnamefont
  {Abdullah}}, \bibinfo {author} {\bibfnamefont {NA}~\bibnamefont {Kurinsky}},
  \bibinfo {author} {\bibfnamefont {C}~\bibnamefont {Stanford}}, \bibinfo
  {author} {\bibfnamefont {L}~\bibnamefont {Cardani}}, \bibinfo {author}
  {\bibfnamefont {G}~\bibnamefont {d’Imperio}}, \bibinfo {author}
  {\bibfnamefont {C}~\bibnamefont {Tomei}}, \bibinfo {author} {\bibfnamefont
  {L}~\bibnamefont {Faoro}}, \bibinfo {author} {\bibfnamefont {LB}~\bibnamefont
  {Ioffe}}, \bibinfo {author} {\bibfnamefont {CH}~\bibnamefont {Liu}},  \emph
  {et~al.},\ }\bibfield  {title} {\enquote {\bibinfo {title} {Correlated charge
  noise and relaxation errors in superconducting qubits},}\ }\href {\doibase
  https://doi.org/10.1038/s41586-021-03557-5} {\bibfield  {journal} {\bibinfo
  {journal} {Nature}\ }\textbf {\bibinfo {volume} {594}},\ \bibinfo {pages}
  {369--373} (\bibinfo {year} {2021})}\BibitemShut {NoStop}%
\bibitem [{\citenamefont {Dial}\ \emph {et~al.}(2016)\citenamefont {Dial},
  \citenamefont {McClure}, \citenamefont {Poletto}, \citenamefont {Keefe},
  \citenamefont {Rothwell}, \citenamefont {Gambetta}, \citenamefont {Abraham},
  \citenamefont {Chow},\ and\ \citenamefont {Steffen}}]{Dial_2016}%
  \BibitemOpen
  \bibfield  {author} {\bibinfo {author} {\bibfnamefont {Oliver}\ \bibnamefont
  {Dial}}, \bibinfo {author} {\bibfnamefont {Douglas~T}\ \bibnamefont
  {McClure}}, \bibinfo {author} {\bibfnamefont {Stefano}\ \bibnamefont
  {Poletto}}, \bibinfo {author} {\bibfnamefont {G~A}\ \bibnamefont {Keefe}},
  \bibinfo {author} {\bibfnamefont {Mary~Beth}\ \bibnamefont {Rothwell}},
  \bibinfo {author} {\bibfnamefont {Jay~M}\ \bibnamefont {Gambetta}}, \bibinfo
  {author} {\bibfnamefont {David~W}\ \bibnamefont {Abraham}}, \bibinfo {author}
  {\bibfnamefont {Jerry~M}\ \bibnamefont {Chow}}, \ and\ \bibinfo {author}
  {\bibfnamefont {Matthias}\ \bibnamefont {Steffen}},\ }\bibfield  {title}
  {\enquote {\bibinfo {title} {Bulk and surface loss in superconducting
  transmon qubits},}\ }\href {\doibase 10.1088/0953-2048/29/4/044001}
  {\bibfield  {journal} {\bibinfo  {journal} {Superconductor Science and
  Technology}\ }\textbf {\bibinfo {volume} {29}},\ \bibinfo {pages} {044001}
  (\bibinfo {year} {2016})}\BibitemShut {NoStop}%
\bibitem [{\citenamefont {Brownnutt}\ \emph {et~al.}(2015)\citenamefont
  {Brownnutt}, \citenamefont {Kumph}, \citenamefont {Rabl},\ and\ \citenamefont
  {Blatt}}]{RevModPhys.87.1419}%
  \BibitemOpen
  \bibfield  {author} {\bibinfo {author} {\bibfnamefont {M.}~\bibnamefont
  {Brownnutt}}, \bibinfo {author} {\bibfnamefont {M.}~\bibnamefont {Kumph}},
  \bibinfo {author} {\bibfnamefont {P.}~\bibnamefont {Rabl}}, \ and\ \bibinfo
  {author} {\bibfnamefont {R.}~\bibnamefont {Blatt}},\ }\bibfield  {title}
  {\enquote {\bibinfo {title} {Ion-trap measurements of electric-field noise
  near surfaces},}\ }\href {\doibase 10.1103/RevModPhys.87.1419} {\bibfield
  {journal} {\bibinfo  {journal} {Rev. Mod. Phys.}\ }\textbf {\bibinfo {volume}
  {87}},\ \bibinfo {pages} {1419--1482} (\bibinfo {year} {2015})}\BibitemShut
  {NoStop}%
\bibitem [{\citenamefont {Parrado-Rodr{\'{i}}guez}\ \emph
  {et~al.}(2021)\citenamefont {Parrado-Rodr{\'{i}}guez}, \citenamefont
  {Ryan-Anderson}, \citenamefont {Bermudez},\ and\ \citenamefont
  {M{\"{u}}ller}}]{ParradoRodriguez2021crosstalk}%
  \BibitemOpen
  \bibfield  {author} {\bibinfo {author} {\bibfnamefont {Pedro}\ \bibnamefont
  {Parrado-Rodr{\'{i}}guez}}, \bibinfo {author} {\bibfnamefont {Ciar{\'{a}}n}\
  \bibnamefont {Ryan-Anderson}}, \bibinfo {author} {\bibfnamefont {Alejandro}\
  \bibnamefont {Bermudez}}, \ and\ \bibinfo {author} {\bibfnamefont {Markus}\
  \bibnamefont {M{\"{u}}ller}},\ }\bibfield  {title} {\enquote {\bibinfo
  {title} {Crosstalk {S}uppression for {F}ault-tolerant {Q}uantum {E}rror
  {C}orrection with {T}rapped {I}ons},}\ }\href {\doibase
  10.22331/q-2021-06-29-487} {\bibfield  {journal} {\bibinfo  {journal}
  {{Quantum}}\ }\textbf {\bibinfo {volume} {5}},\ \bibinfo {pages} {487}
  (\bibinfo {year} {2021})}\BibitemShut {NoStop}%
\bibitem [{\citenamefont {Kuhlmann}\ \emph {et~al.}(2013)\citenamefont
  {Kuhlmann}, \citenamefont {Houel}, \citenamefont {Ludwig}, \citenamefont
  {Greuter}, \citenamefont {Reuter}, \citenamefont {Wieck}, \citenamefont
  {Poggio},\ and\ \citenamefont {Warburton}}]{kuhlmann2013charge}%
  \BibitemOpen
  \bibfield  {author} {\bibinfo {author} {\bibfnamefont {Andreas~V}\
  \bibnamefont {Kuhlmann}}, \bibinfo {author} {\bibfnamefont {Julien}\
  \bibnamefont {Houel}}, \bibinfo {author} {\bibfnamefont {Arne}\ \bibnamefont
  {Ludwig}}, \bibinfo {author} {\bibfnamefont {Lukas}\ \bibnamefont {Greuter}},
  \bibinfo {author} {\bibfnamefont {Dirk}\ \bibnamefont {Reuter}}, \bibinfo
  {author} {\bibfnamefont {Andreas~D}\ \bibnamefont {Wieck}}, \bibinfo {author}
  {\bibfnamefont {Martino}\ \bibnamefont {Poggio}}, \ and\ \bibinfo {author}
  {\bibfnamefont {Richard~J}\ \bibnamefont {Warburton}},\ }\bibfield  {title}
  {\enquote {\bibinfo {title} {Charge noise and spin noise in a semiconductor
  quantum device},}\ }\href {\doibase https://doi.org/10.1038/nphys2688}
  {\bibfield  {journal} {\bibinfo  {journal} {Nature Physics}\ }\textbf
  {\bibinfo {volume} {9}},\ \bibinfo {pages} {570--575} (\bibinfo {year}
  {2013})}\BibitemShut {NoStop}%
\bibitem [{\citenamefont {Yoneda}\ \emph {et~al.}(2022)\citenamefont {Yoneda},
  \citenamefont {Rojas-Arias}, \citenamefont {Stano}, \citenamefont {Takeda},
  \citenamefont {Noiri}, \citenamefont {Nakajima}, \citenamefont {Loss},\ and\
  \citenamefont {Tarucha}}]{yoneda2022noise}%
  \BibitemOpen
  \bibfield  {author} {\bibinfo {author} {\bibfnamefont {J.}~\bibnamefont
  {Yoneda}}, \bibinfo {author} {\bibfnamefont {J.~S.}\ \bibnamefont
  {Rojas-Arias}}, \bibinfo {author} {\bibfnamefont {P.}~\bibnamefont {Stano}},
  \bibinfo {author} {\bibfnamefont {K.}~\bibnamefont {Takeda}}, \bibinfo
  {author} {\bibfnamefont {A.}~\bibnamefont {Noiri}}, \bibinfo {author}
  {\bibfnamefont {T.}~\bibnamefont {Nakajima}}, \bibinfo {author}
  {\bibfnamefont {D.}~\bibnamefont {Loss}}, \ and\ \bibinfo {author}
  {\bibfnamefont {S.}~\bibnamefont {Tarucha}},\ }\bibfield  {title} {\enquote
  {\bibinfo {title} {Noise-correlation spectrum for a pair of spin qubits in
  silicon},}\ }\href {\doibase https://doi.org/10.1038/s41567-023-02238-6}
  {\bibfield  {journal} {\bibinfo  {journal} {Nature Physics}\ }\textbf
  {\bibinfo {volume} {19}},\ \bibinfo {pages} {1793} (\bibinfo {year}
  {2022})},\ \Eprint {http://arxiv.org/abs/2208.14150} {arXiv:2208.14150
  [quant-ph]} \BibitemShut {NoStop}%
\bibitem [{\citenamefont {Rojas-Arias}\ \emph {et~al.}(2023)\citenamefont
  {Rojas-Arias}, \citenamefont {Noiri}, \citenamefont {Stano}, \citenamefont
  {Nakajima}, \citenamefont {Yoneda} \emph {et~al.}}]{rojas2023spatial}%
  \BibitemOpen
  \bibfield  {author} {\bibinfo {author} {\bibfnamefont {Juan~S.}\ \bibnamefont
  {Rojas-Arias}}, \bibinfo {author} {\bibfnamefont {Akito}\ \bibnamefont
  {Noiri}}, \bibinfo {author} {\bibfnamefont {Peter}\ \bibnamefont {Stano}},
  \bibinfo {author} {\bibfnamefont {Takashi}\ \bibnamefont {Nakajima}},
  \bibinfo {author} {\bibfnamefont {Jun}\ \bibnamefont {Yoneda}},  \emph
  {et~al.},\ }\bibfield  {title} {\enquote {\bibinfo {title} {{Spatial noise
  correlations beyond nearest-neighbor in ${}^{28}$Si/SiGe spin qubits}},}\
  }\href {\doibase https://doi.org/10.1103/PhysRevApplied.20.054024} {\bibfield
   {journal} {\bibinfo  {journal} {Phys. Rev. Applied}\ }\textbf {\bibinfo
  {volume} {20}},\ \bibinfo {pages} {054024} (\bibinfo {year} {2023})},\
  \Eprint {http://arxiv.org/abs/2302.11717} {arXiv:2302.11717
  [cond-mat.mes-hall]} \BibitemShut {NoStop}%
\bibitem [{\citenamefont {Wu}\ \emph {et~al.}(2021)\citenamefont {Wu},
  \citenamefont {Liang}, \citenamefont {Tian}, \citenamefont {Yang},
  \citenamefont {Chen}, \citenamefont {Liu}, \citenamefont {Tey},\ and\
  \citenamefont {You}}]{wu2021concise}%
  \BibitemOpen
  \bibfield  {author} {\bibinfo {author} {\bibfnamefont {Xiaoling}\
  \bibnamefont {Wu}}, \bibinfo {author} {\bibfnamefont {Xinhui}\ \bibnamefont
  {Liang}}, \bibinfo {author} {\bibfnamefont {Yaoqi}\ \bibnamefont {Tian}},
  \bibinfo {author} {\bibfnamefont {Fan}\ \bibnamefont {Yang}}, \bibinfo
  {author} {\bibfnamefont {Cheng}\ \bibnamefont {Chen}}, \bibinfo {author}
  {\bibfnamefont {Yong-Chun}\ \bibnamefont {Liu}}, \bibinfo {author}
  {\bibfnamefont {Meng~Khoon}\ \bibnamefont {Tey}}, \ and\ \bibinfo {author}
  {\bibfnamefont {Li}~\bibnamefont {You}},\ }\bibfield  {title} {\enquote
  {\bibinfo {title} {A concise review of {Rydberg} atom based quantum
  computation and quantum simulation},}\ }\href {\doibase
  10.1088/1674-1056/abd76f} {\bibfield  {journal} {\bibinfo  {journal} {Chinese
  Physics B}\ }\textbf {\bibinfo {volume} {30}},\ \bibinfo {pages} {020305}
  (\bibinfo {year} {2021})}\BibitemShut {NoStop}%
\bibitem [{\citenamefont {Terhal}\ \emph {et~al.}(2020)\citenamefont {Terhal},
  \citenamefont {Conrad},\ and\ \citenamefont {Vuillot}}]{Terhal_2020}%
  \BibitemOpen
  \bibfield  {author} {\bibinfo {author} {\bibfnamefont {B~M}\ \bibnamefont
  {Terhal}}, \bibinfo {author} {\bibfnamefont {J}~\bibnamefont {Conrad}}, \
  and\ \bibinfo {author} {\bibfnamefont {C}~\bibnamefont {Vuillot}},\
  }\bibfield  {title} {\enquote {\bibinfo {title} {Towards scalable bosonic
  quantum error correction},}\ }\href {\doibase 10.1088/2058-9565/ab98a5}
  {\bibfield  {journal} {\bibinfo  {journal} {Quantum Science and Technology}\
  }\textbf {\bibinfo {volume} {5}},\ \bibinfo {pages} {043001} (\bibinfo {year}
  {2020})}\BibitemShut {NoStop}%
\bibitem [{\citenamefont {Lai}\ \emph {et~al.}(2018)\citenamefont {Lai},
  \citenamefont {Yang}, \citenamefont {Huang},\ and\ \citenamefont
  {Zhang}}]{PhysRevB.97.054508}%
  \BibitemOpen
  \bibfield  {author} {\bibinfo {author} {\bibfnamefont {Hon-Lam}\ \bibnamefont
  {Lai}}, \bibinfo {author} {\bibfnamefont {Pei-Yun}\ \bibnamefont {Yang}},
  \bibinfo {author} {\bibfnamefont {Yu-Wei}\ \bibnamefont {Huang}}, \ and\
  \bibinfo {author} {\bibfnamefont {Wei-Min}\ \bibnamefont {Zhang}},\
  }\bibfield  {title} {\enquote {\bibinfo {title} {{Exact master equation and
  non-Markovian decoherence dynamics of Majorana zero modes under gate-induced
  charge fluctuations}},}\ }\href {\doibase 10.1103/PhysRevB.97.054508}
  {\bibfield  {journal} {\bibinfo  {journal} {Phys. Rev. B}\ }\textbf {\bibinfo
  {volume} {97}},\ \bibinfo {pages} {054508} (\bibinfo {year}
  {2018})}\BibitemShut {NoStop}%
\bibitem [{\citenamefont {Wolf}\ \emph {et~al.}(2008)\citenamefont {Wolf},
  \citenamefont {Eisert}, \citenamefont {Cubitt},\ and\ \citenamefont
  {Cirac}}]{PhysRevLett.101.150402}%
  \BibitemOpen
  \bibfield  {author} {\bibinfo {author} {\bibfnamefont {M.~M.}\ \bibnamefont
  {Wolf}}, \bibinfo {author} {\bibfnamefont {J.}~\bibnamefont {Eisert}},
  \bibinfo {author} {\bibfnamefont {T.~S.}\ \bibnamefont {Cubitt}}, \ and\
  \bibinfo {author} {\bibfnamefont {J.~I.}\ \bibnamefont {Cirac}},\ }\bibfield
  {title} {\enquote {\bibinfo {title} {{Assessing Non-Markovian Quantum
  Dynamics}},}\ }\href {\doibase 10.1103/PhysRevLett.101.150402} {\bibfield
  {journal} {\bibinfo  {journal} {Phys. Rev. Lett.}\ }\textbf {\bibinfo
  {volume} {101}},\ \bibinfo {pages} {150402} (\bibinfo {year}
  {2008})}\BibitemShut {NoStop}%
\bibitem [{\citenamefont {Ángel Rivas}\ \emph {et~al.}(2014)\citenamefont
  {Ángel Rivas}, \citenamefont {Huelga},\ and\ \citenamefont
  {Plenio}}]{Rivas_2014}%
  \BibitemOpen
  \bibfield  {author} {\bibinfo {author} {\bibnamefont {Ángel Rivas}},
  \bibinfo {author} {\bibfnamefont {Susana~F}\ \bibnamefont {Huelga}}, \ and\
  \bibinfo {author} {\bibfnamefont {Martin~B}\ \bibnamefont {Plenio}},\
  }\bibfield  {title} {\enquote {\bibinfo {title} {Quantum non-markovianity:
  characterization, quantification and detection},}\ }\href {\doibase
  10.1088/0034-4885/77/9/094001} {\bibfield  {journal} {\bibinfo  {journal}
  {Reports on Progress in Physics}\ }\textbf {\bibinfo {volume} {77}},\
  \bibinfo {pages} {094001} (\bibinfo {year} {2014})}\BibitemShut {NoStop}%
\bibitem [{\citenamefont {Berk}\ \emph {et~al.}(2023)\citenamefont {Berk},
  \citenamefont {Milz}, \citenamefont {Pollock},\ and\ \citenamefont
  {Modi}}]{berk2021extracting}%
  \BibitemOpen
  \bibfield  {author} {\bibinfo {author} {\bibfnamefont {Graeme~D}\
  \bibnamefont {Berk}}, \bibinfo {author} {\bibfnamefont {Simon}\ \bibnamefont
  {Milz}}, \bibinfo {author} {\bibfnamefont {Felix~A}\ \bibnamefont {Pollock}},
  \ and\ \bibinfo {author} {\bibfnamefont {Kavan}\ \bibnamefont {Modi}},\
  }\bibfield  {title} {\enquote {\bibinfo {title} {{Extracting quantum
  dynamical resources: Consumption of non-Markovianity for noise reduction}},}\
  }\href {\doibase https://doi.org/10.1038/s41534-023-00774-w} {\bibfield
  {journal} {\bibinfo  {journal} {npj Quantum Information}\ }\textbf {\bibinfo
  {volume} {9}},\ \bibinfo {pages} {104} (\bibinfo {year} {2023})},\ \Eprint
  {http://arxiv.org/abs/2110.02613} {arXiv:2110.02613} \BibitemShut {NoStop}%
\bibitem [{\citenamefont {Chubb}\ and\ \citenamefont
  {Flammia}(2021)}]{chubb2021statistical}%
  \BibitemOpen
  \bibfield  {author} {\bibinfo {author} {\bibfnamefont {Christopher~T}\
  \bibnamefont {Chubb}}\ and\ \bibinfo {author} {\bibfnamefont {Steven~T}\
  \bibnamefont {Flammia}},\ }\bibfield  {title} {\enquote {\bibinfo {title}
  {Statistical mechanical models for quantum codes with correlated noise},}\
  }\href {\doibase https://doi.org/10.4171/AIHPD/105} {\bibfield  {journal}
  {\bibinfo  {journal} {Annales de l’Institut Henri Poincar{\'e} D}\ }\textbf
  {\bibinfo {volume} {8}},\ \bibinfo {pages} {269--321} (\bibinfo {year}
  {2021})}\BibitemShut {NoStop}%
\bibitem [{\citenamefont {Mauron}\ \emph {et~al.}(2023)\citenamefont {Mauron},
  \citenamefont {Farrelly},\ and\ \citenamefont
  {Stace}}]{mauron2023optimization}%
  \BibitemOpen
  \bibfield  {author} {\bibinfo {author} {\bibfnamefont {Caroline}\
  \bibnamefont {Mauron}}, \bibinfo {author} {\bibfnamefont {Terry}\
  \bibnamefont {Farrelly}}, \ and\ \bibinfo {author} {\bibfnamefont
  {Thomas~M.}\ \bibnamefont {Stace}},\ }\href@noop {} {\enquote {\bibinfo
  {title} {Optimization of tensor network codes with reinforcement learning},}\
  } (\bibinfo {year} {2023}),\ \Eprint {http://arxiv.org/abs/2305.11470}
  {arXiv:2305.11470 [quant-ph]} \BibitemShut {NoStop}%
\bibitem [{\citenamefont {Su}\ \emph {et~al.}(2023{\natexlab{a}})\citenamefont
  {Su}, \citenamefont {Cao}, \citenamefont {Hu}, \citenamefont {Yanay},
  \citenamefont {Tahan},\ and\ \citenamefont {Swingle}}]{su2023discovery}%
  \BibitemOpen
  \bibfield  {author} {\bibinfo {author} {\bibfnamefont {Vincent~Paul}\
  \bibnamefont {Su}}, \bibinfo {author} {\bibfnamefont {ChunJun}\ \bibnamefont
  {Cao}}, \bibinfo {author} {\bibfnamefont {Hong-Ye}\ \bibnamefont {Hu}},
  \bibinfo {author} {\bibfnamefont {Yariv}\ \bibnamefont {Yanay}}, \bibinfo
  {author} {\bibfnamefont {Charles}\ \bibnamefont {Tahan}}, \ and\ \bibinfo
  {author} {\bibfnamefont {Brian}\ \bibnamefont {Swingle}},\ }\href@noop {}
  {\enquote {\bibinfo {title} {Discovery of optimal quantum error correcting
  codes via reinforcement learning},}\ } (\bibinfo {year}
  {2023}{\natexlab{a}}),\ \Eprint {http://arxiv.org/abs/2305.06378}
  {arXiv:2305.06378 [quant-ph]} \BibitemShut {NoStop}%
\bibitem [{\citenamefont {Cai}\ \emph {et~al.}(2023)\citenamefont {Cai},
  \citenamefont {Babbush}, \citenamefont {Benjamin}, \citenamefont {Endo},
  \citenamefont {Huggins}, \citenamefont {Li}, \citenamefont {McClean},\ and\
  \citenamefont {O'Brien}}]{cai2023quantum}%
  \BibitemOpen
  \bibfield  {author} {\bibinfo {author} {\bibfnamefont {Zhenyu}\ \bibnamefont
  {Cai}}, \bibinfo {author} {\bibfnamefont {Ryan}\ \bibnamefont {Babbush}},
  \bibinfo {author} {\bibfnamefont {Simon~C.}\ \bibnamefont {Benjamin}},
  \bibinfo {author} {\bibfnamefont {Suguru}\ \bibnamefont {Endo}}, \bibinfo
  {author} {\bibfnamefont {William~J.}\ \bibnamefont {Huggins}}, \bibinfo
  {author} {\bibfnamefont {Ying}\ \bibnamefont {Li}}, \bibinfo {author}
  {\bibfnamefont {Jarrod~R.}\ \bibnamefont {McClean}}, \ and\ \bibinfo {author}
  {\bibfnamefont {Thomas~E.}\ \bibnamefont {O'Brien}},\ }\href@noop {}
  {\enquote {\bibinfo {title} {Quantum error mitigation},}\ } (\bibinfo {year}
  {2023}),\ \Eprint {http://arxiv.org/abs/2210.00921} {arXiv:2210.00921
  [quant-ph]} \BibitemShut {NoStop}%
\bibitem [{\citenamefont {White}\ \emph {et~al.}(2020)\citenamefont {White},
  \citenamefont {Hill}, \citenamefont {Pollock}, \citenamefont {Hollenberg},\
  and\ \citenamefont {Modi}}]{White-NM-2020}%
  \BibitemOpen
  \bibfield  {author} {\bibinfo {author} {\bibfnamefont {G.~A.~L.}\
  \bibnamefont {White}}, \bibinfo {author} {\bibfnamefont {C.~D.}\ \bibnamefont
  {Hill}}, \bibinfo {author} {\bibfnamefont {F.~A.}\ \bibnamefont {Pollock}},
  \bibinfo {author} {\bibfnamefont {L.~C.~L.}\ \bibnamefont {Hollenberg}}, \
  and\ \bibinfo {author} {\bibfnamefont {K.}~\bibnamefont {Modi}},\ }\bibfield
  {title} {\enquote {\bibinfo {title} {{Demonstration of non-Markovian process
  characterisation and control on a quantum processor}},}\ }\href {\doibase
  10.1038/s41467-020-20113-3} {\bibfield  {journal} {\bibinfo  {journal}
  {Nature Communications}\ }\textbf {\bibinfo {volume} {11}},\ \bibinfo {pages}
  {6301} (\bibinfo {year} {2020})},\ \Eprint {http://arxiv.org/abs/2004.14018}
  {arXiv:2004.14018} \BibitemShut {NoStop}%
\bibitem [{\citenamefont {White}\ \emph {et~al.}(2022)\citenamefont {White},
  \citenamefont {Pollock}, \citenamefont {Hollenberg}, \citenamefont {Modi},\
  and\ \citenamefont {Hill}}]{White-MLPT}%
  \BibitemOpen
  \bibfield  {author} {\bibinfo {author} {\bibfnamefont {Gregory A~L}\
  \bibnamefont {White}}, \bibinfo {author} {\bibfnamefont {Felix~A}\
  \bibnamefont {Pollock}}, \bibinfo {author} {\bibfnamefont {Lloyd C~L}\
  \bibnamefont {Hollenberg}}, \bibinfo {author} {\bibfnamefont {Kavan}\
  \bibnamefont {Modi}}, \ and\ \bibinfo {author} {\bibfnamefont {Charles~D}\
  \bibnamefont {Hill}},\ }\bibfield  {title} {\enquote {\bibinfo {title}
  {{Non-Markovian quantum process tomography}},}\ }\href {\doibase
  https://doi.org/10.1103/PRXQuantum.3.020344} {\bibfield  {journal} {\bibinfo
  {journal} {PRX Quantum}\ }\textbf {\bibinfo {volume} {3}},\ \bibinfo {pages}
  {020344} (\bibinfo {year} {2022})}\BibitemShut {NoStop}%
\bibitem [{\citenamefont {White}\ \emph
  {et~al.}(2021{\natexlab{a}})\citenamefont {White}, \citenamefont {Pollock},
  \citenamefont {Hollenberg}, \citenamefont {Hill},\ and\ \citenamefont
  {Modi}}]{white2021many}%
  \BibitemOpen
  \bibfield  {author} {\bibinfo {author} {\bibfnamefont {Gregory A~L}\
  \bibnamefont {White}}, \bibinfo {author} {\bibfnamefont {Felix~A}\
  \bibnamefont {Pollock}}, \bibinfo {author} {\bibfnamefont {Lloyd C~L}\
  \bibnamefont {Hollenberg}}, \bibinfo {author} {\bibfnamefont {Charles~D}\
  \bibnamefont {Hill}}, \ and\ \bibinfo {author} {\bibfnamefont {Kavan}\
  \bibnamefont {Modi}},\ }\bibfield  {title} {\enquote {\bibinfo {title} {From
  many-body to many-time physics},}\ }\href {http://arxiv.org/abs/2107.13934}
  {\bibfield  {journal} {\bibinfo  {journal} {arXiv:2107.13934}\ } (\bibinfo
  {year} {2021}{\natexlab{a}})}\BibitemShut {NoStop}%
\bibitem [{\citenamefont {Giarmatzi}\ \emph {et~al.}(2023)\citenamefont
  {Giarmatzi}, \citenamefont {Jones}, \citenamefont {Gilchrist}, \citenamefont
  {Pakkiam}, \citenamefont {Fedorov},\ and\ \citenamefont
  {Costa}}]{giarmatzi2023multitime}%
  \BibitemOpen
  \bibfield  {author} {\bibinfo {author} {\bibfnamefont {Christina}\
  \bibnamefont {Giarmatzi}}, \bibinfo {author} {\bibfnamefont {Tyler}\
  \bibnamefont {Jones}}, \bibinfo {author} {\bibfnamefont {Alexei}\
  \bibnamefont {Gilchrist}}, \bibinfo {author} {\bibfnamefont {Prasanna}\
  \bibnamefont {Pakkiam}}, \bibinfo {author} {\bibfnamefont {Arkady}\
  \bibnamefont {Fedorov}}, \ and\ \bibinfo {author} {\bibfnamefont {Fabio}\
  \bibnamefont {Costa}},\ }\href@noop {} {\enquote {\bibinfo {title}
  {Multi-time quantum process tomography of a superconducting qubit},}\ }
  (\bibinfo {year} {2023}),\ \Eprint {http://arxiv.org/abs/2308.00750}
  {arXiv:2308.00750 [quant-ph]} \BibitemShut {NoStop}%
\bibitem [{\citenamefont {Guo}\ \emph {et~al.}(2020)\citenamefont {Guo},
  \citenamefont {Modi},\ and\ \citenamefont {Poletti}}]{PhysRevA.102.062414}%
  \BibitemOpen
  \bibfield  {author} {\bibinfo {author} {\bibfnamefont {Chu}\ \bibnamefont
  {Guo}}, \bibinfo {author} {\bibfnamefont {Kavan}\ \bibnamefont {Modi}}, \
  and\ \bibinfo {author} {\bibfnamefont {Dario}\ \bibnamefont {Poletti}},\
  }\bibfield  {title} {\enquote {\bibinfo {title} {{Tensor-network-based
  machine learning of non-Markovian quantum processes}},}\ }\href {\doibase
  10.1103/PhysRevA.102.062414} {\bibfield  {journal} {\bibinfo  {journal}
  {Phys. Rev. A}\ }\textbf {\bibinfo {volume} {102}},\ \bibinfo {pages}
  {062414} (\bibinfo {year} {2020})}\BibitemShut {NoStop}%
\bibitem [{\citenamefont {Guo}(2022)}]{PhysRevA.106.022411}%
  \BibitemOpen
  \bibfield  {author} {\bibinfo {author} {\bibfnamefont {Chu}\ \bibnamefont
  {Guo}},\ }\bibfield  {title} {\enquote {\bibinfo {title} {{Reconstructing
  non-Markovian open quantum evolution from multiple time measurements}},}\
  }\href {\doibase 10.1103/PhysRevA.106.022411} {\bibfield  {journal} {\bibinfo
   {journal} {Phys. Rev. A}\ }\textbf {\bibinfo {volume} {106}},\ \bibinfo
  {pages} {022411} (\bibinfo {year} {2022})}\BibitemShut {NoStop}%
\bibitem [{\citenamefont {Li}\ \emph {et~al.}(2023)\citenamefont {Li},
  \citenamefont {Zheng}, \citenamefont {Meng}, \citenamefont {Zhang},\ and\
  \citenamefont {Yu}}]{Li_NMGST}%
  \BibitemOpen
  \bibfield  {author} {\bibinfo {author} {\bibfnamefont {Ze-Tong}\ \bibnamefont
  {Li}}, \bibinfo {author} {\bibfnamefont {Cong-Cong}\ \bibnamefont {Zheng}},
  \bibinfo {author} {\bibfnamefont {Fan-Xu}\ \bibnamefont {Meng}}, \bibinfo
  {author} {\bibfnamefont {Zai-Chen}\ \bibnamefont {Zhang}}, \ and\ \bibinfo
  {author} {\bibfnamefont {Xu-Tao}\ \bibnamefont {Yu}},\ }\href@noop {}
  {\enquote {\bibinfo {title} {{Non-Markovian Quantum Gate Set Tomography}},}\
  } (\bibinfo {year} {2023}),\ \Eprint {http://arxiv.org/abs/2307.14696}
  {arXiv:2307.14696 [quant-ph]} \BibitemShut {NoStop}%
\bibitem [{\citenamefont {Pollock}\ \emph
  {et~al.}(2018{\natexlab{a}})\citenamefont {Pollock}, \citenamefont
  {Rodr{\'{i}}guez-Rosario}, \citenamefont {Frauenheim}, \citenamefont
  {Paternostro},\ and\ \citenamefont {Modi}}]{Pollock2018a}%
  \BibitemOpen
  \bibfield  {author} {\bibinfo {author} {\bibfnamefont {Felix~A.}\
  \bibnamefont {Pollock}}, \bibinfo {author} {\bibfnamefont {C{\'{e}}sar}\
  \bibnamefont {Rodr{\'{i}}guez-Rosario}}, \bibinfo {author} {\bibfnamefont
  {Thomas}\ \bibnamefont {Frauenheim}}, \bibinfo {author} {\bibfnamefont
  {Mauro}\ \bibnamefont {Paternostro}}, \ and\ \bibinfo {author} {\bibfnamefont
  {Kavan}\ \bibnamefont {Modi}},\ }\bibfield  {title} {\enquote {\bibinfo
  {title} {{Non-Markovian quantum processes: Complete framework and efficient
  characterization}},}\ }\href {\doibase 10.1103/PhysRevA.97.012127} {\bibfield
   {journal} {\bibinfo  {journal} {Physical Review A}\ }\textbf {\bibinfo
  {volume} {97}},\ \bibinfo {pages} {012127} (\bibinfo {year}
  {2018}{\natexlab{a}})},\ \Eprint {http://arxiv.org/abs/1512.00589}
  {arXiv:1512.00589} \BibitemShut {NoStop}%
\bibitem [{\citenamefont {Milz}\ and\ \citenamefont
  {Modi}(2021)}]{Milz2021PRXQ}%
  \BibitemOpen
  \bibfield  {author} {\bibinfo {author} {\bibfnamefont {Simon}\ \bibnamefont
  {Milz}}\ and\ \bibinfo {author} {\bibfnamefont {Kavan}\ \bibnamefont
  {Modi}},\ }\bibfield  {title} {\enquote {\bibinfo {title} {{Quantum
  Stochastic Processes and Quantum non-Markovian Phenomena}},}\ }\href
  {\doibase 10.1103/PRXQuantum.2.030201} {\bibfield  {journal} {\bibinfo
  {journal} {PRX Quantum}\ }\textbf {\bibinfo {volume} {2}},\ \bibinfo {pages}
  {030201} (\bibinfo {year} {2021})},\ \Eprint
  {http://arxiv.org/abs/2012.01894} {arXiv:2012.01894} \BibitemShut {NoStop}%
\bibitem [{\citenamefont {Chiribella}\ \emph {et~al.}(2009)\citenamefont
  {Chiribella}, \citenamefont {D'Ariano},\ and\ \citenamefont
  {Perinotti}}]{Chiribella2009TheoreticalNetworks}%
  \BibitemOpen
  \bibfield  {author} {\bibinfo {author} {\bibfnamefont {Giulio}\ \bibnamefont
  {Chiribella}}, \bibinfo {author} {\bibfnamefont {Giacomo~Mauro}\ \bibnamefont
  {D'Ariano}}, \ and\ \bibinfo {author} {\bibfnamefont {Paolo}\ \bibnamefont
  {Perinotti}},\ }\bibfield  {title} {\enquote {\bibinfo {title} {{Theoretical
  framework for quantum networks}},}\ }\href {\doibase
  10.1103/PhysRevA.80.022339} {\bibfield  {journal} {\bibinfo  {journal}
  {Physical Review A - Atomic, Molecular, and Optical Physics}\ }\textbf
  {\bibinfo {volume} {80}},\ \bibinfo {pages} {1--20} (\bibinfo {year}
  {2009})}\BibitemShut {NoStop}%
\bibitem [{\citenamefont {Costa}\ and\ \citenamefont
  {Shrapnel}(2016)}]{1367-2630-18-6-063032}%
  \BibitemOpen
  \bibfield  {author} {\bibinfo {author} {\bibfnamefont {Fabio}\ \bibnamefont
  {Costa}}\ and\ \bibinfo {author} {\bibfnamefont {Sally}\ \bibnamefont
  {Shrapnel}},\ }\bibfield  {title} {\enquote {\bibinfo {title} {Quantum causal
  modelling},}\ }\href {http://stacks.iop.org/1367-2630/18/i=6/a=063032}
  {\bibfield  {journal} {\bibinfo  {journal} {New Journal of Physics}\ }\textbf
  {\bibinfo {volume} {18}},\ \bibinfo {pages} {063032} (\bibinfo {year}
  {2016})}\BibitemShut {NoStop}%
\bibitem [{\citenamefont {Shrapnel}\ \emph {et~al.}(2018)\citenamefont
  {Shrapnel}, \citenamefont {Costa},\ and\ \citenamefont
  {Milburn}}]{Shrapnel_2018}%
  \BibitemOpen
  \bibfield  {author} {\bibinfo {author} {\bibfnamefont {Sally}\ \bibnamefont
  {Shrapnel}}, \bibinfo {author} {\bibfnamefont {Fabio}\ \bibnamefont {Costa}},
  \ and\ \bibinfo {author} {\bibfnamefont {Gerard}\ \bibnamefont {Milburn}},\
  }\bibfield  {title} {\enquote {\bibinfo {title} {{Updating the Born rule}},}\
  }\href {\doibase 10.1088/1367-2630/aabe12} {\bibfield  {journal} {\bibinfo
  {journal} {New Journal of Physics}\ }\textbf {\bibinfo {volume} {20}},\
  \bibinfo {pages} {053010} (\bibinfo {year} {2018})}\BibitemShut {NoStop}%
\bibitem [{\citenamefont {Milz}\ \emph {et~al.}(2020)\citenamefont {Milz},
  \citenamefont {Sakuldee}, \citenamefont {Pollock},\ and\ \citenamefont
  {Modi}}]{Milz2020}%
  \BibitemOpen
  \bibfield  {author} {\bibinfo {author} {\bibfnamefont {Simon}\ \bibnamefont
  {Milz}}, \bibinfo {author} {\bibfnamefont {Fattah}\ \bibnamefont {Sakuldee}},
  \bibinfo {author} {\bibfnamefont {Felix~A.}\ \bibnamefont {Pollock}}, \ and\
  \bibinfo {author} {\bibfnamefont {Kavan}\ \bibnamefont {Modi}},\ }\bibfield
  {title} {\enquote {\bibinfo {title} {{Kolmogorov extension theorem for
  (quantum) causal modelling and general probabilistic theories}},}\ }\href
  {\doibase 10.22331/q-2020-04-20-255} {\bibfield  {journal} {\bibinfo
  {journal} {Quantum}\ }\textbf {\bibinfo {volume} {4}},\ \bibinfo {pages}
  {255} (\bibinfo {year} {2020})},\ \Eprint {http://arxiv.org/abs/1712.02589}
  {arXiv:1712.02589} \BibitemShut {NoStop}%
\bibitem [{\citenamefont {Aloisio}\ \emph {et~al.}(2023)\citenamefont
  {Aloisio}, \citenamefont {White}, \citenamefont {Hill},\ and\ \citenamefont
  {Modi}}]{aloisio-complexity}%
  \BibitemOpen
  \bibfield  {author} {\bibinfo {author} {\bibfnamefont {I.A.}\ \bibnamefont
  {Aloisio}}, \bibinfo {author} {\bibfnamefont {G.A.L.}\ \bibnamefont {White}},
  \bibinfo {author} {\bibfnamefont {C.D.}\ \bibnamefont {Hill}}, \ and\
  \bibinfo {author} {\bibfnamefont {K.}~\bibnamefont {Modi}},\ }\bibfield
  {title} {\enquote {\bibinfo {title} {Sampling complexity of open quantum
  systems},}\ }\href {\doibase 10.1103/PRXQuantum.4.020310} {\bibfield
  {journal} {\bibinfo  {journal} {PRX Quantum}\ }\textbf {\bibinfo {volume}
  {4}},\ \bibinfo {pages} {020310} (\bibinfo {year} {2023})}\BibitemShut
  {NoStop}%
\bibitem [{\citenamefont {Greenbaum}(2015)}]{greenbaum2015introduction}%
  \BibitemOpen
  \bibfield  {author} {\bibinfo {author} {\bibfnamefont {Daniel}\ \bibnamefont
  {Greenbaum}},\ }\href@noop {} {\enquote {\bibinfo {title} {Introduction to
  quantum gate set tomography},}\ } (\bibinfo {year} {2015}),\ \Eprint
  {http://arxiv.org/abs/1509.02921} {arXiv:1509.02921 [quant-ph]} \BibitemShut
  {NoStop}%
\bibitem [{\citenamefont {Merkel}\ \emph {et~al.}(2013)\citenamefont {Merkel},
  \citenamefont {Gambetta}, \citenamefont {Smolin}, \citenamefont {Poletto},
  \citenamefont {C\'orcoles}, \citenamefont {Johnson}, \citenamefont {Ryan},\
  and\ \citenamefont {Steffen}}]{PhysRevA.87.062119}%
  \BibitemOpen
  \bibfield  {author} {\bibinfo {author} {\bibfnamefont {Seth~T.}\ \bibnamefont
  {Merkel}}, \bibinfo {author} {\bibfnamefont {Jay~M.}\ \bibnamefont
  {Gambetta}}, \bibinfo {author} {\bibfnamefont {John~A.}\ \bibnamefont
  {Smolin}}, \bibinfo {author} {\bibfnamefont {Stefano}\ \bibnamefont
  {Poletto}}, \bibinfo {author} {\bibfnamefont {Antonio~D.}\ \bibnamefont
  {C\'orcoles}}, \bibinfo {author} {\bibfnamefont {Blake~R.}\ \bibnamefont
  {Johnson}}, \bibinfo {author} {\bibfnamefont {Colm~A.}\ \bibnamefont {Ryan}},
  \ and\ \bibinfo {author} {\bibfnamefont {Matthias}\ \bibnamefont {Steffen}},\
  }\bibfield  {title} {\enquote {\bibinfo {title} {Self-consistent quantum
  process tomography},}\ }\href {\doibase 10.1103/PhysRevA.87.062119}
  {\bibfield  {journal} {\bibinfo  {journal} {Phys. Rev. A}\ }\textbf {\bibinfo
  {volume} {87}},\ \bibinfo {pages} {062119} (\bibinfo {year}
  {2013})}\BibitemShut {NoStop}%
\bibitem [{\citenamefont {Blume-Kohout}\ \emph {et~al.}(2017)\citenamefont
  {Blume-Kohout}, \citenamefont {Gamble}, \citenamefont {Nielsen},
  \citenamefont {Rudinger}, \citenamefont {Mizrahi}, \citenamefont {Fortier},\
  and\ \citenamefont {Maunz}}]{RBK2017}%
  \BibitemOpen
  \bibfield  {author} {\bibinfo {author} {\bibfnamefont {Robin}\ \bibnamefont
  {Blume-Kohout}}, \bibinfo {author} {\bibfnamefont {John~King}\ \bibnamefont
  {Gamble}}, \bibinfo {author} {\bibfnamefont {Erik}\ \bibnamefont {Nielsen}},
  \bibinfo {author} {\bibfnamefont {Kenneth}\ \bibnamefont {Rudinger}},
  \bibinfo {author} {\bibfnamefont {Jonathan}\ \bibnamefont {Mizrahi}},
  \bibinfo {author} {\bibfnamefont {Kevin}\ \bibnamefont {Fortier}}, \ and\
  \bibinfo {author} {\bibfnamefont {Peter}\ \bibnamefont {Maunz}},\ }\bibfield
  {title} {\enquote {\bibinfo {title} {{Demonstration of qubit operations below
  a rigorous fault tolerance threshold with gate set tomography}},}\ }\href
  {\doibase 10.1038/ncomms14485} {\bibfield  {journal} {\bibinfo  {journal}
  {Nature Communications}\ }\textbf {\bibinfo {volume} {8}},\ \bibinfo {pages}
  {14485} (\bibinfo {year} {2017})},\ \Eprint {http://arxiv.org/abs/1605.07674}
  {arXiv:1605.07674} \BibitemShut {NoStop}%
\bibitem [{\citenamefont {Nielsen}\ \emph {et~al.}(2021)\citenamefont
  {Nielsen}, \citenamefont {Gamble}, \citenamefont {Rudinger}, \citenamefont
  {Scholten}, \citenamefont {Young},\ and\ \citenamefont
  {Blume-Kohout}}]{Nielsen2021gatesettomography}%
  \BibitemOpen
  \bibfield  {author} {\bibinfo {author} {\bibfnamefont {Erik}\ \bibnamefont
  {Nielsen}}, \bibinfo {author} {\bibfnamefont {John~King}\ \bibnamefont
  {Gamble}}, \bibinfo {author} {\bibfnamefont {Kenneth}\ \bibnamefont
  {Rudinger}}, \bibinfo {author} {\bibfnamefont {Travis}\ \bibnamefont
  {Scholten}}, \bibinfo {author} {\bibfnamefont {Kevin}\ \bibnamefont {Young}},
  \ and\ \bibinfo {author} {\bibfnamefont {Robin}\ \bibnamefont
  {Blume-Kohout}},\ }\bibfield  {title} {\enquote {\bibinfo {title} {Gate {S}et
  {T}omography},}\ }\href {\doibase 10.22331/q-2021-10-05-557} {\bibfield
  {journal} {\bibinfo  {journal} {{Quantum}}\ }\textbf {\bibinfo {volume}
  {5}},\ \bibinfo {pages} {557} (\bibinfo {year} {2021})}\BibitemShut {NoStop}%
\bibitem [{\citenamefont {Zhang}\ \emph {et~al.}(2023)\citenamefont {Zhang},
  \citenamefont {Wu}, \citenamefont {White}, \citenamefont {Xiang},
  \citenamefont {Hu}, \citenamefont {Peng}, \citenamefont {Liu}, \citenamefont
  {Zheng}, \citenamefont {Fu}, \citenamefont {Huang}, \citenamefont {Poletti},
  \citenamefont {Modi}, \citenamefont {Wu}, \citenamefont {Deng},\ and\
  \citenamefont {Guo}}]{zhang2023randomised}%
  \BibitemOpen
  \bibfield  {author} {\bibinfo {author} {\bibfnamefont {Xinfang}\ \bibnamefont
  {Zhang}}, \bibinfo {author} {\bibfnamefont {Zhihao}\ \bibnamefont {Wu}},
  \bibinfo {author} {\bibfnamefont {Gregory A.~L.}\ \bibnamefont {White}},
  \bibinfo {author} {\bibfnamefont {Zhongcheng}\ \bibnamefont {Xiang}},
  \bibinfo {author} {\bibfnamefont {Shun}\ \bibnamefont {Hu}}, \bibinfo
  {author} {\bibfnamefont {Zhihui}\ \bibnamefont {Peng}}, \bibinfo {author}
  {\bibfnamefont {Yong}\ \bibnamefont {Liu}}, \bibinfo {author} {\bibfnamefont
  {Dongning}\ \bibnamefont {Zheng}}, \bibinfo {author} {\bibfnamefont {Xiang}\
  \bibnamefont {Fu}}, \bibinfo {author} {\bibfnamefont {Anqi}\ \bibnamefont
  {Huang}}, \bibinfo {author} {\bibfnamefont {Dario}\ \bibnamefont {Poletti}},
  \bibinfo {author} {\bibfnamefont {Kavan}\ \bibnamefont {Modi}}, \bibinfo
  {author} {\bibfnamefont {Junjie}\ \bibnamefont {Wu}}, \bibinfo {author}
  {\bibfnamefont {Mingtang}\ \bibnamefont {Deng}}, \ and\ \bibinfo {author}
  {\bibfnamefont {Chu}\ \bibnamefont {Guo}},\ }\href@noop {} {\enquote
  {\bibinfo {title} {Randomised benchmarking for characterizing and forecasting
  correlated processes},}\ } (\bibinfo {year} {2023}),\ \Eprint
  {http://arxiv.org/abs/2312.06062} {arXiv:2312.06062 [quant-ph]} \BibitemShut
  {NoStop}%
\bibitem [{\citenamefont {Vidal}(2003)}]{vidal-MPS}%
  \BibitemOpen
  \bibfield  {author} {\bibinfo {author} {\bibfnamefont {Guifré}\ \bibnamefont
  {Vidal}},\ }\bibfield  {title} {\enquote {\bibinfo {title} {{Efficient
  classical simulation of slightly entangled quantum computations}},}\ }\href
  {\doibase 10.1103/PhysRevLett.91.147902} {\bibfield  {journal} {\bibinfo
  {journal} {Physical Review Letters}\ }\textbf {\bibinfo {volume} {91}},\
  \bibinfo {pages} {147902} (\bibinfo {year} {2003})}\BibitemShut {NoStop}%
\bibitem [{\citenamefont {Or{\'{u}}s}(2014)}]{Orus2014}%
  \BibitemOpen
  \bibfield  {author} {\bibinfo {author} {\bibfnamefont {Román}\ \bibnamefont
  {Or{\'{u}}s}},\ }\bibfield  {title} {\enquote {\bibinfo {title} {{A practical
  introduction to tensor networks: Matrix product states and projected
  entangled pair states}},}\ }\href {\doibase 10.1016/j.aop.2014.06.013}
  {\bibfield  {journal} {\bibinfo  {journal} {Annals of Physics}\ }\textbf
  {\bibinfo {volume} {349}},\ \bibinfo {pages} {117--158} (\bibinfo {year}
  {2014})}\BibitemShut {NoStop}%
\bibitem [{\citenamefont {Cirac}\ \emph {et~al.}(2021)\citenamefont {Cirac},
  \citenamefont {P\'erez-Garc\'{\i}a}, \citenamefont {Schuch},\ and\
  \citenamefont {Verstraete}}]{Cirac2020MatrixPS}%
  \BibitemOpen
  \bibfield  {author} {\bibinfo {author} {\bibfnamefont {J.~Ignacio}\
  \bibnamefont {Cirac}}, \bibinfo {author} {\bibfnamefont {David}\ \bibnamefont
  {P\'erez-Garc\'{\i}a}}, \bibinfo {author} {\bibfnamefont {Norbert}\
  \bibnamefont {Schuch}}, \ and\ \bibinfo {author} {\bibfnamefont {Frank}\
  \bibnamefont {Verstraete}},\ }\bibfield  {title} {\enquote {\bibinfo {title}
  {Matrix product states and projected entangled pair states: Concepts,
  symmetries, theorems},}\ }\href {\doibase 10.1103/RevModPhys.93.045003}
  {\bibfield  {journal} {\bibinfo  {journal} {Rev. Mod. Phys.}\ }\textbf
  {\bibinfo {volume} {93}},\ \bibinfo {pages} {045003} (\bibinfo {year}
  {2021})}\BibitemShut {NoStop}%
\bibitem [{\citenamefont {White}(1992)}]{PhysRevLett.69.2863}%
  \BibitemOpen
  \bibfield  {author} {\bibinfo {author} {\bibfnamefont {Steven~R.}\
  \bibnamefont {White}},\ }\bibfield  {title} {\enquote {\bibinfo {title}
  {Density matrix formulation for quantum renormalization groups},}\ }\href
  {\doibase 10.1103/PhysRevLett.69.2863} {\bibfield  {journal} {\bibinfo
  {journal} {Phys. Rev. Lett.}\ }\textbf {\bibinfo {volume} {69}},\ \bibinfo
  {pages} {2863--2866} (\bibinfo {year} {1992})}\BibitemShut {NoStop}%
\bibitem [{\citenamefont {Eisert}(2021)}]{Eisert2021EntanglingPA}%
  \BibitemOpen
  \bibfield  {author} {\bibinfo {author} {\bibfnamefont {J.}~\bibnamefont
  {Eisert}},\ }\bibfield  {title} {\enquote {\bibinfo {title} {Entangling power
  and quantum circuit complexity},}\ }\href {\doibase
  10.1103/PhysRevLett.127.020501} {\bibfield  {journal} {\bibinfo  {journal}
  {Phys. Rev. Lett.}\ }\textbf {\bibinfo {volume} {127}},\ \bibinfo {pages}
  {020501} (\bibinfo {year} {2021})}\BibitemShut {NoStop}%
\bibitem [{\citenamefont {Jozsa}\ and\ \citenamefont
  {Linden}(2003)}]{entanglement-jozsa}%
  \BibitemOpen
  \bibfield  {author} {\bibinfo {author} {\bibfnamefont {Richard}\ \bibnamefont
  {Jozsa}}\ and\ \bibinfo {author} {\bibfnamefont {Noah}\ \bibnamefont
  {Linden}},\ }\bibfield  {title} {\enquote {\bibinfo {title} {{On the role of
  entanglement in quantum-computational speed-up}},}\ }\href {\doibase
  10.1098/rspa.2002.1097} {\bibfield  {journal} {\bibinfo  {journal}
  {Proceedings of the Royal Society A: Mathematical, Physical and Engineering
  Sciences}\ }\textbf {\bibinfo {volume} {459}},\ \bibinfo {pages} {2011}
  (\bibinfo {year} {2003})}\BibitemShut {NoStop}%
\bibitem [{\citenamefont {Li}\ \emph {et~al.}(2018)\citenamefont {Li},
  \citenamefont {Wu}, \citenamefont {Ying}, \citenamefont {Sun},\ and\
  \citenamefont {Yang}}]{china-sim}%
  \BibitemOpen
  \bibfield  {author} {\bibinfo {author} {\bibfnamefont {Riling}\ \bibnamefont
  {Li}}, \bibinfo {author} {\bibfnamefont {Bujiao}\ \bibnamefont {Wu}},
  \bibinfo {author} {\bibfnamefont {Mingsheng}\ \bibnamefont {Ying}}, \bibinfo
  {author} {\bibfnamefont {Xiaoming}\ \bibnamefont {Sun}}, \ and\ \bibinfo
  {author} {\bibfnamefont {Guangwen}\ \bibnamefont {Yang}},\ }\href@noop {}
  {\enquote {\bibinfo {title} {{Quantum Supremacy Circuit Simulation on Sunway
  TaihuLight}},}\ } (\bibinfo {year} {2018}),\ \Eprint
  {http://arxiv.org/abs/1804.04797} {arXiv:1804.04797 [quant-ph]} \BibitemShut
  {NoStop}%
\bibitem [{\citenamefont {Pan}\ and\ \citenamefont
  {Zhang}(2021)}]{pan2021simulating}%
  \BibitemOpen
  \bibfield  {author} {\bibinfo {author} {\bibfnamefont {Feng}\ \bibnamefont
  {Pan}}\ and\ \bibinfo {author} {\bibfnamefont {Pan}\ \bibnamefont {Zhang}},\
  }\href@noop {} {\enquote {\bibinfo {title} {Simulating the sycamore quantum
  supremacy circuits},}\ } (\bibinfo {year} {2021}),\ \Eprint
  {http://arxiv.org/abs/2103.03074} {arXiv:2103.03074 [quant-ph]} \BibitemShut
  {NoStop}%
\bibitem [{\citenamefont {Akhtar}\ \emph {et~al.}(2022)\citenamefont {Akhtar},
  \citenamefont {Hu},\ and\ \citenamefont {You}}]{akhtar2022scalable}%
  \BibitemOpen
  \bibfield  {author} {\bibinfo {author} {\bibfnamefont {Ahmed~A.}\
  \bibnamefont {Akhtar}}, \bibinfo {author} {\bibfnamefont {Hong-Ye}\
  \bibnamefont {Hu}}, \ and\ \bibinfo {author} {\bibfnamefont {Yi-Zhuang}\
  \bibnamefont {You}},\ }\href@noop {} {\enquote {\bibinfo {title} {Scalable
  and flexible classical shadow tomography with tensor networks},}\ } (\bibinfo
  {year} {2022}),\ \Eprint {http://arxiv.org/abs/2209.02093} {arXiv:2209.02093
  [quant-ph]} \BibitemShut {NoStop}%
\bibitem [{\citenamefont {Bertoni}\ \emph {et~al.}(2022)\citenamefont
  {Bertoni}, \citenamefont {Haferkamp}, \citenamefont {Hinsche}, \citenamefont
  {Ioannou}, \citenamefont {Eisert},\ and\ \citenamefont
  {Pashayan}}]{bertoni2022shallow}%
  \BibitemOpen
  \bibfield  {author} {\bibinfo {author} {\bibfnamefont {Christian}\
  \bibnamefont {Bertoni}}, \bibinfo {author} {\bibfnamefont {Jonas}\
  \bibnamefont {Haferkamp}}, \bibinfo {author} {\bibfnamefont {Marcel}\
  \bibnamefont {Hinsche}}, \bibinfo {author} {\bibfnamefont {Marios}\
  \bibnamefont {Ioannou}}, \bibinfo {author} {\bibfnamefont {Jens}\
  \bibnamefont {Eisert}}, \ and\ \bibinfo {author} {\bibfnamefont {Hakop}\
  \bibnamefont {Pashayan}},\ }\href@noop {} {\enquote {\bibinfo {title}
  {Shallow shadows: Expectation estimation using low-depth random clifford
  circuits},}\ } (\bibinfo {year} {2022}),\ \Eprint
  {http://arxiv.org/abs/2209.12924} {arXiv:2209.12924 [quant-ph]} \BibitemShut
  {NoStop}%
\bibitem [{\citenamefont {Torlai}\ \emph {et~al.}(2023)\citenamefont {Torlai},
  \citenamefont {Wood}, \citenamefont {Acharya}, \citenamefont {Carleo},
  \citenamefont {Carrasquilla},\ and\ \citenamefont
  {Aolita}}]{torlai2020quantum}%
  \BibitemOpen
  \bibfield  {author} {\bibinfo {author} {\bibfnamefont {Giacomo}\ \bibnamefont
  {Torlai}}, \bibinfo {author} {\bibfnamefont {Christopher~J}\ \bibnamefont
  {Wood}}, \bibinfo {author} {\bibfnamefont {Atithi}\ \bibnamefont {Acharya}},
  \bibinfo {author} {\bibfnamefont {Giuseppe}\ \bibnamefont {Carleo}}, \bibinfo
  {author} {\bibfnamefont {Juan}\ \bibnamefont {Carrasquilla}}, \ and\ \bibinfo
  {author} {\bibfnamefont {Leandro}\ \bibnamefont {Aolita}},\ }\bibfield
  {title} {\enquote {\bibinfo {title} {Quantum process tomography with
  unsupervised learning and tensor networks},}\ }\href {\doibase
  https://doi.org/10.1038/s41467-023-38332-9} {\bibfield  {journal} {\bibinfo
  {journal} {Nature Communications}\ }\textbf {\bibinfo {volume} {14}},\
  \bibinfo {pages} {2858} (\bibinfo {year} {2023})}\BibitemShut {NoStop}%
\bibitem [{\citenamefont {Wilde}\ \emph {et~al.}(2022)\citenamefont {Wilde},
  \citenamefont {Kshetrimayum}, \citenamefont {Roth}, \citenamefont
  {Hangleiter}, \citenamefont {Sweke},\ and\ \citenamefont
  {Eisert}}]{wilde2022scalably}%
  \BibitemOpen
  \bibfield  {author} {\bibinfo {author} {\bibfnamefont {Frederik}\
  \bibnamefont {Wilde}}, \bibinfo {author} {\bibfnamefont {Augustine}\
  \bibnamefont {Kshetrimayum}}, \bibinfo {author} {\bibfnamefont {Ingo}\
  \bibnamefont {Roth}}, \bibinfo {author} {\bibfnamefont {Dominik}\
  \bibnamefont {Hangleiter}}, \bibinfo {author} {\bibfnamefont {Ryan}\
  \bibnamefont {Sweke}}, \ and\ \bibinfo {author} {\bibfnamefont {Jens}\
  \bibnamefont {Eisert}},\ }\href@noop {} {\enquote {\bibinfo {title} {Scalably
  learning quantum many-body hamiltonians from dynamical data},}\ } (\bibinfo
  {year} {2022}),\ \Eprint {http://arxiv.org/abs/2209.14328} {arXiv:2209.14328
  [quant-ph]} \BibitemShut {NoStop}%
\bibitem [{\citenamefont {White}(2024)}]{PTT_github_repo}%
  \BibitemOpen
  \bibfield  {author} {\bibinfo {author} {\bibfnamefont {Gregory A~L}\
  \bibnamefont {White}},\ }\href
  {https://github.com/gwhitequantum/process-tensor-network-tomography}
  {\enquote {\bibinfo {title} {{Process Tensor Network Tomography}},}\ }
  (\bibinfo {year} {2024})\BibitemShut {NoStop}%
\bibitem [{\citenamefont {Milz}\ \emph {et~al.}(2021)\citenamefont {Milz},
  \citenamefont {Spee}, \citenamefont {Xu}, \citenamefont {Pollock},
  \citenamefont {Modi},\ and\ \citenamefont {Gühne}}]{milz2021genuine}%
  \BibitemOpen
  \bibfield  {author} {\bibinfo {author} {\bibfnamefont {Simon}\ \bibnamefont
  {Milz}}, \bibinfo {author} {\bibfnamefont {Cornelia}\ \bibnamefont {Spee}},
  \bibinfo {author} {\bibfnamefont {Zhen-Peng}\ \bibnamefont {Xu}}, \bibinfo
  {author} {\bibfnamefont {Felix~A.}\ \bibnamefont {Pollock}}, \bibinfo
  {author} {\bibfnamefont {Kavan}\ \bibnamefont {Modi}}, \ and\ \bibinfo
  {author} {\bibfnamefont {Otfried}\ \bibnamefont {Gühne}},\ }\bibfield
  {title} {\enquote {\bibinfo {title} {{Genuine multipartite entanglement in
  time}},}\ }\href {\doibase 10.21468/SciPostPhys.10.6.141} {\bibfield
  {journal} {\bibinfo  {journal} {SciPost Phys.}\ }\textbf {\bibinfo {volume}
  {10}},\ \bibinfo {pages} {141} (\bibinfo {year} {2021})}\BibitemShut
  {NoStop}%
\bibitem [{\citenamefont {McKay}\ \emph {et~al.}(2017)\citenamefont {McKay},
  \citenamefont {Wood}, \citenamefont {Sheldon}, \citenamefont {Chow},\ and\
  \citenamefont {Gambetta}}]{McKay_2017}%
  \BibitemOpen
  \bibfield  {author} {\bibinfo {author} {\bibfnamefont {David~C.}\
  \bibnamefont {McKay}}, \bibinfo {author} {\bibfnamefont {Christopher~J.}\
  \bibnamefont {Wood}}, \bibinfo {author} {\bibfnamefont {Sarah}\ \bibnamefont
  {Sheldon}}, \bibinfo {author} {\bibfnamefont {Jerry~M.}\ \bibnamefont
  {Chow}}, \ and\ \bibinfo {author} {\bibfnamefont {Jay~M.}\ \bibnamefont
  {Gambetta}},\ }\bibfield  {title} {\enquote {\bibinfo {title} {Efficient
  <mml:math
  xmlns:mml="http://www.w3.org/1998/math/mathml"><mml:mi>z</mml:mi></mml:math>
  gates for quantum computing},}\ }\href {\doibase 10.1103/physreva.96.022330}
  {\bibfield  {journal} {\bibinfo  {journal} {Physical Review A}\ }\textbf
  {\bibinfo {volume} {96}} (\bibinfo {year} {2017}),\
  10.1103/physreva.96.022330}\BibitemShut {NoStop}%
\bibitem [{\citenamefont {Dutta}\ and\ \citenamefont
  {Horn}(1981)}]{RevModPhys.53.497}%
  \BibitemOpen
  \bibfield  {author} {\bibinfo {author} {\bibfnamefont {P.}~\bibnamefont
  {Dutta}}\ and\ \bibinfo {author} {\bibfnamefont {P.~M.}\ \bibnamefont
  {Horn}},\ }\bibfield  {title} {\enquote {\bibinfo {title} {Low-frequency
  fluctuations in solids: $1/f$ noise},}\ }\href {\doibase
  10.1103/RevModPhys.53.497} {\bibfield  {journal} {\bibinfo  {journal} {Rev.
  Mod. Phys.}\ }\textbf {\bibinfo {volume} {53}},\ \bibinfo {pages} {497--516}
  (\bibinfo {year} {1981})}\BibitemShut {NoStop}%
\bibitem [{\citenamefont {West}\ and\ \citenamefont
  {Shlesinger}(1989)}]{west1989ubiquity}%
  \BibitemOpen
  \bibfield  {author} {\bibinfo {author} {\bibfnamefont {Bruce~J}\ \bibnamefont
  {West}}\ and\ \bibinfo {author} {\bibfnamefont {Michael~F}\ \bibnamefont
  {Shlesinger}},\ }\bibfield  {title} {\enquote {\bibinfo {title} {On the
  ubiquity of $1/f$ noise},}\ }\href {\doibase
  https://doi.org/10.1142/S0217979289000609} {\bibfield  {journal} {\bibinfo
  {journal} {International Journal of Modern Physics B}\ }\textbf {\bibinfo
  {volume} {3}},\ \bibinfo {pages} {795--819} (\bibinfo {year}
  {1989})}\BibitemShut {NoStop}%
\bibitem [{\citenamefont {Hausdorff}\ and\ \citenamefont
  {Peng}(1996)}]{PhysRevE.54.2154}%
  \BibitemOpen
  \bibfield  {author} {\bibinfo {author} {\bibfnamefont {Jeffrey~M.}\
  \bibnamefont {Hausdorff}}\ and\ \bibinfo {author} {\bibfnamefont {C.-K.}\
  \bibnamefont {Peng}},\ }\bibfield  {title} {\enquote {\bibinfo {title}
  {Multiscaled randomness: A possible source of $1/f$ noise in biology},}\
  }\href {\doibase 10.1103/PhysRevE.54.2154} {\bibfield  {journal} {\bibinfo
  {journal} {Phys. Rev. E}\ }\textbf {\bibinfo {volume} {54}},\ \bibinfo
  {pages} {2154--2157} (\bibinfo {year} {1996})}\BibitemShut {NoStop}%
\bibitem [{\citenamefont {Paladino}\ \emph {et~al.}(2014)\citenamefont
  {Paladino}, \citenamefont {Galperin}, \citenamefont {Falci},\ and\
  \citenamefont {Altshuler}}]{paladino20141}%
  \BibitemOpen
  \bibfield  {author} {\bibinfo {author} {\bibfnamefont {E.}~\bibnamefont
  {Paladino}}, \bibinfo {author} {\bibfnamefont {Y.~M.}\ \bibnamefont
  {Galperin}}, \bibinfo {author} {\bibfnamefont {G.}~\bibnamefont {Falci}}, \
  and\ \bibinfo {author} {\bibfnamefont {B.~L.}\ \bibnamefont {Altshuler}},\
  }\bibfield  {title} {\enquote {\bibinfo {title} {$1/f$ noise: Implications
  for solid-state quantum information},}\ }\href {\doibase
  10.1103/RevModPhys.86.361} {\bibfield  {journal} {\bibinfo  {journal} {Rev.
  Mod. Phys.}\ }\textbf {\bibinfo {volume} {86}},\ \bibinfo {pages} {361--418}
  (\bibinfo {year} {2014})}\BibitemShut {NoStop}%
\bibitem [{\citenamefont {Aquino}\ \emph {et~al.}(2023)\citenamefont {Aquino},
  \citenamefont {Alberto}, \citenamefont {de~Sousa},\ and\ \citenamefont
  {Rogério}}]{aquino2023model}%
  \BibitemOpen
  \bibfield  {author} {\bibinfo {author} {\bibfnamefont {Nava}\ \bibnamefont
  {Aquino}}, \bibinfo {author} {\bibfnamefont {José}\ \bibnamefont {Alberto}},
  \bibinfo {author} {\bibnamefont {de~Sousa}}, \ and\ \bibinfo {author}
  {\bibnamefont {Rogério}},\ }\href@noop {} {\enquote {\bibinfo {title} {Model
  for $1/f$ flux noise in superconducting aluminum devices: Impact of external
  magnetic fields},}\ } (\bibinfo {year} {2023}),\ \Eprint
  {http://arxiv.org/abs/2302.12316} {arXiv:2302.12316 [quant-ph]} \BibitemShut
  {NoStop}%
\bibitem [{\citenamefont {Zhang}\ \emph {et~al.}(2022)\citenamefont {Zhang},
  \citenamefont {Majumder}, \citenamefont {Leung}, \citenamefont {Crain},
  \citenamefont {Wang}, \citenamefont {Fang}, \citenamefont {Debroy},
  \citenamefont {Kim},\ and\ \citenamefont {Brown}}]{PhysRevApplied.17.034074}%
  \BibitemOpen
  \bibfield  {author} {\bibinfo {author} {\bibfnamefont {Bichen}\ \bibnamefont
  {Zhang}}, \bibinfo {author} {\bibfnamefont {Swarnadeep}\ \bibnamefont
  {Majumder}}, \bibinfo {author} {\bibfnamefont {Pak~Hong}\ \bibnamefont
  {Leung}}, \bibinfo {author} {\bibfnamefont {Stephen}\ \bibnamefont {Crain}},
  \bibinfo {author} {\bibfnamefont {Ye}~\bibnamefont {Wang}}, \bibinfo {author}
  {\bibfnamefont {Chao}\ \bibnamefont {Fang}}, \bibinfo {author} {\bibfnamefont
  {Dripto~M.}\ \bibnamefont {Debroy}}, \bibinfo {author} {\bibfnamefont
  {Jungsang}\ \bibnamefont {Kim}}, \ and\ \bibinfo {author} {\bibfnamefont
  {Kenneth~R.}\ \bibnamefont {Brown}},\ }\bibfield  {title} {\enquote {\bibinfo
  {title} {Hidden inverses: Coherent error cancellation at the circuit
  level},}\ }\href {\doibase 10.1103/PhysRevApplied.17.034074} {\bibfield
  {journal} {\bibinfo  {journal} {Phys. Rev. Appl.}\ }\textbf {\bibinfo
  {volume} {17}},\ \bibinfo {pages} {034074} (\bibinfo {year}
  {2022})}\BibitemShut {NoStop}%
\bibitem [{\citenamefont {Gullans}\ \emph {et~al.}(2023)\citenamefont
  {Gullans}, \citenamefont {Caranti}, \citenamefont {Mills},\ and\
  \citenamefont {Petta}}]{gullans2023compressed}%
  \BibitemOpen
  \bibfield  {author} {\bibinfo {author} {\bibfnamefont {M.~J.}\ \bibnamefont
  {Gullans}}, \bibinfo {author} {\bibfnamefont {M.}~\bibnamefont {Caranti}},
  \bibinfo {author} {\bibfnamefont {A.~R.}\ \bibnamefont {Mills}}, \ and\
  \bibinfo {author} {\bibfnamefont {J.~R.}\ \bibnamefont {Petta}},\ }\href@noop
  {} {\enquote {\bibinfo {title} {Compressed gate characterization for quantum
  devices with time-correlated noise},}\ } (\bibinfo {year} {2023}),\ \Eprint
  {http://arxiv.org/abs/2307.14432} {arXiv:2307.14432 [quant-ph]} \BibitemShut
  {NoStop}%
\bibitem [{\citenamefont {Pollock}\ \emph
  {et~al.}(2018{\natexlab{b}})\citenamefont {Pollock}, \citenamefont
  {Rodr{\'{i}}guez-Rosario}, \citenamefont {Frauenheim}, \citenamefont
  {Paternostro},\ and\ \citenamefont {Modi}}]{Pollock2018}%
  \BibitemOpen
  \bibfield  {author} {\bibinfo {author} {\bibfnamefont {Felix~A.}\
  \bibnamefont {Pollock}}, \bibinfo {author} {\bibfnamefont {C{\'{e}}sar}\
  \bibnamefont {Rodr{\'{i}}guez-Rosario}}, \bibinfo {author} {\bibfnamefont
  {Thomas}\ \bibnamefont {Frauenheim}}, \bibinfo {author} {\bibfnamefont
  {Mauro}\ \bibnamefont {Paternostro}}, \ and\ \bibinfo {author} {\bibfnamefont
  {Kavan}\ \bibnamefont {Modi}},\ }\bibfield  {title} {\enquote {\bibinfo
  {title} {{Operational Markov Condition for Quantum Processes}},}\ }\href
  {\doibase 10.1103/PhysRevLett.120.040405} {\bibfield  {journal} {\bibinfo
  {journal} {Physical Review Letters}\ }\textbf {\bibinfo {volume} {120}},\
  \bibinfo {pages} {040405} (\bibinfo {year} {2018}{\natexlab{b}})},\ \Eprint
  {http://arxiv.org/abs/1801.09811} {arXiv:1801.09811} \BibitemShut {NoStop}%
\bibitem [{\citenamefont {Bengtsson}\ and\ \citenamefont
  {Zyczkowski}(2006)}]{bengtsson_zyczkowski_2006}%
  \BibitemOpen
  \bibfield  {author} {\bibinfo {author} {\bibfnamefont {Ingemar}\ \bibnamefont
  {Bengtsson}}\ and\ \bibinfo {author} {\bibfnamefont {Karol}\ \bibnamefont
  {Zyczkowski}},\ }\href {\doibase 10.1017/CBO9780511535048} {\emph {\bibinfo
  {title} {Geometry of Quantum States: An Introduction to Quantum
  Entanglement}}}\ (\bibinfo  {publisher} {Cambridge University Press},\
  \bibinfo {year} {2006})\BibitemShut {NoStop}%
\bibitem [{\citenamefont {Piveteau}\ \emph {et~al.}(2023)\citenamefont
  {Piveteau}, \citenamefont {Chubb},\ and\ \citenamefont
  {Renes}}]{piveteau2023tensor}%
  \BibitemOpen
  \bibfield  {author} {\bibinfo {author} {\bibfnamefont {Christophe}\
  \bibnamefont {Piveteau}}, \bibinfo {author} {\bibfnamefont {Christopher~T.}\
  \bibnamefont {Chubb}}, \ and\ \bibinfo {author} {\bibfnamefont {Joseph~M.}\
  \bibnamefont {Renes}},\ }\href@noop {} {\enquote {\bibinfo {title} {Tensor
  network decoding beyond 2d},}\ } (\bibinfo {year} {2023}),\ \Eprint
  {http://arxiv.org/abs/2310.10722} {arXiv:2310.10722 [quant-ph]} \BibitemShut
  {NoStop}%
\bibitem [{\citenamefont {Wilde}(2013)}]{wilde_2013}%
  \BibitemOpen
  \bibfield  {author} {\bibinfo {author} {\bibfnamefont {Mark~M.}\ \bibnamefont
  {Wilde}},\ }\href {\doibase 10.1017/CBO9781139525343} {\emph {\bibinfo
  {title} {Quantum Information Theory}}}\ (\bibinfo  {publisher} {Cambridge
  University Press},\ \bibinfo {year} {2013})\BibitemShut {NoStop}%
\bibitem [{\citenamefont {Huang}\ \emph {et~al.}(2019)\citenamefont {Huang},
  \citenamefont {Doherty},\ and\ \citenamefont {Flammia}}]{PhysRevA.99.022313}%
  \BibitemOpen
  \bibfield  {author} {\bibinfo {author} {\bibfnamefont {Eric}\ \bibnamefont
  {Huang}}, \bibinfo {author} {\bibfnamefont {Andrew~C.}\ \bibnamefont
  {Doherty}}, \ and\ \bibinfo {author} {\bibfnamefont {Steven}\ \bibnamefont
  {Flammia}},\ }\bibfield  {title} {\enquote {\bibinfo {title} {Performance of
  quantum error correction with coherent errors},}\ }\href {\doibase
  10.1103/PhysRevA.99.022313} {\bibfield  {journal} {\bibinfo  {journal} {Phys.
  Rev. A}\ }\textbf {\bibinfo {volume} {99}},\ \bibinfo {pages} {022313}
  (\bibinfo {year} {2019})}\BibitemShut {NoStop}%
\bibitem [{\citenamefont {Cross}\ \emph {et~al.}(2019)\citenamefont {Cross},
  \citenamefont {Bishop}, \citenamefont {Sheldon}, \citenamefont {Nation},\
  and\ \citenamefont {Gambetta}}]{Cross-QV}%
  \BibitemOpen
  \bibfield  {author} {\bibinfo {author} {\bibfnamefont {Andrew~W.}\
  \bibnamefont {Cross}}, \bibinfo {author} {\bibfnamefont {Lev~S.}\
  \bibnamefont {Bishop}}, \bibinfo {author} {\bibfnamefont {Sarah}\
  \bibnamefont {Sheldon}}, \bibinfo {author} {\bibfnamefont {Paul~D.}\
  \bibnamefont {Nation}}, \ and\ \bibinfo {author} {\bibfnamefont {Jay~M.}\
  \bibnamefont {Gambetta}},\ }\bibfield  {title} {\enquote {\bibinfo {title}
  {{Validating quantum computers using randomized model circuits}},}\ }\href
  {\doibase 10.1103/PhysRevA.100.032328} {\bibfield  {journal} {\bibinfo
  {journal} {Physical Review A}\ }\textbf {\bibinfo {volume} {100}},\ \bibinfo
  {pages} {032328} (\bibinfo {year} {2019})},\ \Eprint
  {http://arxiv.org/abs/1811.12926} {arXiv:1811.12926} \BibitemShut {NoStop}%
\bibitem [{\citenamefont {Viola}\ \emph {et~al.}(1999)\citenamefont {Viola},
  \citenamefont {Knill},\ and\ \citenamefont {Lloyd}}]{PhysRevLett.82.2417}%
  \BibitemOpen
  \bibfield  {author} {\bibinfo {author} {\bibfnamefont {Lorenza}\ \bibnamefont
  {Viola}}, \bibinfo {author} {\bibfnamefont {Emanuel}\ \bibnamefont {Knill}},
  \ and\ \bibinfo {author} {\bibfnamefont {Seth}\ \bibnamefont {Lloyd}},\
  }\bibfield  {title} {\enquote {\bibinfo {title} {Dynamical decoupling of open
  quantum systems},}\ }\href {\doibase 10.1103/PhysRevLett.82.2417} {\bibfield
  {journal} {\bibinfo  {journal} {Phys. Rev. Lett.}\ }\textbf {\bibinfo
  {volume} {82}},\ \bibinfo {pages} {2417--2421} (\bibinfo {year}
  {1999})}\BibitemShut {NoStop}%
\bibitem [{\citenamefont {Khodjasteh}\ and\ \citenamefont
  {Lidar}(2007)}]{PhysRevA.75.062310}%
  \BibitemOpen
  \bibfield  {author} {\bibinfo {author} {\bibfnamefont {Kaveh}\ \bibnamefont
  {Khodjasteh}}\ and\ \bibinfo {author} {\bibfnamefont {Daniel~A.}\
  \bibnamefont {Lidar}},\ }\bibfield  {title} {\enquote {\bibinfo {title}
  {Performance of deterministic dynamical decoupling schemes: Concatenated and
  periodic pulse sequences},}\ }\href {\doibase 10.1103/PhysRevA.75.062310}
  {\bibfield  {journal} {\bibinfo  {journal} {Phys. Rev. A}\ }\textbf {\bibinfo
  {volume} {75}},\ \bibinfo {pages} {062310} (\bibinfo {year}
  {2007})}\BibitemShut {NoStop}%
\bibitem [{\citenamefont {Kuo}\ and\ \citenamefont
  {Lidar}(2011)}]{PhysRevA.84.042329}%
  \BibitemOpen
  \bibfield  {author} {\bibinfo {author} {\bibfnamefont {Wan-Jung}\
  \bibnamefont {Kuo}}\ and\ \bibinfo {author} {\bibfnamefont {Daniel~A.}\
  \bibnamefont {Lidar}},\ }\bibfield  {title} {\enquote {\bibinfo {title}
  {Quadratic dynamical decoupling: Universality proof and error analysis},}\
  }\href {\doibase 10.1103/PhysRevA.84.042329} {\bibfield  {journal} {\bibinfo
  {journal} {Phys. Rev. A}\ }\textbf {\bibinfo {volume} {84}},\ \bibinfo
  {pages} {042329} (\bibinfo {year} {2011})}\BibitemShut {NoStop}%
\bibitem [{\citenamefont {Quiroz}\ and\ \citenamefont
  {Lidar}(2013)}]{PhysRevA.88.052306}%
  \BibitemOpen
  \bibfield  {author} {\bibinfo {author} {\bibfnamefont {Gregory}\ \bibnamefont
  {Quiroz}}\ and\ \bibinfo {author} {\bibfnamefont {Daniel~A.}\ \bibnamefont
  {Lidar}},\ }\bibfield  {title} {\enquote {\bibinfo {title} {Optimized
  dynamical decoupling via genetic algorithms},}\ }\href {\doibase
  10.1103/PhysRevA.88.052306} {\bibfield  {journal} {\bibinfo  {journal} {Phys.
  Rev. A}\ }\textbf {\bibinfo {volume} {88}},\ \bibinfo {pages} {052306}
  (\bibinfo {year} {2013})}\BibitemShut {NoStop}%
\bibitem [{\citenamefont {Biercuk}\ \emph {et~al.}(2009)\citenamefont
  {Biercuk}, \citenamefont {Uys}, \citenamefont {VanDevender}, \citenamefont
  {Shiga}, \citenamefont {Itano},\ and\ \citenamefont
  {Bollinger}}]{biercuk2009optimized}%
  \BibitemOpen
  \bibfield  {author} {\bibinfo {author} {\bibfnamefont {Michael~J.}\
  \bibnamefont {Biercuk}}, \bibinfo {author} {\bibfnamefont {Hermann}\
  \bibnamefont {Uys}}, \bibinfo {author} {\bibfnamefont {Aaron~P.}\
  \bibnamefont {VanDevender}}, \bibinfo {author} {\bibfnamefont {Nobuyasu}\
  \bibnamefont {Shiga}}, \bibinfo {author} {\bibfnamefont {Wayne~M.}\
  \bibnamefont {Itano}}, \ and\ \bibinfo {author} {\bibfnamefont {John~J.}\
  \bibnamefont {Bollinger}},\ }\bibfield  {title} {\enquote {\bibinfo {title}
  {Optimized dynamical decoupling in a model quantum memory},}\ }\href
  {\doibase 10.1038/nature07951} {\bibfield  {journal} {\bibinfo  {journal}
  {Nature}\ }\textbf {\bibinfo {volume} {458}},\ \bibinfo {pages} {996--1000}
  (\bibinfo {year} {2009})}\BibitemShut {NoStop}%
\bibitem [{\citenamefont {Khodjasteh}\ and\ \citenamefont
  {Viola}(2009)}]{PhysRevLett.102.080501}%
  \BibitemOpen
  \bibfield  {author} {\bibinfo {author} {\bibfnamefont {Kaveh}\ \bibnamefont
  {Khodjasteh}}\ and\ \bibinfo {author} {\bibfnamefont {Lorenza}\ \bibnamefont
  {Viola}},\ }\bibfield  {title} {\enquote {\bibinfo {title} {Dynamically
  error-corrected gates for universal quantum computation},}\ }\href {\doibase
  10.1103/PhysRevLett.102.080501} {\bibfield  {journal} {\bibinfo  {journal}
  {Phys. Rev. Lett.}\ }\textbf {\bibinfo {volume} {102}},\ \bibinfo {pages}
  {080501} (\bibinfo {year} {2009})}\BibitemShut {NoStop}%
\bibitem [{\citenamefont {Gullion}\ \emph {et~al.}(1990)\citenamefont
  {Gullion}, \citenamefont {Baker},\ and\ \citenamefont
  {Conradi}}]{gullion1990new}%
  \BibitemOpen
  \bibfield  {author} {\bibinfo {author} {\bibfnamefont {Terry}\ \bibnamefont
  {Gullion}}, \bibinfo {author} {\bibfnamefont {David~B}\ \bibnamefont
  {Baker}}, \ and\ \bibinfo {author} {\bibfnamefont {Mark~S}\ \bibnamefont
  {Conradi}},\ }\bibfield  {title} {\enquote {\bibinfo {title} {{New,
  compensated Carr-Purcell sequences}},}\ }\href {\doibase
  https://doi.org/10.1016/0022-2364(90)90331-3} {\bibfield  {journal} {\bibinfo
   {journal} {Journal of Magnetic Resonance (1969)}\ }\textbf {\bibinfo
  {volume} {89}},\ \bibinfo {pages} {479--484} (\bibinfo {year}
  {1990})}\BibitemShut {NoStop}%
\bibitem [{\citenamefont {He}\ \emph {et~al.}(2019)\citenamefont {He},
  \citenamefont {Gorman}, \citenamefont {Keith}, \citenamefont {Kranz},
  \citenamefont {Keizer},\ and\ \citenamefont {Simmons}}]{He2019}%
  \BibitemOpen
  \bibfield  {author} {\bibinfo {author} {\bibfnamefont {Y.}~\bibnamefont
  {He}}, \bibinfo {author} {\bibfnamefont {S.~K.}\ \bibnamefont {Gorman}},
  \bibinfo {author} {\bibfnamefont {D.}~\bibnamefont {Keith}}, \bibinfo
  {author} {\bibfnamefont {L.}~\bibnamefont {Kranz}}, \bibinfo {author}
  {\bibfnamefont {J.~G.}\ \bibnamefont {Keizer}}, \ and\ \bibinfo {author}
  {\bibfnamefont {M.~Y.}\ \bibnamefont {Simmons}},\ }\bibfield  {title}
  {\enquote {\bibinfo {title} {{A two-qubit gate between phosphorus donor
  electrons in silicon}},}\ }\href {\doibase 10.1038/s41586-019-1381-2}
  {\bibfield  {journal} {\bibinfo  {journal} {Nature}\ }\textbf {\bibinfo
  {volume} {571}},\ \bibinfo {pages} {371--375} (\bibinfo {year}
  {2019})}\BibitemShut {NoStop}%
\bibitem [{\citenamefont {Vidal}(2008)}]{PhysRevLett.101.110501}%
  \BibitemOpen
  \bibfield  {author} {\bibinfo {author} {\bibfnamefont {G.}~\bibnamefont
  {Vidal}},\ }\bibfield  {title} {\enquote {\bibinfo {title} {Class of quantum
  many-body states that can be efficiently simulated},}\ }\href {\doibase
  10.1103/PhysRevLett.101.110501} {\bibfield  {journal} {\bibinfo  {journal}
  {Phys. Rev. Lett.}\ }\textbf {\bibinfo {volume} {101}},\ \bibinfo {pages}
  {110501} (\bibinfo {year} {2008})}\BibitemShut {NoStop}%
\bibitem [{\citenamefont {Dowling}\ \emph {et~al.}(2023)\citenamefont
  {Dowling}, \citenamefont {Modi}, \citenamefont {Muñoz}, \citenamefont
  {Singh},\ and\ \citenamefont {White}}]{dowling2023process}%
  \BibitemOpen
  \bibfield  {author} {\bibinfo {author} {\bibfnamefont {Neil}\ \bibnamefont
  {Dowling}}, \bibinfo {author} {\bibfnamefont {Kavan}\ \bibnamefont {Modi}},
  \bibinfo {author} {\bibfnamefont {Roberto~N.}\ \bibnamefont {Muñoz}},
  \bibinfo {author} {\bibfnamefont {Sukhbinder}\ \bibnamefont {Singh}}, \ and\
  \bibinfo {author} {\bibfnamefont {Gregory A.~L.}\ \bibnamefont {White}},\
  }\href@noop {} {\enquote {\bibinfo {title} {Process tree: Efficient
  representation of quantum processes with complex long-range memory},}\ }
  (\bibinfo {year} {2023}),\ \Eprint {http://arxiv.org/abs/2312.04624}
  {arXiv:2312.04624 [quant-ph]} \BibitemShut {NoStop}%
\bibitem [{\citenamefont {Wallman}\ \emph {et~al.}(2015)\citenamefont
  {Wallman}, \citenamefont {Granade}, \citenamefont {Harper},\ and\
  \citenamefont {Flammia}}]{noise-coherence-2015}%
  \BibitemOpen
  \bibfield  {author} {\bibinfo {author} {\bibfnamefont {J.}~\bibnamefont
  {Wallman}}, \bibinfo {author} {\bibfnamefont {C.}~\bibnamefont {Granade}},
  \bibinfo {author} {\bibfnamefont {R.}~\bibnamefont {Harper}}, \ and\ \bibinfo
  {author} {\bibfnamefont {S.~T.}\ \bibnamefont {Flammia}},\ }\bibfield
  {title} {\enquote {\bibinfo {title} {{Estimating the coherence of noise}},}\
  }\href {\doibase 10.1088/1367-2630/17/11/113020} {\bibfield  {journal}
  {\bibinfo  {journal} {New Journal of Physics}\ }\textbf {\bibinfo {volume}
  {17}},\ \bibinfo {pages} {113020} (\bibinfo {year} {2015})}\BibitemShut
  {NoStop}%
\bibitem [{\citenamefont {Blume-Kohout}\ \emph {et~al.}(2022)\citenamefont
  {Blume-Kohout}, \citenamefont {da~Silva}, \citenamefont {Nielsen},
  \citenamefont {Proctor}, \citenamefont {Rudinger}, \citenamefont {Sarovar},\
  and\ \citenamefont {Young}}]{blume2022taxonomy}%
  \BibitemOpen
  \bibfield  {author} {\bibinfo {author} {\bibfnamefont {Robin}\ \bibnamefont
  {Blume-Kohout}}, \bibinfo {author} {\bibfnamefont {Marcus~P.}\ \bibnamefont
  {da~Silva}}, \bibinfo {author} {\bibfnamefont {Erik}\ \bibnamefont
  {Nielsen}}, \bibinfo {author} {\bibfnamefont {Timothy}\ \bibnamefont
  {Proctor}}, \bibinfo {author} {\bibfnamefont {Kenneth}\ \bibnamefont
  {Rudinger}}, \bibinfo {author} {\bibfnamefont {Mohan}\ \bibnamefont
  {Sarovar}}, \ and\ \bibinfo {author} {\bibfnamefont {Kevin}\ \bibnamefont
  {Young}},\ }\bibfield  {title} {\enquote {\bibinfo {title} {A taxonomy of
  small markovian errors},}\ }\href {\doibase 10.1103/PRXQuantum.3.020335}
  {\bibfield  {journal} {\bibinfo  {journal} {PRX Quantum}\ }\textbf {\bibinfo
  {volume} {3}},\ \bibinfo {pages} {020335} (\bibinfo {year}
  {2022})}\BibitemShut {NoStop}%
\bibitem [{\citenamefont {Sarovar}\ \emph {et~al.}(2020)\citenamefont
  {Sarovar}, \citenamefont {Proctor}, \citenamefont {Rudinger}, \citenamefont
  {Young}, \citenamefont {Nielsen},\ and\ \citenamefont
  {Blume-Kohout}}]{Sarovar2020detectingcrosstalk}%
  \BibitemOpen
  \bibfield  {author} {\bibinfo {author} {\bibfnamefont {Mohan}\ \bibnamefont
  {Sarovar}}, \bibinfo {author} {\bibfnamefont {Timothy}\ \bibnamefont
  {Proctor}}, \bibinfo {author} {\bibfnamefont {Kenneth}\ \bibnamefont
  {Rudinger}}, \bibinfo {author} {\bibfnamefont {Kevin}\ \bibnamefont {Young}},
  \bibinfo {author} {\bibfnamefont {Erik}\ \bibnamefont {Nielsen}}, \ and\
  \bibinfo {author} {\bibfnamefont {Robin}\ \bibnamefont {Blume-Kohout}},\
  }\bibfield  {title} {\enquote {\bibinfo {title} {Detecting crosstalk errors
  in quantum information processors},}\ }\href {\doibase
  10.22331/q-2020-09-11-321} {\bibfield  {journal} {\bibinfo  {journal}
  {{Quantum}}\ }\textbf {\bibinfo {volume} {4}},\ \bibinfo {pages} {321}
  (\bibinfo {year} {2020})}\BibitemShut {NoStop}%
\bibitem [{\citenamefont {McKay}\ \emph {et~al.}(2020)\citenamefont {McKay},
  \citenamefont {Cross}, \citenamefont {Wood},\ and\ \citenamefont
  {Gambetta}}]{mckay2020correlated}%
  \BibitemOpen
  \bibfield  {author} {\bibinfo {author} {\bibfnamefont {David~C.}\
  \bibnamefont {McKay}}, \bibinfo {author} {\bibfnamefont {Andrew~W.}\
  \bibnamefont {Cross}}, \bibinfo {author} {\bibfnamefont {Christopher~J.}\
  \bibnamefont {Wood}}, \ and\ \bibinfo {author} {\bibfnamefont {Jay~M.}\
  \bibnamefont {Gambetta}},\ }\href@noop {} {\enquote {\bibinfo {title}
  {Correlated randomized benchmarking},}\ } (\bibinfo {year} {2020}),\ \Eprint
  {http://arxiv.org/abs/2003.02354} {arXiv:2003.02354 [quant-ph]} \BibitemShut
  {NoStop}%
\bibitem [{\citenamefont {Su}\ \emph {et~al.}(2023{\natexlab{b}})\citenamefont
  {Su}, \citenamefont {Huang}, \citenamefont {Stuyck}, \citenamefont {Feng},
  \citenamefont {Gilbert}, \citenamefont {Evans}, \citenamefont {Lim},
  \citenamefont {Hudson}, \citenamefont {Chan}, \citenamefont {Huang},
  \citenamefont {Itoh}, \citenamefont {Harper}, \citenamefont {Bartlett},
  \citenamefont {Yang}, \citenamefont {Laucht}, \citenamefont {Saraiva},
  \citenamefont {Tanttu},\ and\ \citenamefont {Dzurak}}]{su2023characterizing}%
  \BibitemOpen
  \bibfield  {author} {\bibinfo {author} {\bibfnamefont {R.~Y.}\ \bibnamefont
  {Su}}, \bibinfo {author} {\bibfnamefont {J.~Y.}\ \bibnamefont {Huang}},
  \bibinfo {author} {\bibfnamefont {N.~Dumoulin.}\ \bibnamefont {Stuyck}},
  \bibinfo {author} {\bibfnamefont {M.~K.}\ \bibnamefont {Feng}}, \bibinfo
  {author} {\bibfnamefont {W.}~\bibnamefont {Gilbert}}, \bibinfo {author}
  {\bibfnamefont {T.~J.}\ \bibnamefont {Evans}}, \bibinfo {author}
  {\bibfnamefont {W.~H.}\ \bibnamefont {Lim}}, \bibinfo {author} {\bibfnamefont
  {F.~E.}\ \bibnamefont {Hudson}}, \bibinfo {author} {\bibfnamefont {K.~W.}\
  \bibnamefont {Chan}}, \bibinfo {author} {\bibfnamefont {W.}~\bibnamefont
  {Huang}}, \bibinfo {author} {\bibfnamefont {Kohei~M.}\ \bibnamefont {Itoh}},
  \bibinfo {author} {\bibfnamefont {R.}~\bibnamefont {Harper}}, \bibinfo
  {author} {\bibfnamefont {S.~D.}\ \bibnamefont {Bartlett}}, \bibinfo {author}
  {\bibfnamefont {C.~H.}\ \bibnamefont {Yang}}, \bibinfo {author}
  {\bibfnamefont {A.}~\bibnamefont {Laucht}}, \bibinfo {author} {\bibfnamefont
  {A.}~\bibnamefont {Saraiva}}, \bibinfo {author} {\bibfnamefont
  {T.}~\bibnamefont {Tanttu}}, \ and\ \bibinfo {author} {\bibfnamefont {A.~S.}\
  \bibnamefont {Dzurak}},\ }\href@noop {} {\enquote {\bibinfo {title}
  {{Characterizing non-Markovian Quantum Process by Fast Bayesian
  Tomography}},}\ } (\bibinfo {year} {2023}{\natexlab{b}}),\ \Eprint
  {http://arxiv.org/abs/2307.12452} {arXiv:2307.12452 [quant-ph]} \BibitemShut
  {NoStop}%
\bibitem [{\citenamefont {Proctor}\ \emph {et~al.}(2020)\citenamefont
  {Proctor}, \citenamefont {Revelle}, \citenamefont {Nielsen}, \citenamefont
  {Rudinger}, \citenamefont {Lobser}, \citenamefont {Maunz}, \citenamefont
  {Blume-Kohout},\ and\ \citenamefont {Young}}]{proctor2020detecting}%
  \BibitemOpen
  \bibfield  {author} {\bibinfo {author} {\bibfnamefont {Timothy}\ \bibnamefont
  {Proctor}}, \bibinfo {author} {\bibfnamefont {Melissa}\ \bibnamefont
  {Revelle}}, \bibinfo {author} {\bibfnamefont {Erik}\ \bibnamefont {Nielsen}},
  \bibinfo {author} {\bibfnamefont {Kenneth}\ \bibnamefont {Rudinger}},
  \bibinfo {author} {\bibfnamefont {Daniel}\ \bibnamefont {Lobser}}, \bibinfo
  {author} {\bibfnamefont {Peter}\ \bibnamefont {Maunz}}, \bibinfo {author}
  {\bibfnamefont {Robin}\ \bibnamefont {Blume-Kohout}}, \ and\ \bibinfo
  {author} {\bibfnamefont {Kevin}\ \bibnamefont {Young}},\ }\bibfield  {title}
  {\enquote {\bibinfo {title} {{Detecting and tracking drift in quantum
  information processors}},}\ }\href {\doibase 10.1038/s41467-020-19074-4}
  {\bibfield  {journal} {\bibinfo  {journal} {Nature Communications}\ }\textbf
  {\bibinfo {volume} {11}},\ \bibinfo {pages} {5396} (\bibinfo {year}
  {2020})},\ \Eprint {http://arxiv.org/abs/1907.13608} {arXiv:1907.13608}
  \BibitemShut {NoStop}%
\bibitem [{\citenamefont {Varbanov}\ \emph {et~al.}(2020)\citenamefont
  {Varbanov}, \citenamefont {Battistel}, \citenamefont {Tarasinski},
  \citenamefont {Ostroukh}, \citenamefont {O'Brien}, \citenamefont {DiCarlo},\
  and\ \citenamefont {Terhal}}]{varbanov2020leakage}%
  \BibitemOpen
  \bibfield  {author} {\bibinfo {author} {\bibfnamefont {Boris~Mihailov}\
  \bibnamefont {Varbanov}}, \bibinfo {author} {\bibfnamefont {Francesco}\
  \bibnamefont {Battistel}}, \bibinfo {author} {\bibfnamefont {Brian~Michael}\
  \bibnamefont {Tarasinski}}, \bibinfo {author} {\bibfnamefont
  {Viacheslav~Petrovych}\ \bibnamefont {Ostroukh}}, \bibinfo {author}
  {\bibfnamefont {Thomas~Eugene}\ \bibnamefont {O'Brien}}, \bibinfo {author}
  {\bibfnamefont {Leonardo}\ \bibnamefont {DiCarlo}}, \ and\ \bibinfo {author}
  {\bibfnamefont {Barbara~Maria}\ \bibnamefont {Terhal}},\ }\bibfield  {title}
  {\enquote {\bibinfo {title} {Leakage detection for a transmon-based surface
  code},}\ }\href {\doibase https://doi.org/10.1038/s41534-020-00330-w}
  {\bibfield  {journal} {\bibinfo  {journal} {npj Quantum Information}\
  }\textbf {\bibinfo {volume} {6}},\ \bibinfo {pages} {102} (\bibinfo {year}
  {2020})}\BibitemShut {NoStop}%
\bibitem [{\citenamefont {Strikis}\ \emph {et~al.}(2019)\citenamefont
  {Strikis}, \citenamefont {Datta},\ and\ \citenamefont
  {Knee}}]{strikis2019quantum}%
  \BibitemOpen
  \bibfield  {author} {\bibinfo {author} {\bibfnamefont {Armands}\ \bibnamefont
  {Strikis}}, \bibinfo {author} {\bibfnamefont {Animesh}\ \bibnamefont
  {Datta}}, \ and\ \bibinfo {author} {\bibfnamefont {George~C}\ \bibnamefont
  {Knee}},\ }\bibfield  {title} {\enquote {\bibinfo {title} {Quantum leakage
  detection using a model-independent dimension witness},}\ }\href {\doibase
  https://doi.org/10.1103/PhysRevA.99.032328} {\bibfield  {journal} {\bibinfo
  {journal} {Physical Review A}\ }\textbf {\bibinfo {volume} {99}},\ \bibinfo
  {pages} {032328} (\bibinfo {year} {2019})}\BibitemShut {NoStop}%
\bibitem [{\citenamefont {Chiribella}\ \emph {et~al.}(2010)\citenamefont
  {Chiribella}, \citenamefont {D'Ariano},\ and\ \citenamefont
  {Perinotti}}]{PhysRevA.81.062348}%
  \BibitemOpen
  \bibfield  {author} {\bibinfo {author} {\bibfnamefont {Giulio}\ \bibnamefont
  {Chiribella}}, \bibinfo {author} {\bibfnamefont {Giacomo~Mauro}\ \bibnamefont
  {D'Ariano}}, \ and\ \bibinfo {author} {\bibfnamefont {Paolo}\ \bibnamefont
  {Perinotti}},\ }\bibfield  {title} {\enquote {\bibinfo {title} {Probabilistic
  theories with purification},}\ }\href {\doibase 10.1103/PhysRevA.81.062348}
  {\bibfield  {journal} {\bibinfo  {journal} {Phys. Rev. A}\ }\textbf {\bibinfo
  {volume} {81}},\ \bibinfo {pages} {062348} (\bibinfo {year}
  {2010})}\BibitemShut {NoStop}%
\bibitem [{\citenamefont {Berk}\ \emph {et~al.}(2021)\citenamefont {Berk},
  \citenamefont {Garner}, \citenamefont {Yadin}, \citenamefont {Modi},\ and\
  \citenamefont {Pollock}}]{berk}%
  \BibitemOpen
  \bibfield  {author} {\bibinfo {author} {\bibfnamefont {Graeme~D.}\
  \bibnamefont {Berk}}, \bibinfo {author} {\bibfnamefont {Andrew J.~P.}\
  \bibnamefont {Garner}}, \bibinfo {author} {\bibfnamefont {Benjamin}\
  \bibnamefont {Yadin}}, \bibinfo {author} {\bibfnamefont {Kavan}\ \bibnamefont
  {Modi}}, \ and\ \bibinfo {author} {\bibfnamefont {Felix~A.}\ \bibnamefont
  {Pollock}},\ }\bibfield  {title} {\enquote {\bibinfo {title} {{Resource
  theories of multi-time processes: A window into quantum non-Markovianity}},}\
  }\href {\doibase 10.22331/q-2021-04-20-435} {\bibfield  {journal} {\bibinfo
  {journal} {Quantum}\ }\textbf {\bibinfo {volume} {5}},\ \bibinfo {pages}
  {435} (\bibinfo {year} {2021})}\BibitemShut {NoStop}%
\bibitem [{Note1()}]{Note1}%
  \BibitemOpen
  \bibinfo {note} {The $G_0$ gate must not simply have its target be the
  identity, it must be a perfect implementation of the identity superoperator.
  Experimentally, this amounts to doing literally nothing.}\BibitemShut {Stop}%
\bibitem [{\citenamefont {White}\ \emph
  {et~al.}(2021{\natexlab{b}})\citenamefont {White}, \citenamefont {Hill},\
  and\ \citenamefont {Hollenberg}}]{white-POST}%
  \BibitemOpen
  \bibfield  {author} {\bibinfo {author} {\bibfnamefont {G.~A.~L.}\
  \bibnamefont {White}}, \bibinfo {author} {\bibfnamefont {C.~D.}\ \bibnamefont
  {Hill}}, \ and\ \bibinfo {author} {\bibfnamefont {L.~C.~L.}\ \bibnamefont
  {Hollenberg}},\ }\bibfield  {title} {\enquote {\bibinfo {title} {Performance
  optimization for drift-robust fidelity improvement of two-qubit gates},}\
  }\href {\doibase 10.1103/PhysRevApplied.15.014023} {\bibfield  {journal}
  {\bibinfo  {journal} {Physical Review Applied}\ }\textbf {\bibinfo {volume}
  {15}},\ \bibinfo {pages} {014023} (\bibinfo {year} {2021}{\natexlab{b}})},\
  \Eprint {http://arxiv.org/abs/1911.12096} {arXiv:1911.12096} \BibitemShut
  {NoStop}%
\bibitem [{\citenamefont {Dehollain}\ \emph {et~al.}(2016)\citenamefont
  {Dehollain}, \citenamefont {Muhonen}, \citenamefont {Blume-Kohout},
  \citenamefont {Rudinger}, \citenamefont {Gamble}, \citenamefont {Nielsen},
  \citenamefont {Laucht}, \citenamefont {Simmons}, \citenamefont {Kalra},
  \citenamefont {Dzurak},\ and\ \citenamefont {Morello}}]{Dehollain_2016}%
  \BibitemOpen
  \bibfield  {author} {\bibinfo {author} {\bibfnamefont {Juan~P}\ \bibnamefont
  {Dehollain}}, \bibinfo {author} {\bibfnamefont {Juha~T}\ \bibnamefont
  {Muhonen}}, \bibinfo {author} {\bibfnamefont {Robin}\ \bibnamefont
  {Blume-Kohout}}, \bibinfo {author} {\bibfnamefont {Kenneth~M}\ \bibnamefont
  {Rudinger}}, \bibinfo {author} {\bibfnamefont {John~King}\ \bibnamefont
  {Gamble}}, \bibinfo {author} {\bibfnamefont {Erik}\ \bibnamefont {Nielsen}},
  \bibinfo {author} {\bibfnamefont {Arne}\ \bibnamefont {Laucht}}, \bibinfo
  {author} {\bibfnamefont {Stephanie}\ \bibnamefont {Simmons}}, \bibinfo
  {author} {\bibfnamefont {Rachpon}\ \bibnamefont {Kalra}}, \bibinfo {author}
  {\bibfnamefont {Andrew~S}\ \bibnamefont {Dzurak}}, \ and\ \bibinfo {author}
  {\bibfnamefont {Andrea}\ \bibnamefont {Morello}},\ }\bibfield  {title}
  {\enquote {\bibinfo {title} {Optimization of a solid-state electron spin
  qubit using gate set tomography},}\ }\href {\doibase
  10.1088/1367-2630/18/10/103018} {\bibfield  {journal} {\bibinfo  {journal}
  {New Journal of Physics}\ }\textbf {\bibinfo {volume} {18}},\ \bibinfo
  {pages} {103018} (\bibinfo {year} {2016})}\BibitemShut {NoStop}%
\bibitem [{\citenamefont {Kim}\ \emph {et~al.}(2015)\citenamefont {Kim},
  \citenamefont {Ward}, \citenamefont {Simmons}, \citenamefont {Gamble},
  \citenamefont {Blume-Kohout}, \citenamefont {Nielsen}, \citenamefont
  {Savage}, \citenamefont {Lagally}, \citenamefont {Friesen}, \citenamefont
  {Coppersmith},\ and\ \citenamefont {Eriksson}}]{kim_microwave-driven_2015}%
  \BibitemOpen
  \bibfield  {author} {\bibinfo {author} {\bibfnamefont {Dohun}\ \bibnamefont
  {Kim}}, \bibinfo {author} {\bibfnamefont {D.~R.}\ \bibnamefont {Ward}},
  \bibinfo {author} {\bibfnamefont {C.~B.}\ \bibnamefont {Simmons}}, \bibinfo
  {author} {\bibfnamefont {John~King}\ \bibnamefont {Gamble}}, \bibinfo
  {author} {\bibfnamefont {Robin}\ \bibnamefont {Blume-Kohout}}, \bibinfo
  {author} {\bibfnamefont {Erik}\ \bibnamefont {Nielsen}}, \bibinfo {author}
  {\bibfnamefont {D.~E.}\ \bibnamefont {Savage}}, \bibinfo {author}
  {\bibfnamefont {M.~G.}\ \bibnamefont {Lagally}}, \bibinfo {author}
  {\bibfnamefont {Mark}\ \bibnamefont {Friesen}}, \bibinfo {author}
  {\bibfnamefont {S.~N.}\ \bibnamefont {Coppersmith}}, \ and\ \bibinfo {author}
  {\bibfnamefont {M.~A.}\ \bibnamefont {Eriksson}},\ }\bibfield  {title}
  {\enquote {\bibinfo {title} {Microwave-driven coherent operation of a
  semiconductor quantum dot charge qubit},}\ }\href {\doibase
  10.1038/nnano.2014.336} {\bibfield  {journal} {\bibinfo  {journal} {Nature
  Nanotechnology}\ }\textbf {\bibinfo {volume} {10}},\ \bibinfo {pages}
  {243--247} (\bibinfo {year} {2015})}\BibitemShut {NoStop}%
\bibitem [{\citenamefont {Taranto}\ \emph {et~al.}(2019)\citenamefont
  {Taranto}, \citenamefont {Pollock}, \citenamefont {Milz}, \citenamefont
  {Tomamichel},\ and\ \citenamefont {Modi}}]{taranto1}%
  \BibitemOpen
  \bibfield  {author} {\bibinfo {author} {\bibfnamefont {Philip}\ \bibnamefont
  {Taranto}}, \bibinfo {author} {\bibfnamefont {Felix~A}\ \bibnamefont
  {Pollock}}, \bibinfo {author} {\bibfnamefont {Simon}\ \bibnamefont {Milz}},
  \bibinfo {author} {\bibfnamefont {Marco}\ \bibnamefont {Tomamichel}}, \ and\
  \bibinfo {author} {\bibfnamefont {Kavan}\ \bibnamefont {Modi}},\ }\bibfield
  {title} {\enquote {\bibinfo {title} {{Quantum Markov Order}},}\ }\href
  {\doibase 10.1103/PhysRevLett.122.140401} {\bibfield  {journal} {\bibinfo
  {journal} {Physical Review Letters}\ }\textbf {\bibinfo {volume} {122}},\
  \bibinfo {pages} {140401} (\bibinfo {year} {2019})}\BibitemShut {NoStop}%
\bibitem [{\citenamefont {Kingma}\ and\ \citenamefont
  {Ba}(2017)}]{kingma2014adam}%
  \BibitemOpen
  \bibfield  {author} {\bibinfo {author} {\bibfnamefont {Diederik~P.}\
  \bibnamefont {Kingma}}\ and\ \bibinfo {author} {\bibfnamefont {Jimmy}\
  \bibnamefont {Ba}},\ }\href@noop {} {\enquote {\bibinfo {title} {Adam: A
  method for stochastic optimization},}\ } (\bibinfo {year} {2017}),\ \Eprint
  {http://arxiv.org/abs/1412.6980} {arXiv:1412.6980 [cs.LG]} \BibitemShut
  {NoStop}%
\bibitem [{\citenamefont {Gray}(2018)}]{quimb}%
  \BibitemOpen
  \bibfield  {author} {\bibinfo {author} {\bibfnamefont {Johnnie}\ \bibnamefont
  {Gray}},\ }\bibfield  {title} {\enquote {\bibinfo {title} {quimb: A python
  package for quantum information and many-body calculations},}\ }\href
  {\doibase 10.21105/joss.00819} {\bibfield  {journal} {\bibinfo  {journal}
  {Journal of Open Source Software}\ }\textbf {\bibinfo {volume} {3}},\
  \bibinfo {pages} {819} (\bibinfo {year} {2018})}\BibitemShut {NoStop}%
\bibitem [{\citenamefont {Bradbury}\ \emph {et~al.}(2018)\citenamefont
  {Bradbury}, \citenamefont {Frostig}, \citenamefont {Hawkins}, \citenamefont
  {Johnson}, \citenamefont {Leary}, \citenamefont {Maclaurin}, \citenamefont
  {Necula}, \citenamefont {Paszke}, \citenamefont {Vander{P}las}, \citenamefont
  {Wanderman-{M}ilne},\ and\ \citenamefont {Zhang}}]{jax2018github}%
  \BibitemOpen
  \bibfield  {author} {\bibinfo {author} {\bibfnamefont {James}\ \bibnamefont
  {Bradbury}}, \bibinfo {author} {\bibfnamefont {Roy}\ \bibnamefont {Frostig}},
  \bibinfo {author} {\bibfnamefont {Peter}\ \bibnamefont {Hawkins}}, \bibinfo
  {author} {\bibfnamefont {Matthew~James}\ \bibnamefont {Johnson}}, \bibinfo
  {author} {\bibfnamefont {Chris}\ \bibnamefont {Leary}}, \bibinfo {author}
  {\bibfnamefont {Dougal}\ \bibnamefont {Maclaurin}}, \bibinfo {author}
  {\bibfnamefont {George}\ \bibnamefont {Necula}}, \bibinfo {author}
  {\bibfnamefont {Adam}\ \bibnamefont {Paszke}}, \bibinfo {author}
  {\bibfnamefont {Jake}\ \bibnamefont {Vander{P}las}}, \bibinfo {author}
  {\bibfnamefont {Skye}\ \bibnamefont {Wanderman-{M}ilne}}, \ and\ \bibinfo
  {author} {\bibfnamefont {Qiao}\ \bibnamefont {Zhang}},\ }\href
  {http://github.com/google/jax} {\enquote {\bibinfo {title} {{JAX}: composable
  transformations of {P}ython+{N}um{P}y programs},}\ } (\bibinfo {year}
  {2018})\BibitemShut {NoStop}%
\bibitem [{\citenamefont {Werschnik}\ and\ \citenamefont
  {Gross}(2007)}]{werschnik2007quantum}%
  \BibitemOpen
  \bibfield  {author} {\bibinfo {author} {\bibfnamefont {J}~\bibnamefont
  {Werschnik}}\ and\ \bibinfo {author} {\bibfnamefont {EKU}\ \bibnamefont
  {Gross}},\ }\bibfield  {title} {\enquote {\bibinfo {title} {Quantum optimal
  control theory},}\ }\href@noop {} {\bibfield  {journal} {\bibinfo  {journal}
  {Journal of Physics B: Atomic, Molecular and Optical Physics}\ }\textbf
  {\bibinfo {volume} {40}},\ \bibinfo {pages} {R175} (\bibinfo {year}
  {2007})}\BibitemShut {NoStop}%
\bibitem [{\citenamefont {Dong}\ \emph {et~al.}(2021)\citenamefont {Dong},
  \citenamefont {Zhuang}, \citenamefont {Economou},\ and\ \citenamefont
  {Barnes}}]{PRXQuantum.2.030333}%
  \BibitemOpen
  \bibfield  {author} {\bibinfo {author} {\bibfnamefont {Wenzheng}\
  \bibnamefont {Dong}}, \bibinfo {author} {\bibfnamefont {Fei}\ \bibnamefont
  {Zhuang}}, \bibinfo {author} {\bibfnamefont {Sophia~E.}\ \bibnamefont
  {Economou}}, \ and\ \bibinfo {author} {\bibfnamefont {Edwin}\ \bibnamefont
  {Barnes}},\ }\bibfield  {title} {\enquote {\bibinfo {title} {Doubly geometric
  quantum control},}\ }\href {\doibase 10.1103/PRXQuantum.2.030333} {\bibfield
  {journal} {\bibinfo  {journal} {PRX Quantum}\ }\textbf {\bibinfo {volume}
  {2}},\ \bibinfo {pages} {030333} (\bibinfo {year} {2021})}\BibitemShut
  {NoStop}%
\bibitem [{\citenamefont {Veitia}\ and\ \citenamefont {van
  Enk}(2020)}]{veitia2020testing}%
  \BibitemOpen
  \bibfield  {author} {\bibinfo {author} {\bibfnamefont {Andrzej}\ \bibnamefont
  {Veitia}}\ and\ \bibinfo {author} {\bibfnamefont {Steven~J.}\ \bibnamefont
  {van Enk}},\ }\href@noop {} {\enquote {\bibinfo {title} {Testing the
  context-independence of quantum gates},}\ } (\bibinfo {year} {2020}),\
  \Eprint {http://arxiv.org/abs/1810.05945} {arXiv:1810.05945 [quant-ph]}
  \BibitemShut {NoStop}%
\bibitem [{\citenamefont {Rudinger}\ \emph {et~al.}(2019)\citenamefont
  {Rudinger}, \citenamefont {Proctor}, \citenamefont {Langharst}, \citenamefont
  {Sarovar}, \citenamefont {Young},\ and\ \citenamefont
  {Blume-Kohout}}]{PhysRevX.9.021045}%
  \BibitemOpen
  \bibfield  {author} {\bibinfo {author} {\bibfnamefont {Kenneth}\ \bibnamefont
  {Rudinger}}, \bibinfo {author} {\bibfnamefont {Timothy}\ \bibnamefont
  {Proctor}}, \bibinfo {author} {\bibfnamefont {Dylan}\ \bibnamefont
  {Langharst}}, \bibinfo {author} {\bibfnamefont {Mohan}\ \bibnamefont
  {Sarovar}}, \bibinfo {author} {\bibfnamefont {Kevin}\ \bibnamefont {Young}},
  \ and\ \bibinfo {author} {\bibfnamefont {Robin}\ \bibnamefont
  {Blume-Kohout}},\ }\bibfield  {title} {\enquote {\bibinfo {title} {Probing
  context-dependent errors in quantum processors},}\ }\href {\doibase
  10.1103/PhysRevX.9.021045} {\bibfield  {journal} {\bibinfo  {journal} {Phys.
  Rev. X}\ }\textbf {\bibinfo {volume} {9}},\ \bibinfo {pages} {021045}
  (\bibinfo {year} {2019})}\BibitemShut {NoStop}%
\bibitem [{\citenamefont {Babu}\ \emph {et~al.}(2023)\citenamefont {Babu},
  \citenamefont {Orell}, \citenamefont {Vadimov}, \citenamefont {Teixeira},
  \citenamefont {M\"ott\"onen},\ and\ \citenamefont
  {Silveri}}]{PhysRevResearch.5.043161}%
  \BibitemOpen
  \bibfield  {author} {\bibinfo {author} {\bibfnamefont {Aravind~P.}\
  \bibnamefont {Babu}}, \bibinfo {author} {\bibfnamefont {Tuure}\ \bibnamefont
  {Orell}}, \bibinfo {author} {\bibfnamefont {Vasilii}\ \bibnamefont
  {Vadimov}}, \bibinfo {author} {\bibfnamefont {Wallace}\ \bibnamefont
  {Teixeira}}, \bibinfo {author} {\bibfnamefont {Mikko}\ \bibnamefont
  {M\"ott\"onen}}, \ and\ \bibinfo {author} {\bibfnamefont {Matti}\
  \bibnamefont {Silveri}},\ }\bibfield  {title} {\enquote {\bibinfo {title}
  {Quantum error correction under numerically exact open-quantum-system
  dynamics},}\ }\href {\doibase 10.1103/PhysRevResearch.5.043161} {\bibfield
  {journal} {\bibinfo  {journal} {Phys. Rev. Res.}\ }\textbf {\bibinfo {volume}
  {5}},\ \bibinfo {pages} {043161} (\bibinfo {year} {2023})}\BibitemShut
  {NoStop}%
\bibitem [{\citenamefont {Strikis}\ \emph {et~al.}(2021)\citenamefont
  {Strikis}, \citenamefont {Qin}, \citenamefont {Chen}, \citenamefont
  {Benjamin},\ and\ \citenamefont {Li}}]{PRXQuantum.2.040330}%
  \BibitemOpen
  \bibfield  {author} {\bibinfo {author} {\bibfnamefont {Armands}\ \bibnamefont
  {Strikis}}, \bibinfo {author} {\bibfnamefont {Dayue}\ \bibnamefont {Qin}},
  \bibinfo {author} {\bibfnamefont {Yanzhu}\ \bibnamefont {Chen}}, \bibinfo
  {author} {\bibfnamefont {Simon~C.}\ \bibnamefont {Benjamin}}, \ and\ \bibinfo
  {author} {\bibfnamefont {Ying}\ \bibnamefont {Li}},\ }\bibfield  {title}
  {\enquote {\bibinfo {title} {Learning-based quantum error mitigation},}\
  }\href {\doibase 10.1103/PRXQuantum.2.040330} {\bibfield  {journal} {\bibinfo
   {journal} {PRX Quantum}\ }\textbf {\bibinfo {volume} {2}},\ \bibinfo {pages}
  {040330} (\bibinfo {year} {2021})}\BibitemShut {NoStop}%
\bibitem [{\citenamefont {Quek}\ \emph {et~al.}(2023)\citenamefont {Quek},
  \citenamefont {França}, \citenamefont {Khatri}, \citenamefont {Meyer},\ and\
  \citenamefont {Eisert}}]{quek2023exponentially}%
  \BibitemOpen
  \bibfield  {author} {\bibinfo {author} {\bibfnamefont {Yihui}\ \bibnamefont
  {Quek}}, \bibinfo {author} {\bibfnamefont {Daniel~Stilck}\ \bibnamefont
  {França}}, \bibinfo {author} {\bibfnamefont {Sumeet}\ \bibnamefont
  {Khatri}}, \bibinfo {author} {\bibfnamefont {Johannes~Jakob}\ \bibnamefont
  {Meyer}}, \ and\ \bibinfo {author} {\bibfnamefont {Jens}\ \bibnamefont
  {Eisert}},\ }\href@noop {} {\enquote {\bibinfo {title} {Exponentially tighter
  bounds on limitations of quantum error mitigation},}\ } (\bibinfo {year}
  {2023}),\ \Eprint {http://arxiv.org/abs/2210.11505} {arXiv:2210.11505
  [quant-ph]} \BibitemShut {NoStop}%
\bibitem [{\citenamefont {Chen}\ \emph {et~al.}(2024)\citenamefont {Chen},
  \citenamefont {Zhang}, \citenamefont {Jiang},\ and\ \citenamefont
  {Flammia}}]{chen2024efficientselfconsistentlearninggate}%
  \BibitemOpen
  \bibfield  {author} {\bibinfo {author} {\bibfnamefont {Senrui}\ \bibnamefont
  {Chen}}, \bibinfo {author} {\bibfnamefont {Zhihan}\ \bibnamefont {Zhang}},
  \bibinfo {author} {\bibfnamefont {Liang}\ \bibnamefont {Jiang}}, \ and\
  \bibinfo {author} {\bibfnamefont {Steven~T.}\ \bibnamefont {Flammia}},\
  }\href {https://arxiv.org/abs/2410.03906} {\enquote {\bibinfo {title}
  {Efficient self-consistent learning of gate set pauli noise},}\ } (\bibinfo
  {year} {2024}),\ \Eprint {http://arxiv.org/abs/2410.03906} {arXiv:2410.03906
  [quant-ph]} \BibitemShut {NoStop}%
\bibitem [{\citenamefont {Hockings}\ \emph {et~al.}(2024)\citenamefont
  {Hockings}, \citenamefont {Doherty},\ and\ \citenamefont
  {Harper}}]{hockings2024scalablenoisecharacterisationsyndrome}%
  \BibitemOpen
  \bibfield  {author} {\bibinfo {author} {\bibfnamefont {Evan~T.}\ \bibnamefont
  {Hockings}}, \bibinfo {author} {\bibfnamefont {Andrew~C.}\ \bibnamefont
  {Doherty}}, \ and\ \bibinfo {author} {\bibfnamefont {Robin}\ \bibnamefont
  {Harper}},\ }\href {https://arxiv.org/abs/2404.06545} {\enquote {\bibinfo
  {title} {Scalable noise characterisation of syndrome extraction circuits with
  averaged circuit eigenvalue sampling},}\ } (\bibinfo {year} {2024}),\ \Eprint
  {http://arxiv.org/abs/2404.06545} {arXiv:2404.06545 [quant-ph]} \BibitemShut
  {NoStop}%
\bibitem [{\citenamefont {Evans}\ \emph {et~al.}(2022)\citenamefont {Evans},
  \citenamefont {Huang}, \citenamefont {Yoneda}, \citenamefont {Harper},
  \citenamefont {Tanttu}, \citenamefont {Chan}, \citenamefont {Hudson},
  \citenamefont {Itoh}, \citenamefont {Saraiva}, \citenamefont {Yang},
  \citenamefont {Dzurak},\ and\ \citenamefont
  {Bartlett}}]{PhysRevApplied.17.024068}%
  \BibitemOpen
  \bibfield  {author} {\bibinfo {author} {\bibfnamefont {T.J.}\ \bibnamefont
  {Evans}}, \bibinfo {author} {\bibfnamefont {W.}~\bibnamefont {Huang}},
  \bibinfo {author} {\bibfnamefont {J.}~\bibnamefont {Yoneda}}, \bibinfo
  {author} {\bibfnamefont {R.}~\bibnamefont {Harper}}, \bibinfo {author}
  {\bibfnamefont {T.}~\bibnamefont {Tanttu}}, \bibinfo {author} {\bibfnamefont
  {K.W.}\ \bibnamefont {Chan}}, \bibinfo {author} {\bibfnamefont {F.E.}\
  \bibnamefont {Hudson}}, \bibinfo {author} {\bibfnamefont {K.M.}\ \bibnamefont
  {Itoh}}, \bibinfo {author} {\bibfnamefont {A.}~\bibnamefont {Saraiva}},
  \bibinfo {author} {\bibfnamefont {C.H.}\ \bibnamefont {Yang}}, \bibinfo
  {author} {\bibfnamefont {A.S.}\ \bibnamefont {Dzurak}}, \ and\ \bibinfo
  {author} {\bibfnamefont {S.D.}\ \bibnamefont {Bartlett}},\ }\bibfield
  {title} {\enquote {\bibinfo {title} {Fast bayesian tomography of a two-qubit
  gate set in silicon},}\ }\href {\doibase 10.1103/PhysRevApplied.17.024068}
  {\bibfield  {journal} {\bibinfo  {journal} {Phys. Rev. Appl.}\ }\textbf
  {\bibinfo {volume} {17}},\ \bibinfo {pages} {024068} (\bibinfo {year}
  {2022})}\BibitemShut {NoStop}%
\bibitem [{\citenamefont {Brieger}\ \emph {et~al.}(2023)\citenamefont
  {Brieger}, \citenamefont {Roth},\ and\ \citenamefont
  {Kliesch}}]{Brieger_2023}%
  \BibitemOpen
  \bibfield  {author} {\bibinfo {author} {\bibfnamefont {Raphael}\ \bibnamefont
  {Brieger}}, \bibinfo {author} {\bibfnamefont {Ingo}\ \bibnamefont {Roth}}, \
  and\ \bibinfo {author} {\bibfnamefont {Martin}\ \bibnamefont {Kliesch}},\
  }\bibfield  {title} {\enquote {\bibinfo {title} {Compressive gate set
  tomography},}\ }\href {\doibase 10.1103/PRXQuantum.4.010325} {\bibfield
  {journal} {\bibinfo  {journal} {PRX Quantum}\ }\textbf {\bibinfo {volume}
  {4}},\ \bibinfo {pages} {010325} (\bibinfo {year} {2023})}\BibitemShut
  {NoStop}%
\bibitem [{\citenamefont {Gentile}\ \emph {et~al.}(2021)\citenamefont
  {Gentile}, \citenamefont {Flynn}, \citenamefont {Knauer}, \citenamefont
  {Wiebe}, \citenamefont {Paesani}, \citenamefont {Granade}, \citenamefont
  {Rarity}, \citenamefont {Santagati},\ and\ \citenamefont
  {Laing}}]{gentile2021learning}%
  \BibitemOpen
  \bibfield  {author} {\bibinfo {author} {\bibfnamefont {Antonio~A}\
  \bibnamefont {Gentile}}, \bibinfo {author} {\bibfnamefont {Brian}\
  \bibnamefont {Flynn}}, \bibinfo {author} {\bibfnamefont {Sebastian}\
  \bibnamefont {Knauer}}, \bibinfo {author} {\bibfnamefont {Nathan}\
  \bibnamefont {Wiebe}}, \bibinfo {author} {\bibfnamefont {Stefano}\
  \bibnamefont {Paesani}}, \bibinfo {author} {\bibfnamefont {Christopher~E}\
  \bibnamefont {Granade}}, \bibinfo {author} {\bibfnamefont {John~G}\
  \bibnamefont {Rarity}}, \bibinfo {author} {\bibfnamefont {Raffaele}\
  \bibnamefont {Santagati}}, \ and\ \bibinfo {author} {\bibfnamefont {Anthony}\
  \bibnamefont {Laing}},\ }\bibfield  {title} {\enquote {\bibinfo {title}
  {Learning models of quantum systems from experiments},}\ }\href {\doibase
  https://doi.org/10.1038/s41567-021-01201-7} {\bibfield  {journal} {\bibinfo
  {journal} {Nature Physics}\ }\textbf {\bibinfo {volume} {17}},\ \bibinfo
  {pages} {837--843} (\bibinfo {year} {2021})}\BibitemShut {NoStop}%
\bibitem [{\citenamefont {Meyer}\ \emph {et~al.}(2023)\citenamefont {Meyer},
  \citenamefont {Khatri}, \citenamefont {França}, \citenamefont {Eisert},\
  and\ \citenamefont {Faist}}]{meyer2023quantum}%
  \BibitemOpen
  \bibfield  {author} {\bibinfo {author} {\bibfnamefont {Johannes~Jakob}\
  \bibnamefont {Meyer}}, \bibinfo {author} {\bibfnamefont {Sumeet}\
  \bibnamefont {Khatri}}, \bibinfo {author} {\bibfnamefont {Daniel~Stilck}\
  \bibnamefont {França}}, \bibinfo {author} {\bibfnamefont {Jens}\
  \bibnamefont {Eisert}}, \ and\ \bibinfo {author} {\bibfnamefont {Philippe}\
  \bibnamefont {Faist}},\ }\href@noop {} {\enquote {\bibinfo {title} {Quantum
  metrology in the finite-sample regime},}\ } (\bibinfo {year} {2023}),\
  \Eprint {http://arxiv.org/abs/2307.06370} {arXiv:2307.06370 [quant-ph]}
  \BibitemShut {NoStop}%
\bibitem [{\citenamefont {Resch}\ and\ \citenamefont
  {Karpuzcu}(2021)}]{resch2021benchmarking}%
  \BibitemOpen
  \bibfield  {author} {\bibinfo {author} {\bibfnamefont {Salonik}\ \bibnamefont
  {Resch}}\ and\ \bibinfo {author} {\bibfnamefont {Ulya~R}\ \bibnamefont
  {Karpuzcu}},\ }\bibfield  {title} {\enquote {\bibinfo {title} {Benchmarking
  quantum computers and the impact of quantum noise},}\ }\href {\doibase
  https://doi.org/10.1145/3464420} {\bibfield  {journal} {\bibinfo  {journal}
  {ACM Computing Surveys (CSUR)}\ }\textbf {\bibinfo {volume} {54}},\ \bibinfo
  {pages} {142} (\bibinfo {year} {2021})}\BibitemShut {NoStop}%
\bibitem [{\citenamefont {Krantz}\ \emph {et~al.}(2019)\citenamefont {Krantz},
  \citenamefont {Kjaergaard}, \citenamefont {Yan}, \citenamefont {Orlando},
  \citenamefont {Gustavsson},\ and\ \citenamefont
  {Oliver}}]{krantz2019quantum}%
  \BibitemOpen
  \bibfield  {author} {\bibinfo {author} {\bibfnamefont {P.}~\bibnamefont
  {Krantz}}, \bibinfo {author} {\bibfnamefont {M.}~\bibnamefont {Kjaergaard}},
  \bibinfo {author} {\bibfnamefont {F.}~\bibnamefont {Yan}}, \bibinfo {author}
  {\bibfnamefont {T.~P.}\ \bibnamefont {Orlando}}, \bibinfo {author}
  {\bibfnamefont {S.}~\bibnamefont {Gustavsson}}, \ and\ \bibinfo {author}
  {\bibfnamefont {W.~D.}\ \bibnamefont {Oliver}},\ }\bibfield  {title}
  {\enquote {\bibinfo {title} {A quantum engineer's guide to superconducting
  qubits},}\ }\href {\doibase 10.1063/1.5089550} {\bibfield  {journal}
  {\bibinfo  {journal} {Applied Physics Reviews}\ }\textbf {\bibinfo {volume}
  {6}},\ \bibinfo {pages} {021318} (\bibinfo {year} {2019})}\BibitemShut
  {NoStop}%
\bibitem [{\citenamefont {Wei}\ \emph {et~al.}(2022{\natexlab{b}})\citenamefont
  {Wei}, \citenamefont {Magesan}, \citenamefont {Lauer}, \citenamefont
  {Srinivasan}, \citenamefont {Bogorin}, \citenamefont {Carnevale},
  \citenamefont {Keefe}, \citenamefont {Kim}, \citenamefont {Klaus},
  \citenamefont {Landers}, \citenamefont {Sundaresan}, \citenamefont {Wang},
  \citenamefont {Zhang}, \citenamefont {Steffen}, \citenamefont {Dial},
  \citenamefont {McKay},\ and\ \citenamefont
  {Kandala}}]{PhysRevLett.129.060501}%
  \BibitemOpen
  \bibfield  {author} {\bibinfo {author} {\bibfnamefont {K.~X.}\ \bibnamefont
  {Wei}}, \bibinfo {author} {\bibfnamefont {E.}~\bibnamefont {Magesan}},
  \bibinfo {author} {\bibfnamefont {I.}~\bibnamefont {Lauer}}, \bibinfo
  {author} {\bibfnamefont {S.}~\bibnamefont {Srinivasan}}, \bibinfo {author}
  {\bibfnamefont {D.~F.}\ \bibnamefont {Bogorin}}, \bibinfo {author}
  {\bibfnamefont {S.}~\bibnamefont {Carnevale}}, \bibinfo {author}
  {\bibfnamefont {G.~A.}\ \bibnamefont {Keefe}}, \bibinfo {author}
  {\bibfnamefont {Y.}~\bibnamefont {Kim}}, \bibinfo {author} {\bibfnamefont
  {D.}~\bibnamefont {Klaus}}, \bibinfo {author} {\bibfnamefont
  {W.}~\bibnamefont {Landers}}, \bibinfo {author} {\bibfnamefont
  {N.}~\bibnamefont {Sundaresan}}, \bibinfo {author} {\bibfnamefont
  {C.}~\bibnamefont {Wang}}, \bibinfo {author} {\bibfnamefont {E.~J.}\
  \bibnamefont {Zhang}}, \bibinfo {author} {\bibfnamefont {M.}~\bibnamefont
  {Steffen}}, \bibinfo {author} {\bibfnamefont {O.~E.}\ \bibnamefont {Dial}},
  \bibinfo {author} {\bibfnamefont {D.~C.}\ \bibnamefont {McKay}}, \ and\
  \bibinfo {author} {\bibfnamefont {A.}~\bibnamefont {Kandala}},\ }\bibfield
  {title} {\enquote {\bibinfo {title} {Hamiltonian engineering with multicolor
  drives for fast entangling gates and quantum crosstalk cancellation},}\
  }\href {\doibase 10.1103/PhysRevLett.129.060501} {\bibfield  {journal}
  {\bibinfo  {journal} {Phys. Rev. Lett.}\ }\textbf {\bibinfo {volume} {129}},\
  \bibinfo {pages} {060501} (\bibinfo {year} {2022}{\natexlab{b}})}\BibitemShut
  {NoStop}%
\bibitem [{\citenamefont {M\o{}lmer}\ and\ \citenamefont
  {S\o{}rensen}(1999)}]{PhysRevLett.82.1835}%
  \BibitemOpen
  \bibfield  {author} {\bibinfo {author} {\bibfnamefont {Klaus}\ \bibnamefont
  {M\o{}lmer}}\ and\ \bibinfo {author} {\bibfnamefont {Anders}\ \bibnamefont
  {S\o{}rensen}},\ }\bibfield  {title} {\enquote {\bibinfo {title}
  {Multiparticle entanglement of hot trapped ions},}\ }\href {\doibase
  10.1103/PhysRevLett.82.1835} {\bibfield  {journal} {\bibinfo  {journal}
  {Phys. Rev. Lett.}\ }\textbf {\bibinfo {volume} {82}},\ \bibinfo {pages}
  {1835--1838} (\bibinfo {year} {1999})}\BibitemShut {NoStop}%
\bibitem [{\citenamefont {Haferkamp}\ \emph {et~al.}(2020)\citenamefont
  {Haferkamp}, \citenamefont {Hangleiter}, \citenamefont {Eisert},\ and\
  \citenamefont {Gluza}}]{PhysRevResearch.2.013010}%
  \BibitemOpen
  \bibfield  {author} {\bibinfo {author} {\bibfnamefont {Jonas}\ \bibnamefont
  {Haferkamp}}, \bibinfo {author} {\bibfnamefont {Dominik}\ \bibnamefont
  {Hangleiter}}, \bibinfo {author} {\bibfnamefont {Jens}\ \bibnamefont
  {Eisert}}, \ and\ \bibinfo {author} {\bibfnamefont {Marek}\ \bibnamefont
  {Gluza}},\ }\bibfield  {title} {\enquote {\bibinfo {title} {Contracting
  projected entangled pair states is average-case hard},}\ }\href {\doibase
  10.1103/PhysRevResearch.2.013010} {\bibfield  {journal} {\bibinfo  {journal}
  {Phys. Rev. Res.}\ }\textbf {\bibinfo {volume} {2}},\ \bibinfo {pages}
  {013010} (\bibinfo {year} {2020})}\BibitemShut {NoStop}%
\bibitem [{\citenamefont {Daniel}\ and\ \citenamefont
  {Gray}(2018)}]{daniel2018opt}%
  \BibitemOpen
  \bibfield  {author} {\bibinfo {author} {\bibfnamefont {G}~\bibnamefont
  {Daniel}}\ and\ \bibinfo {author} {\bibfnamefont {Johnnie}\ \bibnamefont
  {Gray}},\ }\bibfield  {title} {\enquote {\bibinfo {title} {Opt\_einsum: a
  python package for optimizing contraction order for einsum-like
  expressions},}\ }\href {\doibase https://doi.org/10.21105/joss.00753}
  {\bibfield  {journal} {\bibinfo  {journal} {Journal of Open Source Software}\
  }\textbf {\bibinfo {volume} {3}},\ \bibinfo {pages} {753} (\bibinfo {year}
  {2018})}\BibitemShut {NoStop}%
\end{thebibliography}
\end{document}